\documentclass[aps,prd,reprint,superscriptaddress]{revtex4-1}
 \pdfoutput=1

\usepackage{url}
\usepackage{SIunits}
\usepackage[english]{babel}
\usepackage{graphicx}
\RequirePackage{lineno}

\usepackage{subfigure}
\usepackage{amsmath}
\usepackage{enumitem}

\begin{document}

\title{Performance of two Askaryan Radio Array stations and first results in the search for ultra-high energy neutrinos}

\author{Allison, P.}
\affiliation{Dept. of Physics, The Ohio State University, 191 West Woodruff Avenue, Columbus, OH, USA}
\affiliation{Center for Cosmology and Astro-Particle Physics, The Ohio State University, 191 West Woodruff Avenue, Columbus, OH, USA}

\author{Bard, R.}
\affiliation{Dept. of Physics, Univ. of Maryland, College Park, MD, USA}
\author{Beatty, J. J.}
\affiliation{Dept. of Physics, The Ohio State University, 191 West Woodruff Avenue, Columbus, OH, USA}
\affiliation{Center for Cosmology and Astro-Particle Physics, The Ohio State University, 191 West Woodruff Avenue, Columbus, OH, USA}
\affiliation{Department of Astronomy, The Ohio State University, 4055 McPherson Laboratory, 140 West 18th Avenue, Columbus, OH, USA}
\author{Besson, D. Z.}
\affiliation{Dept. of Physics and Astronomy, Univ. of Kansas, Lawrence, KS, USA}
\affiliation{National Research Nuclear University, Moscow Engineering Physics Institute, Moscow, Russia}
\author{Bora, C.}
\affiliation{Dept. of Physics and Astronomy, Univ. of Nebraska-Lincoln, NE, USA}

\author{Chen, C.-C.}
\affiliation{Dept. of Physics, Grad. Inst. of Astrophys., \& Leung Center for Cosmology and Particle Astrophysics, National Taiwan Univ., Taipei, Taiwan}
\author{Chen, C.-H.}
\affiliation{Dept. of Physics, Grad. Inst. of Astrophys., \& Leung Center for Cosmology and Particle Astrophysics, National Taiwan Univ., Taipei, Taiwan}
\author{Chen, P.}
\affiliation{Dept. of Physics, Grad. Inst. of Astrophys., \& Leung Center for Cosmology and Particle Astrophysics, National Taiwan Univ., Taipei, Taiwan}
\author{Christenson, A.}
\affiliation{Dept. of Physics and Wisconsin IceCube Particle Astrophysics Center, University of Wisconsin, Madison, WI, USA}
\author{Connolly, A.}
\affiliation{Dept. of Physics, The Ohio State University, 191 West Woodruff Avenue, Columbus, OH, USA}
\affiliation{Center for Cosmology and Astro-Particle Physics, The Ohio State University, 191 West Woodruff Avenue, Columbus, OH, USA}

\author{Davies, J.}
\affiliation{Dept. of Physics and Astronomy, Univ. College London, London, United Kingdom}
\author{Duvernois, M.}
\affiliation{Dept. of Physics and Wisconsin IceCube Particle Astrophysics Center, University of Wisconsin, Madison, WI, USA}

\author{Fox, B.}
\affiliation{Dept. of Physics and Astronomy, Univ. of Hawaii, Manoa, HI, USA}

\author{Gaior, R.}
\affiliation{Dept. of Physics, Chiba University, Tokyo, Japan}
\author{Gorham, P. W.}
\affiliation{Dept. of Physics and Astronomy, Univ. of Hawaii, Manoa, HI, USA}

\author{Hanson, K.}
\affiliation{Universit\'{e} libre de Bruxelles - Interuniversity Institute for High Energies (IIHE), Belgium}
\author{Haugen, J.}
\affiliation{Dept. of Physics and Wisconsin IceCube Particle Astrophysics Center, University of Wisconsin, Madison, WI, USA}
\author{Hill, B.}
\affiliation{Dept. of Physics and Astronomy, Univ. of Hawaii, Manoa, HI, USA}
\author{Hoffman, K. D.}
\affiliation{Dept. of Physics, Univ. of Maryland, College Park, MD, USA}
\author{Hong, E.}
\affiliation{Dept. of Physics, The Ohio State University, 191 West Woodruff Avenue, Columbus, OH, USA}
\affiliation{Center for Cosmology and Astro-Particle Physics, The Ohio State University, 191 West Woodruff Avenue, Columbus, OH, USA}
\author{Hsu, S.-Y.}
\affiliation{Dept. of Physics, Grad. Inst. of Astrophys., \& Leung Center for Cosmology and Particle Astrophysics, National Taiwan Univ., Taipei, Taiwan}
\author{Hu, L.}
\affiliation{Dept. of Physics, Grad. Inst. of Astrophys., \& Leung Center for Cosmology and Particle Astrophysics, National Taiwan Univ., Taipei, Taiwan}
\author{Huang, J.-J.}
\affiliation{Dept. of Physics, Grad. Inst. of Astrophys., \& Leung Center for Cosmology and Particle Astrophysics, National Taiwan Univ., Taipei, Taiwan}
\author{Huang, M.-H. A.}
\affiliation{Dept. of Physics, Grad. Inst. of Astrophys., \& Leung Center for Cosmology and Particle Astrophysics, National Taiwan Univ., Taipei, Taiwan}

\author{Ishihara, A.}
\affiliation{Dept. of Physics, Chiba University, Tokyo, Japan}

\author{Karle, A.}
\affiliation{Dept. of Physics and Wisconsin IceCube Particle Astrophysics Center, University of Wisconsin, Madison, WI, USA}
\author{Kelley, J. L.}
\affiliation{Dept. of Physics and Wisconsin IceCube Particle Astrophysics Center, University of Wisconsin, Madison, WI, USA}
\author{Kennedy, D.}
\affiliation{Dept. of Physics and Astronomy, Univ. of Kansas, Lawrence, KS, USA}
\author{Kravchenko, I.}
\affiliation{Dept. of Physics and Astronomy, Univ. of Nebraska-Lincoln, NE, USA}
\author{Kuwabara, T.}
\affiliation{Dept. of Physics, Chiba University, Tokyo, Japan}

\author{Landsman, H.}
\affiliation{Weizmann Institute of Science, Rehovot, Israel}
\author{Laundrie, A.}
\affiliation{Dept. of Physics and Wisconsin IceCube Particle Astrophysics Center, University of Wisconsin, Madison, WI, USA}
\author{Li, C.-J.}
\affiliation{Dept. of Physics, Grad. Inst. of Astrophys., \& Leung Center for Cosmology and Particle Astrophysics, National Taiwan Univ., Taipei, Taiwan}
\author{Liu, T. C.}
\affiliation{Dept. of Physics, Grad. Inst. of Astrophys., \& Leung Center for Cosmology and Particle Astrophysics, National Taiwan Univ., Taipei, Taiwan}
\author{Lu, M.-Y.}
\affiliation{Dept. of Physics and Wisconsin IceCube Particle Astrophysics Center, University of Wisconsin, Madison, WI, USA}

\author{Macchiarulo, L.}
\affiliation{Dept. of Physics and Astronomy, Univ. of Hawaii, Manoa, HI, USA}
\author{Mase, K.}
\affiliation{Dept. of Physics, Chiba University, Tokyo, Japan}
\author{Meures, T.}
\email{thomas.meures@icecube.wisc.edu}
\affiliation{Universit\'{e} libre de Bruxelles - Interuniversity Institute for High Energies (IIHE), Belgium}
\author{Meyhandan, R.}
\affiliation{Dept. of Physics and Astronomy, Univ. of Hawaii, Manoa, HI, USA}
\author{Miki, C.}
\affiliation{Dept. of Physics and Astronomy, Univ. of Hawaii, Manoa, HI, USA}
\author{Morse, R.}
\affiliation{Dept. of Physics and Astronomy, Univ. of Hawaii, Manoa, HI, USA}

\author{Nam, J.}
\affiliation{Dept. of Physics, Grad. Inst. of Astrophys., \& Leung Center for Cosmology and Particle Astrophysics, National Taiwan Univ., Taipei, Taiwan}
\author{Nichol, R. J.}
\affiliation{Dept. of Physics and Astronomy, Univ. College London, London, United Kingdom}
\author{Nir, G.}
\affiliation{Weizmann Institute of Science, Rehovot, Israel}
\author{Novikov, A.}
\affiliation{National Research Nuclear University, Moscow Engineering Physics Institute, Moscow, Russia}

\author{O'Murchadha, A.}
\affiliation{Universit\'{e} libre de Bruxelles - Interuniversity Institute for High Energies (IIHE), Belgium}

\author{Pfendner, C.}
\affiliation{Dept. of Physics, The Ohio State University, 191 West Woodruff Avenue, Columbus, OH, USA}
\affiliation{Center for Cosmology and Astro-Particle Physics, The Ohio State University, 191 West Woodruff Avenue, Columbus, OH, USA}

\author{Ratzlaff, K.}
\affiliation{Instrumentation Design Laboratory, Univ. of Kansas, Lawrence, KS, USA}
\author{Relich, M.}
\affiliation{Dept. of Physics, Chiba University, Tokyo, Japan}
\author{Richman, M.}
\affiliation{Dept. of Physics, Univ. of Maryland, College Park, MD, USA}
\author{Ritter, L.}
\affiliation{Dept. of Physics and Astronomy, Univ. of Hawaii, Manoa, HI, USA}
\author{Rotter, B.}
\affiliation{Dept. of Physics and Astronomy, Univ. of Hawaii, Manoa, HI, USA}

\author{Sandstrom, P.}
\affiliation{Dept. of Physics and Wisconsin IceCube Particle Astrophysics Center, University of Wisconsin, Madison, WI, USA}
\author{Schellin, P.}
\affiliation{Dept. of Physics, The Ohio State University, 191 West Woodruff Avenue, Columbus, OH, USA}
\affiliation{Center for Cosmology and Astro-Particle Physics, The Ohio State University, 191 West Woodruff Avenue, Columbus, OH, USA}
\author{Shultz, A.}
\affiliation{Dept. of Physics and Astronomy, Univ. of Nebraska-Lincoln, NE, USA}
\author{Seckel, D.}
\affiliation{Dept. of Physics and Astronomy, Univ. of Delaware, Newark, DE, USA}
\author{Shiao, Y.-S.}
\affiliation{Dept. of Physics, Grad. Inst. of Astrophys., \& Leung Center for Cosmology and Particle Astrophysics, National Taiwan Univ., Taipei, Taiwan}
\author{Stockham, J.}
\affiliation{Dept. of Physics and Astronomy, Univ. of Kansas, Lawrence, KS, USA}
\author{Stockham, M.}
\affiliation{Dept. of Physics and Astronomy, Univ. of Kansas, Lawrence, KS, USA}

\author{Touart, J.}
\affiliation{Dept. of Physics, Univ. of Maryland, College Park, MD, USA}

\author{Varner, G. S.}
\affiliation{Dept. of Physics and Astronomy, Univ. of Hawaii, Manoa, HI, USA}

\author{Wang, M.-Z.}
\affiliation{Dept. of Physics, Grad. Inst. of Astrophys., \& Leung Center for Cosmology and Particle Astrophysics, National Taiwan Univ., Taipei, Taiwan}
\author{Wang, S.-H.}
\affiliation{Dept. of Physics, Grad. Inst. of Astrophys., \& Leung Center for Cosmology and Particle Astrophysics, National Taiwan Univ., Taipei, Taiwan}

\author{Yang, Y.}
\affiliation{Universit\'{e} libre de Bruxelles - Interuniversity Institute for High Energies (IIHE), Belgium}
\author{Yoshida, S.}
\affiliation{Dept. of Physics, Chiba University, Tokyo, Japan}
\author{Young, R.}
\affiliation{Instrumentation Design Laboratory, Univ. of Kansas, Lawrence, KS, USA}

\collaboration{The ARA collaboration}

\date{\today}

\begin{abstract}
Ultra-high energy neutrinos are interesting messenger particles since, if detected, they can transmit exclusive information about ultra-high energy processes in the Universe. These particles, with energies above $\unit{10^{16}}{\electronvolt}$, interact very rarely. Therefore, detectors that instrument several gigatons of matter are needed to discover them. The ARA detector is currently being constructed at South Pole. It is designed to use the Askaryan effect, the emission of radio waves from neutrino-induced cascades in the South Pole ice, to detect neutrino interactions at very high energies. With antennas distributed among 37 widely-separated stations in the ice, such interactions can be observed in a volume of several hundred cubic kilometers. Currently 3 deep ARA stations are deployed in the ice, of which two have been taking data since the beginning of 2013. In this publication, the ARA detector ``as-built'' and calibrations are described. Data reduction methods used to distinguish the rare radio signals from overwhelming backgrounds of thermal and anthropogenic origin are presented. Using data from only two stations over a short exposure time of $10$ months, a neutrino flux limit of $\unit{1.5\times10^{-6}}{\giga\electronvolt/\centi\meter^{2}/\second/\steradian}$ is calculated for a particle energy of $\unit{10^{18}}{\electronvolt}$, which offers promise for the full ARA detector.
\end{abstract}


\maketitle

\section{Introduction}

In 1966 Greisen, Zatsepin and Kuzmin predicted an interaction of ultra-high energy cosmic rays (UHECRs) with the recently discovered cosmic microwave background radiation \cite{Greisen1966,Zatsepin1966}.
In such interactions, pions are produced resonantly, subsequently decaying into neutrinos, as first postulated by Berezinsky and Zatsepin in 1968 \cite{Beresinsky1968}. Due to the contribution of the Delta resonance to the cross section for this interaction, UHECRs are unable to reach us from sources on cosmological distance scales, i. e. beyond tens of Mpc. This implies a sharp cutoff in the cosmic ray spectrum at an energy of around $\unit{10^{19.5}}{\electronvolt}$, which has been confirmed by the largest cosmic ray air shower detectors, Telescope Array \cite{Tinyakov2014} and the Pierre Auger Observatory \cite{Abraham2010}. However, the lack of UHECRs arriving at Earth could be due to a number of underlying reasons dependent on their properties. The mass composition, for example, plays a crucial role in determining the dominant energy loss for UHECRs \cite{Allard2006}. Furthermore, this energy loss is influenced by the distribution of sources and the primary energy spectrum of cosmic rays.

As a consequence, the neutrino flux depends strongly on all three parameters: the UHECR composition, their energy spectrum and their source distribution. Hence, a measurement of this flux can be used to place constraints on those parameters. Moreover, due to the Greisen-Zatsepin-Kuzmin (GZK) effect and similar absorption mechanisms for HE gamma rays, neutrinos are the only feasible known particles for the study of UHE sources more distant than a few tens of Mpc. Neutrinos, chargeless and only weakly interacting with extremely low cross sections, arrive at detectors unscattered and undeflected by intervening particles and fields, and may be correlated to UHE sources at the furthest distances in the Universe.

Many other UHE neutrino source models have been offered. Recent IceCube results challenge some of these models that predict larger neutrino fluxes at EeV energies, given the somewhat smaller-than-predicted number of neutrinos measured by IceCube with energies in the PeV range \cite{Aartsen:2013dsm}. Nevertheless, other theoretical models predict EeV neutrino fluxes that comply with IceCube limits, e.g. from pulsars \cite{Fang:2013vla}, blazars \cite{Murase:2014foa} or from the afterglow of gamma-ray bursts \cite{Waxman:1999ai,PhysRevD.76.123001}.  A more detailed discussion of such theoretical models will not be presented in the framework of this experimental paper, but may be found in the references.


The expected flux of GZK neutrinos at Earth from different cosmic ray models is very low \cite{AhlersMin2012} and, in combination with the low interaction cross section \cite{Gandhi1996}, leads to an interaction rate of less than 1 GZK neutrino per gigaton of matter per year. Therefore, large detectors, covering several hundred cubic kilometers of water equivalent matter are needed to record neutrino events in sufficient quantity to investigate their flux.

The large attenuation length of Antarctic ice to radiofrequency waves, of $\mathcal{O}(\unit{1}{\kilo\meter})$, opens the possibility to space detectors on a comparable scale and to utilize coherent radio emission from neutrino induced cascades in radio transparent media; the so-called Askaryan effect \cite{Askaryan1962,Askaryan1965}, which has been verified in various experiments \cite{AskaryanSand2001,AskaryanSalt2005,AskaryanIce2007}. In the interactions of high energy neutrinos with electrons or nuclear matter, electromagnetic (EM) cascades are produced which build up a net negative charge of roughly $20\%$ close to the shower maximum. This imbalance originates mainly from Compton scattering of cascade photons on atomic electrons. Smaller contributions are added by other ionizing effects such as positron annihilation with atomic electrons \cite{ZHS1992,AlvarezMuniz1997}. The net charge acts as a moving current and emits electromagnetic waves, which become coherent at wavelengths comparable to the lateral cascade dimensions. This is valid in the radio regime. In the case of coherent emission, the strength of the EM far field is proportional to the cascade energy. The frequency spectrum of the Askaryan signal depends strongly on the observation angle. As described in \cite{ZHS1992}, the strongest signal is observed at the Cherenkov angle for frequencies around $\unit{1}{\giga\hertz}$. The signal distribution around this angle can be approximated by a narrow Gaussian distribution. At lower frequencies the signal is weaker but the angular distribution is much broader. This trade-off needs to be considered when designing a neutrino detector utilizing this emission.

Askaryan radiation is a consequence of a neutrino of any flavor interacting in a radio transparent medium such as ice. However, as the effect ultimately comes from the induced EM cascade, the detection efficiency is strongly dependent on neutrino flavor. Charged-current $\nu_e$ interactions and $\nu_e+e^-$ elastic scatters convert a large fraction of the neutrino energy into the EM cascade. Neutral-current interactions of all $\nu$ flavors may initiate hadronic cascades which receive, on average, only 20\% of the neutrino energy. These hadronic cascades themselves will produce EM sub-cascades due principally to decaying $\pi^0$ particles, but a fraction of energy is lost to hadrons. On the other hand very high energy EM cascades undergo elongation via the Landau-Pomeranchuk-Midgal (LPM) effect \cite{Landau1953,Migdal1956} which does not affect hadronic cascades.  Radiation emitted by the LPM-elongated EM cascades is strongly beamed at the Cherenkov angle and thus is less likely to intersect detection antennas. Therefore, despite their stronger signal emission, $\nu_e$ are not necessarily the dominant detected flavor given an astrophysical expectation of $(\nu_e:\nu_\mu:\nu_\tau) = (1:1:1)$ and most of the acceptance is expected to arise from hadronic cascades.




\section{The instrument} 

The Askaryan Radio Array (ARA) is a neutrino detector, currently under construction at the geographic South Pole next to the IceCube experiment. It is designed to utilize the Askaryan effect to detect interactions of GZK neutrinos in the South Polar ice sheet. At the chosen site, thousands of square kilometers of ice with a thickness of about $\unit{3}{\kilo\meter}$ are available to act as a radio transparent detector medium and to allow for the construction of a $\mathcal{O}(\unit{100}{\giga\ton})$ detector. Due to its low temperature, between $\unit{-55}{\celsius}$ and  $\unit{-30}{\celsius}$ in the top $\unit{2}{\kilo\meter}$ \cite{Price11062002}, the South Pole ice sheet has low radio attenuation. On average, an attenuation length of $\unit{820}{\meter}$ integrated over the top $\unit{2}{\kilo\meter}$ of ice has been measured for frequencies around \unit{300}{\mega\hertz} \cite{Allison2012457}. Furthermore, the Amundsen-Scott station provides the infrastructure to support large projects such as the ARA experiment.

\subsection{General design}     \label{sc_generalDT}

\begin{figure}[t]
\includegraphics[width=0.9\columnwidth]{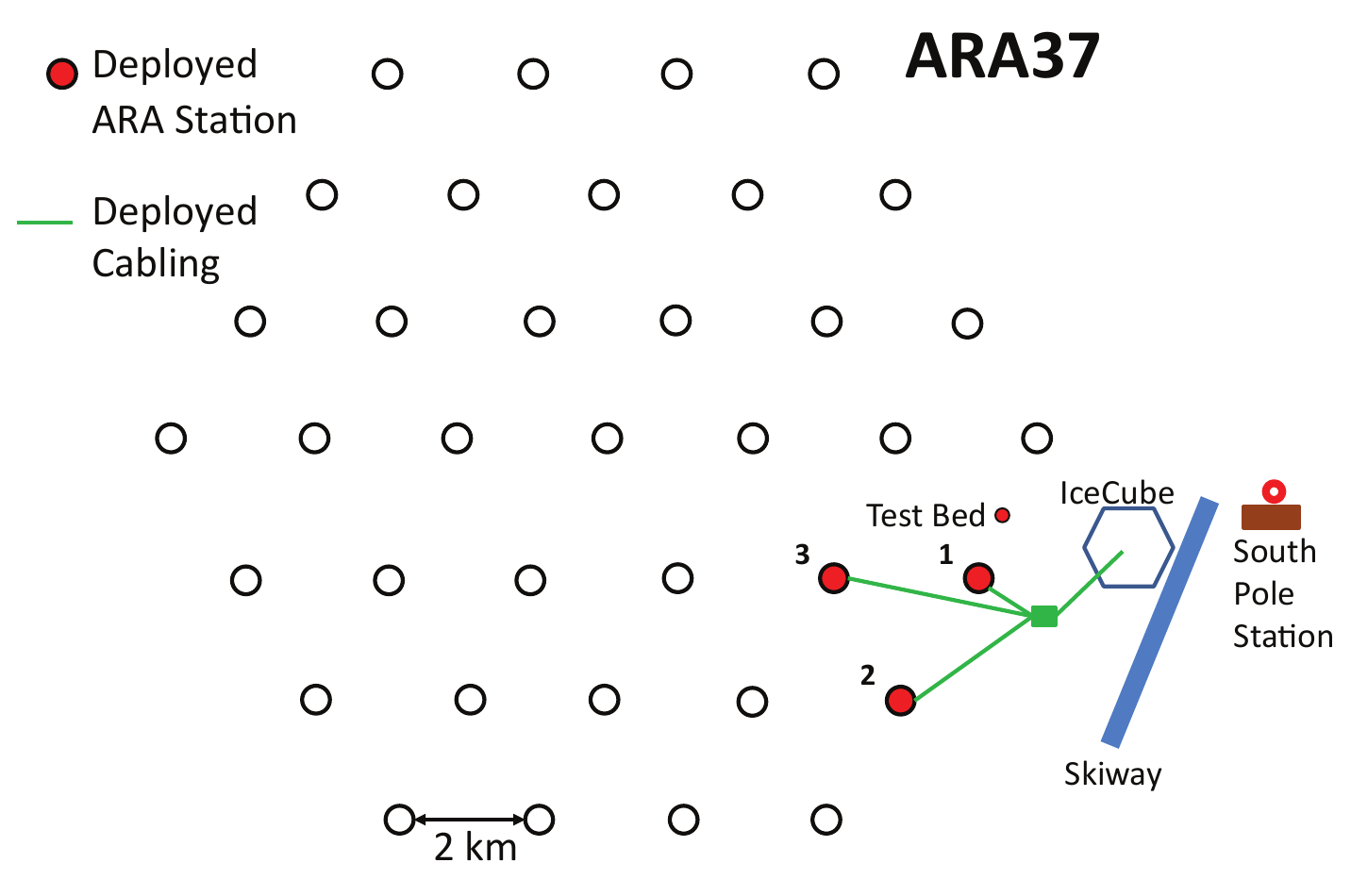} 
\caption{An area map of the planned ARA detector at South Pole. The stations are indicated by the black circles. Red filled circles denote the currently deployed stations.\label{fig_arageom}} 
\end{figure}

The ARA detector baseline consists of 37 antenna clusters (``stations'') spaced by $\unit{2}{\kilo\meter}$ in a hexagonal grid (Figure \ref{fig_arageom}). Each station is designed to operate as an autonomous neutrino detector and simulations have shown a grid spacing of $\unit{2}{\kilo\meter}$ to nearly maximize the array's effective area at neutrino energies of $\unit{10^{18}}{\electronvolt}$ \cite{Allison2012457}.

\begin{figure}[t]
\includegraphics[width=\columnwidth]{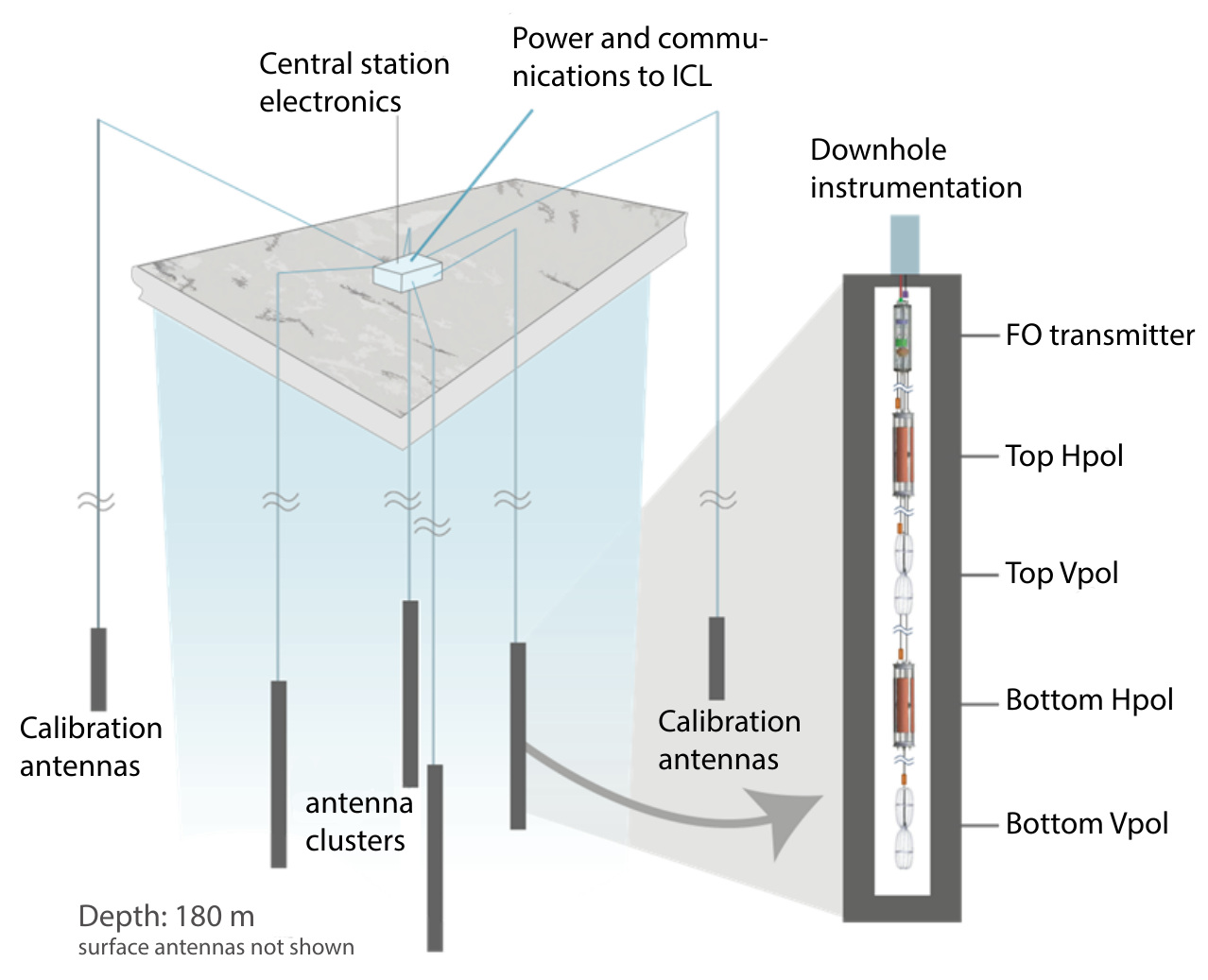} 
\caption{The baseline design of an ARA station with a zoom illustrating the string details and a view of the deployed antennas of both polarization.\label{fig_arastation}}
\end{figure}

Each station comprises 16 measurement antennas, deployed on strings in groups of 4 at the bottom of $\unit{200}{\meter}$ deep holes. In the baseline design, these antennas form a $\unit{20}{\meter} \times \unit{20}{\meter} \times \unit{20}{\meter}$ cube (Figure \ref{fig_arastation}). This design is in the process of being optimized based on analysis results from the first ARA stations and simulations. Each hole contains two antennas of horizontal and two antennas of vertical polarization, all recording data between $\unit{150}{\mega\hertz}$ and $\unit{850}{\mega\hertz}$. Two separated polarizations are chosen to be able to determine the polarization of the incoming signal, which is important for neutrino reconstruction. The antenna names are composed of the string number as \textit{D\#}, their position on the string (\textit{T} for top, \textit{B} for bottom) and their polarization (\textit{V} for vertical, \textit{H} for horizontal).

The antennas are deployed at depths between $\unit{170}{\meter}$ and $\unit{190}{\meter}$ to minimize the effects of ray-tracing in the ice. Due to the depth-dependence of the temperature and density of the South Pole ice sheet, the index of refraction changes with depth \cite{Kravchenko2004}. This effect is strongest in the top $\unit{200}{\meter}$, starting from an index of $1.35$ at the surface and changing to a value of $1.78$ for the deep ice at a depth of roughly \unit{200}{\meter}. As described in \cite{Allison2012457,TestBed2014}, this causes the path of radio rays to be bent downwards rendering vertex reconstructions difficult. Moreover, a shadowed area is produced where signals cannot reach shallowly deployed antennas, thus reducing effective neutrino volume. Therefore, a deep deployment of the antennas is favorable. 

In addition to the receiver channels, 4 calibration transmitter antennas are deployed on two extra strings, D5 and D6. These pulsers can transmit transient signals or continuous broadband noise for calibration of the station timing, geometry, and signal efficiency. Transient emissions are tied to a GPS clock which allows separation of them from other recorded RF signals by timing. The pulsers are positioned at a distance of approximately $\unit{40}{\meter}$ to the station core at a similar depth as the measurement antennas. Each hole contains one antenna of each polarization. 

The antennas used in ARA are birdcage dipoles for the vertical polarization (Vpol) and ferrite loaded quad-slot antennas for the horizontal polarization (Hpol). Given the drilled antenna holes with a diameter of only $\unit{15}{\centi\meter}$, the design of Hpol antennas with reasonable sensitivity down to $\unit{150}{\mega\hertz}$ is very challenging. Slotted copper cylinders show reasonable low-frequency performance with a voltage standing wave ratio below 3, for frequencies above $\unit{300}{\mega\hertz}$ \cite{Allison2012457}. This can be further improved by adding ferrite material in the cylinder core.
 
\begin{figure}[t]
\subfigure{
\includegraphics[height=145pt]{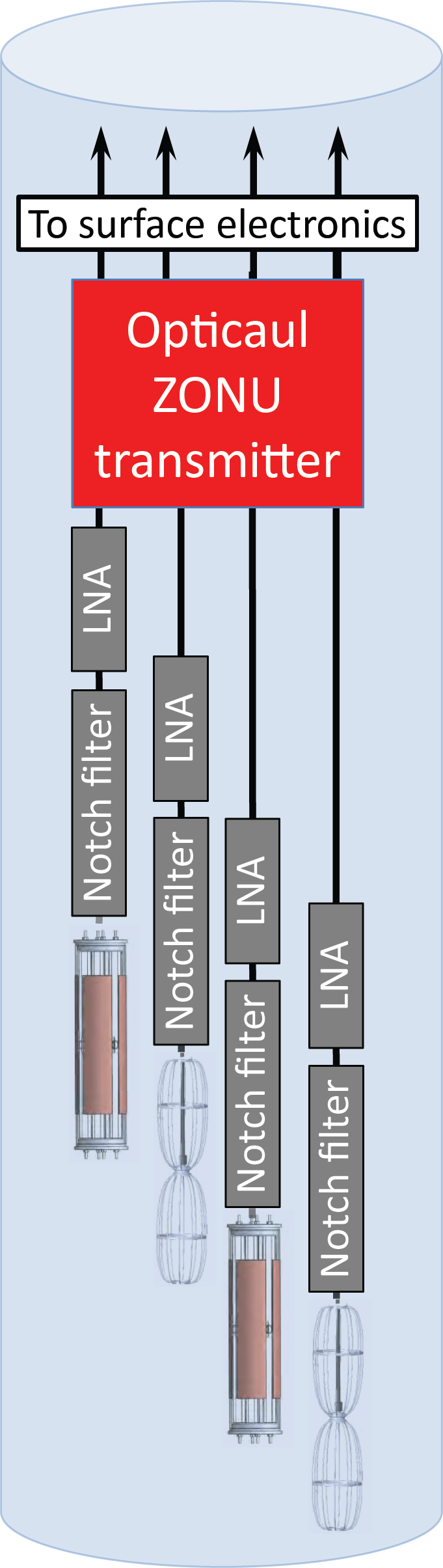} 
\label{fig_signalChain}
}
\subfigure{
\includegraphics[height=140pt]{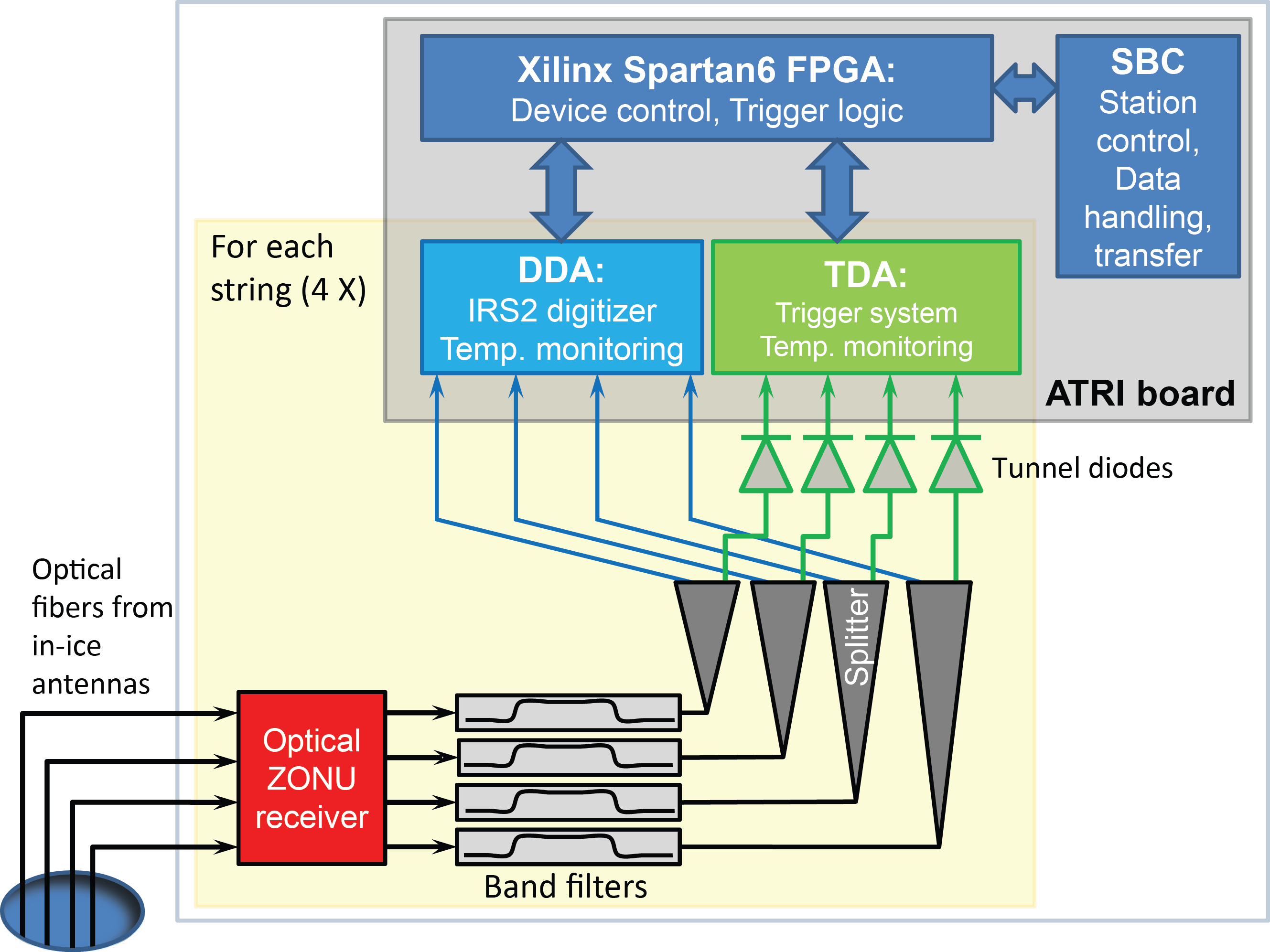}
\label{fig_DAQsceme} 
}
\caption{Left: The components of the down-hole signal chain on each string in the ARA stations. Right: The surface Data acquisition system of the ARA stations, showing the most important components. Components framed in yellow are common to all strings.}
\end{figure}

The signals recorded by the antennas are first filtered by a bandpass and notch filter, to reject frequencies out of band as well as narrow-band communications. After filtering, signals are amplified by Low Noise Amplifiers (LNAs) and transmitted analog to the surface through fiber cables via optical Zonu links (Figure~\ref{fig_signalChain}). At the surface, signals are filtered again, split and fed to the trigger system as well as to the digitization system (Figure~\ref{fig_DAQsceme}). A first calibration of the full signal chain and antenna response after deployment in the ice can be found in Section \ref{ch_scCalibration}.

In the trigger system, the signal is processed by an integrating tunnel diode, producing energy envelopes of the incoming waveforms, which can be processed in the trigger electronics mounted on the Triggering Daughter board for ARA (TDA). On this board, the signal is read into the digital electronics and logic, implemented in an FPGA. This logic then determines whether the event satisfies the trigger condition. This condition is currently a simple multiplicity trigger, requiring signal on 3 out of 8 channels of one polarization. Investigations are being performed to replace this with a smarter algorithm which provides more efficient background rejection and better signal retention.

The digitization system is located on the Digitizing Daughter board for ARA (DDA). In this system the data are sampled by the IRS2 ASIC, a digitization chip capable of sampling data at a rate of $\unit{4}{\giga S \per \second}$. In the ARA detector, the sampling speed is tuned to $\unit{3.2}{\giga S \per \second}$. The IRS2 chip contains 8 channels each with a 32k-element Switched Capacitor Array (SCA). The 32k elements are further subdivided into 512 randomly write-addressable blocks of 64 samples each. Analog sampling is continuous and is stopped by an external trigger to signal the start of digitization and readout of the analog storage blocks of interest.  Performance of this early version of SCAs with deep analog storage buffers shows promise for multichannel high-speed, low-power samplers.  Power consumption is in the range of $\unit{20}{\milli\watt}$ per channel. In principle deadtimeless operation is possible due to the deep analog buffer, however, noise issues related to simultaneous readout and digitization have prevented operation in this mode to date. The calibration of this digitizer is presented in Appendix \ref{ch_digiCal}.

Both the TDA and DDA boards are mounted on the ARA Triggering and Readout Interface (ATRI) which provides all logic for the data acquisition systems in a single Spartan-6 FPGA. This FPGA is programmed by, and exchanges its data with, an Intel atom-based Single Board Computer (SBC) which handles the data transfer to storage on disks in the IceCube Laboratory (ICL). Currently an event rate of $\unit{5}{\hertz}$ can be accommodated on a USB link between the FPGA and the SBC. In December 2015 a much higher bandwidth PCI Express bus was installed to increase the acceptable station trigger rate.

The ARA detector in its current form consists of 3 stations, of which two were taking data in the year 2013 and produced the data for this analysis.

\subsection{Performance of the ARA stations A2 and A3}   \label{sc_ARA23}

\begin{figure}[t]
\centering
\includegraphics[width=0.9\columnwidth]{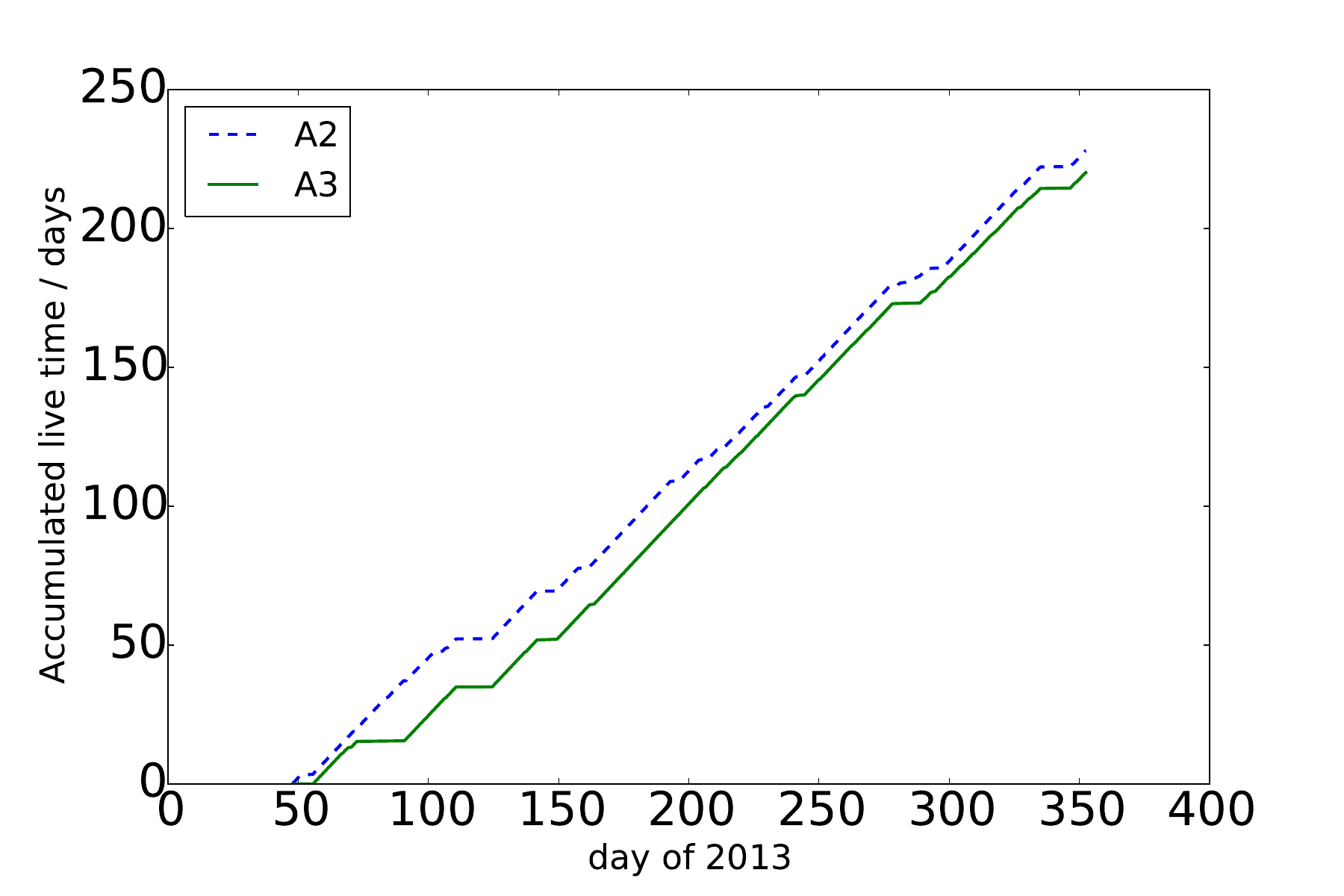} 
\caption{The cumulative live time of stations A2 and A3 in 2013. Horizontal line segments indicate extended downtimes of the detectors.}
\label{fig_ARA23Life} 
\end{figure} 

The first data from the deep ARA detector have been recorded by stations A2 and A3. Station A1 could not deliver data in the year 2013 due to an issue in the communications system. This problem was repaired in the 2015-2016 Antarctic summer season and all 3 stations are now operational.

The positions of A2 and A3, as embedded in the full ARA37 design, are shown in Figure \ref{fig_arageom}. Their structure closely follows the ARA baseline design as described in Section \ref{sc_generalDT}. After being deployed in February 2013, the stations recorded data for 10 months until the end of that year (Figure \ref{fig_ARA23Life}). Due to various infrastructural issues and optimizations which interrupted the detector operation, there were several extended periods of down time, sometimes lasting for days. Therefore, the two detectors were only running $75\%$ of the time and correspondingly accumulated about $228$ (A2) and $220$ (A3) days of live time during those months. Meanwhile, station operation has become more stable through debugging and optimization of the data acquisition (DAQ) firmware and software. In addition to that, new monitoring tools have been developed which allow us to identify and solve problems within a few hours. This resulted in a significant rise in live time for the year 2014. The dead time during operations due to digitizer occupancy and the limited data transfer bandwidth is very small and less than $1\%$ of the total run-time.

\begin{figure}[t]
\centering \subfigure[ Vpol pulser signal]{
\includegraphics[width=0.465\columnwidth]{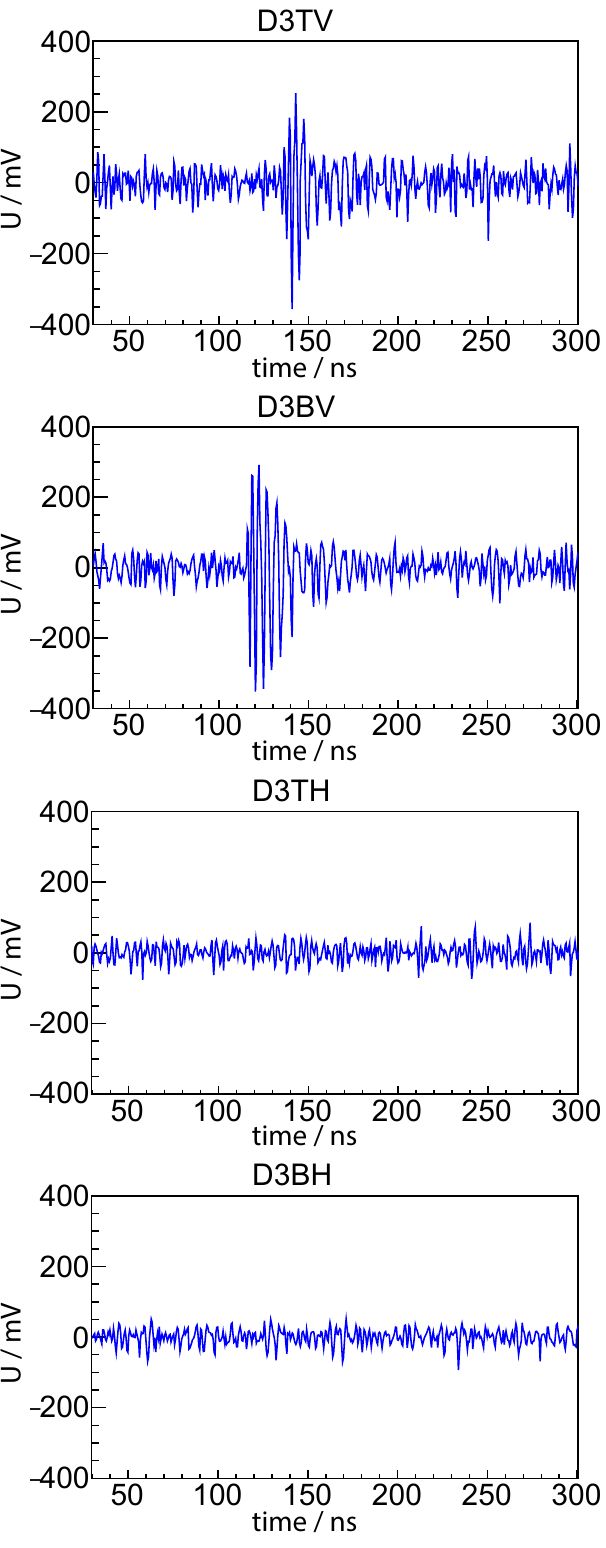}
}
\centering \subfigure[ Hpol pulser signal]{
\includegraphics[width=0.465\columnwidth]{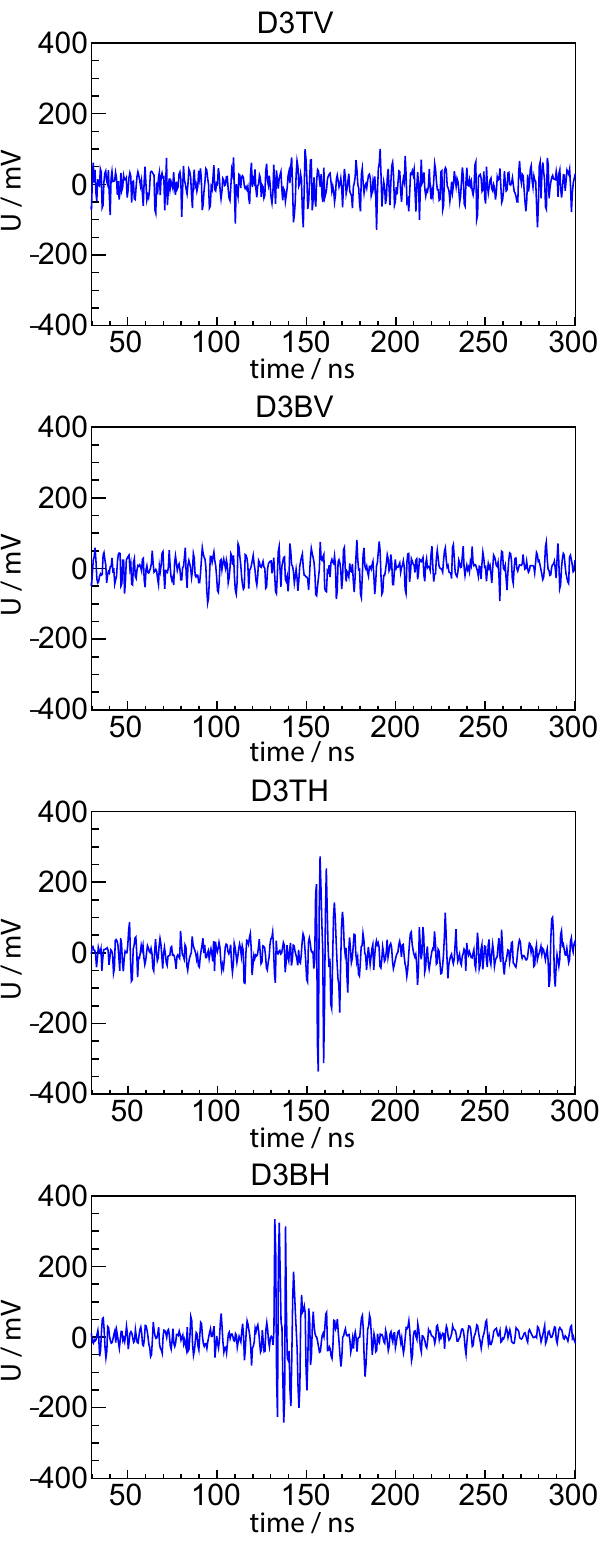}
}
\caption{Signals emitted by a Vpol (a) and an Hpol (b) pulser, as recorded by station A3 string D3. As is evident, the polarization separation is very clean and only the antennas of the emitted polarization show a response.}
\label{fig_calPulserD5}
\end{figure}

\begin{figure}
\centering 
\includegraphics[width=0.9\columnwidth]{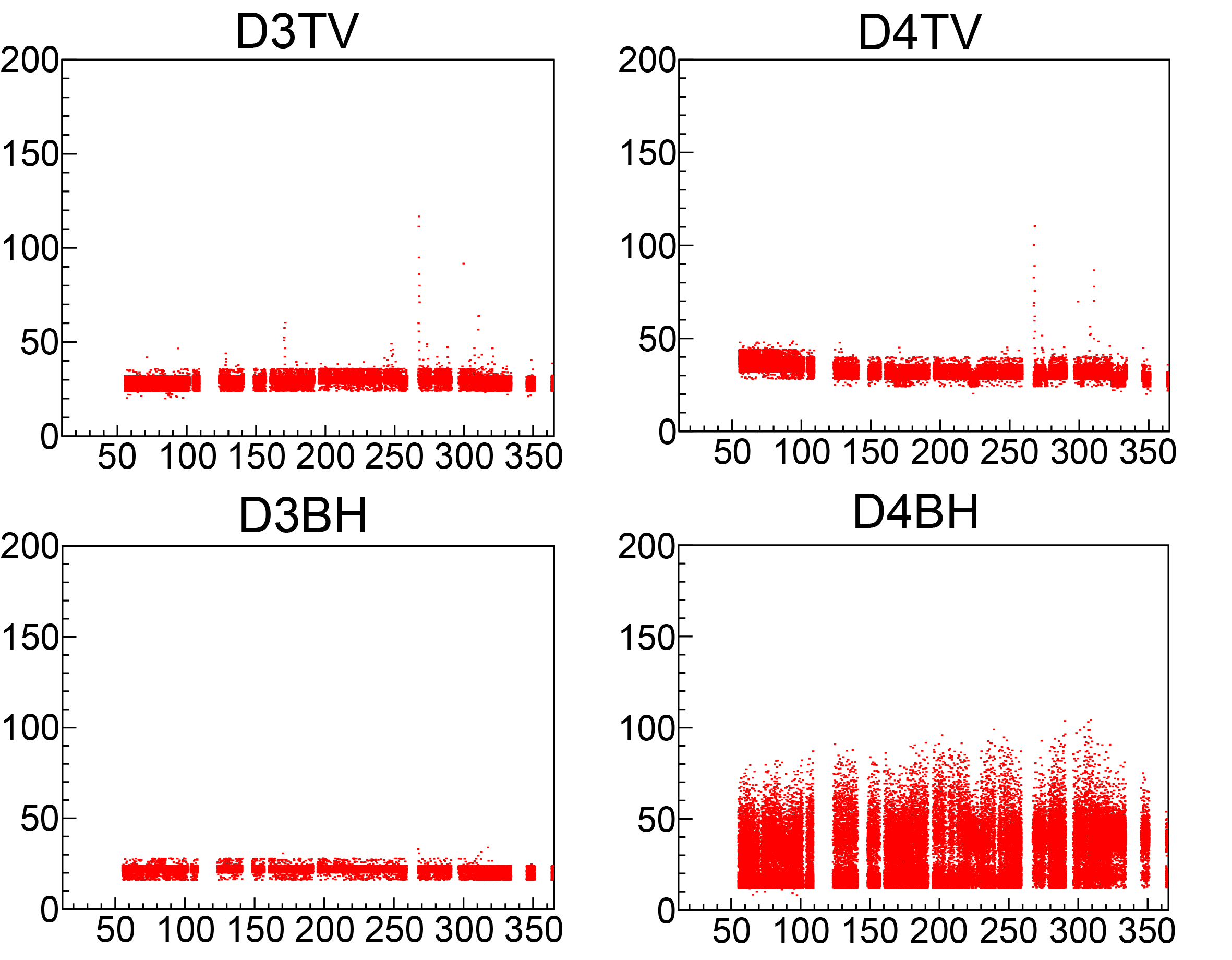} 
\caption{The RMS of events recorded by 4 selected measurement channels from A2 in 2013, plotted in mV versus the day of 2013. The large variation observed in channel D4BH is indicative of failure of that particular channel.} \label{fig_stationRMS2} 
\end{figure}

Of the 32 deep in-ice measurement channels in the two stations, 31 are fully operational. The bottom Hpol channel on string 4 (D4BH) in A2 shows strong noise fluctuations which are believed to be due to a damaged LNA. Figure \ref{fig_calPulserD5} shows pulser waveforms of both polarizations as recorded by A3. One can see the clear absence of a signal in the antennas polarized perpendicular to the emitted signal, which indicates that the two polarizations are very well-separated. The RMS of the background noise on the two stations is relatively stable throughout the year, as shown for A2 in Figure \ref{fig_stationRMS2}.

\subsection{Calibration of station geometry and timing}	\label{ch_geomCal}

After having achieved a stable timing from the digitizer chip, the systematic errors in this timing and the precise positions of the antennas in a station need to be determined, to allow for accurate vertex reconstructions. The antennas are suspended on four strings in four vertical holes, connected by stiff cables. Their XY-coordinates can thus be assumed to coincide within one hole. Furthermore, the vertical distances and the cable delays have been measured and are assumed to be correct with a negligible error. Parameters which still need to be calibrated are the position of each string and the relative time delay between them. One string has to be chosen as perfectly positioned and the rotation of the station around this string needs to be fixed to obtain a well determined coordinate system. With these assumptions there remain 17 parameters of positions and cable delays to be calibrated (neglecting uncertainties in the index-of-refraction model).

\begin{figure}[t!] \centering
\includegraphics[width=0.9\columnwidth]{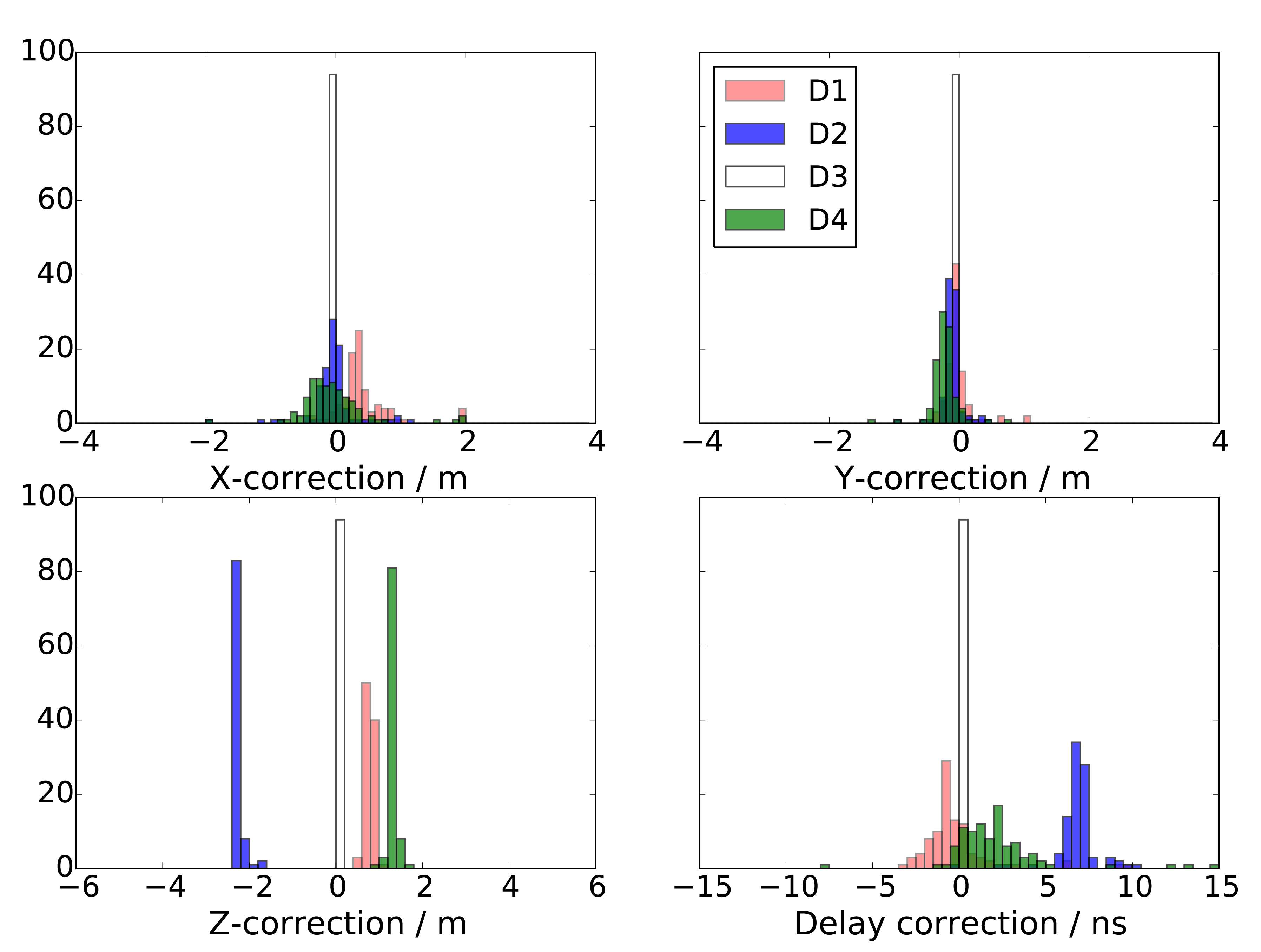}
\caption{Results of the fits for geometrical and timing calibration of the strings in station A3. Note that string D3 has been used as the reference and is fixed in the $\chi^2$-minimization of this calibration.} \label{fig_fitResultsARA03} 
\end{figure}

Such calibration is performed by using calibration pulser signals and determining the arrival time difference between signals on different antennas. For this quantity and the geometrical positions of the antennas one can set up an equation for each possible antenna pair and for all 4 pulsers. Considering the two polarizations, a system of $28$ independent equations can be constructed. For these equations we can set up a $\chi^2$ value as 
\begin{eqnarray} 
\chi^2 &=&\sum\left[ \ c^2 (dt_{ki,\mathrm{ref}}^2 - dt_{kj,\mathrm{ref}}^2) \right. \nonumber \\
& &\ \ \ \left. + x_k \cdot 2 x_{ij} + y_k \cdot 2 y_{ij} + z_k \cdot 2 z_{ij} \right. \nonumber \\ 
& & \ \ \ \ \ \left. - t_{k,\mathrm{ref}} \cdot 2 c^2 dt_{kij} - r_i^2 + r_j^2 \ \right]. \label{eq_fitChi2} 
\end{eqnarray} 

\begin{figure}[t] \centering
\subfigure[]{
\includegraphics[width=0.9\columnwidth]{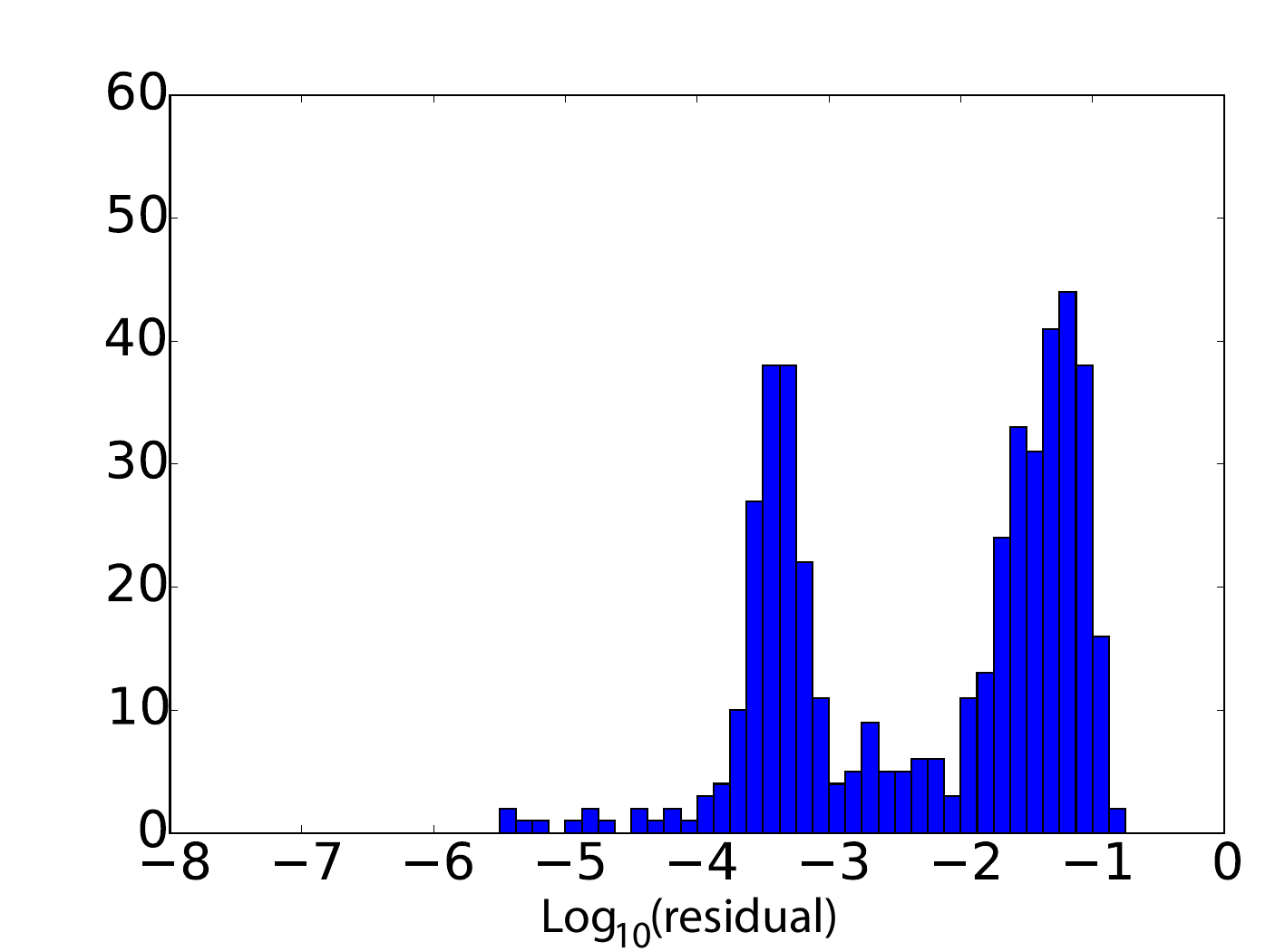}
\label{ARA03impResA} } \subfigure[]{
\includegraphics[width=0.9\columnwidth]{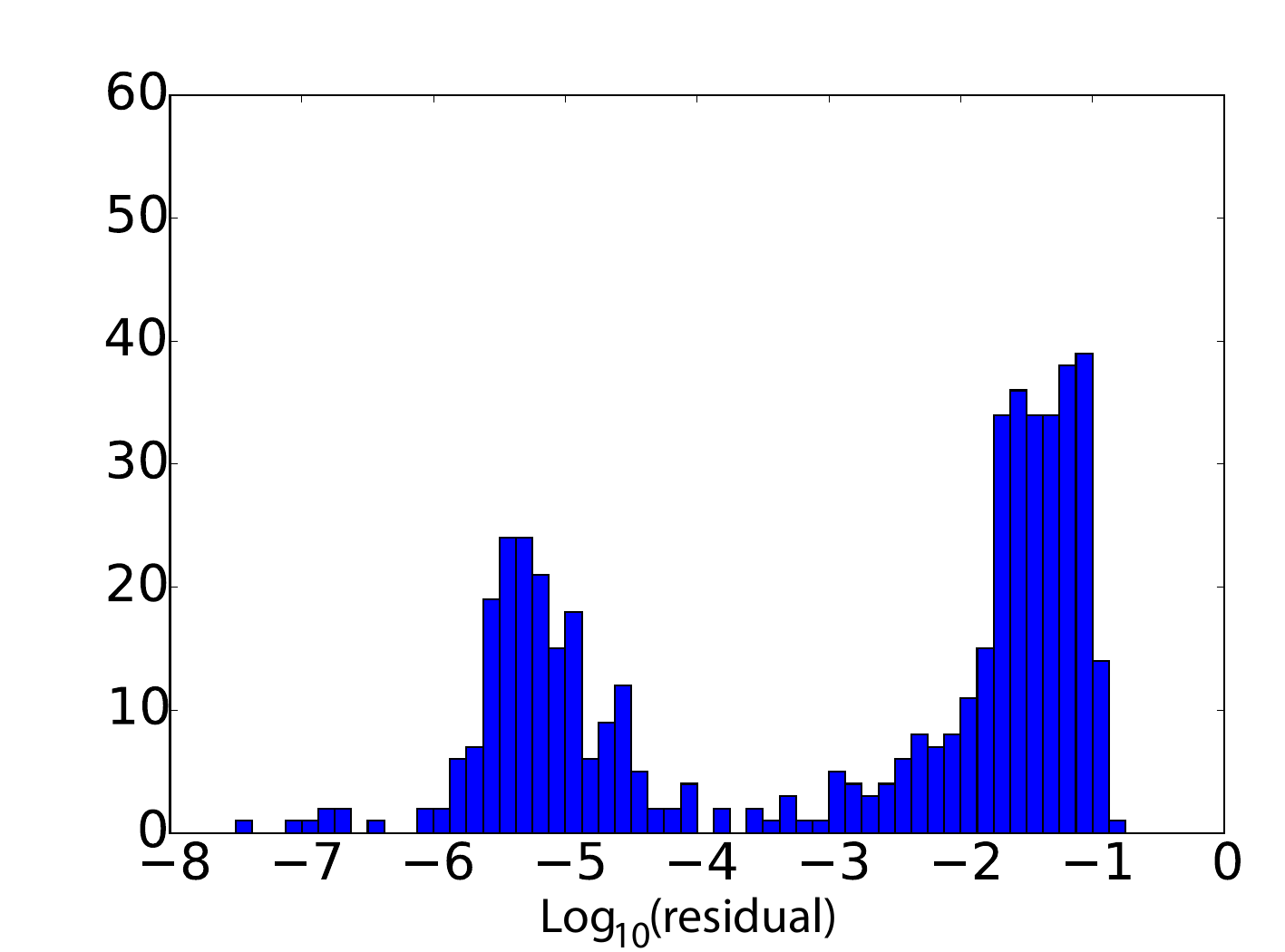}
\label{ARA03impResB} } 
\caption{The residual of the rooftop pulser reconstruction with A3 (a) before and (b) after the geometrical calibration. The bimodal distributions contain noise events with a residual of roughly $10^{-1.5}$ and signal with lower residuals which migrate to smaller values after calibration.} 
\label{ARA03RoofPComparisionRes} 
\end{figure}

Here we use $c=\unit{0.3/1.755}{\meter\per\nano\second}$ as the speed of light in ice at the average antenna depth, taken from \cite{Kravchenko2004}. The coordinates of the pulsers are denoted by $x,y,z$ and the arrival time difference by $dt$, with the subscripts $k$ for the used pulser and $i,j$ for the respective antennas. The parameter $r$ indicates the distance of the antenna to the station center. The presented $\chi^2$ is closely related to the equations which are used for the reconstruction algorithm, described in Section \ref{sc_reconstruction}. Standard minimizer tools are used to minimize the $\chi^2$ using multiple and different seed values for the input parameters. The average of the outcomes is taken as the final result. Figure \ref{fig_fitResultsARA03} shows how different parameters are constrained by this method for station A3. It becomes apparent that the current geometrical setup is strongest in determining the string depth while the constraints on the X- and Y-position, and the inter-string delays are relatively weak.
\begin{figure}[t] \centering \subfigure[]{
\includegraphics[width=0.9\columnwidth]{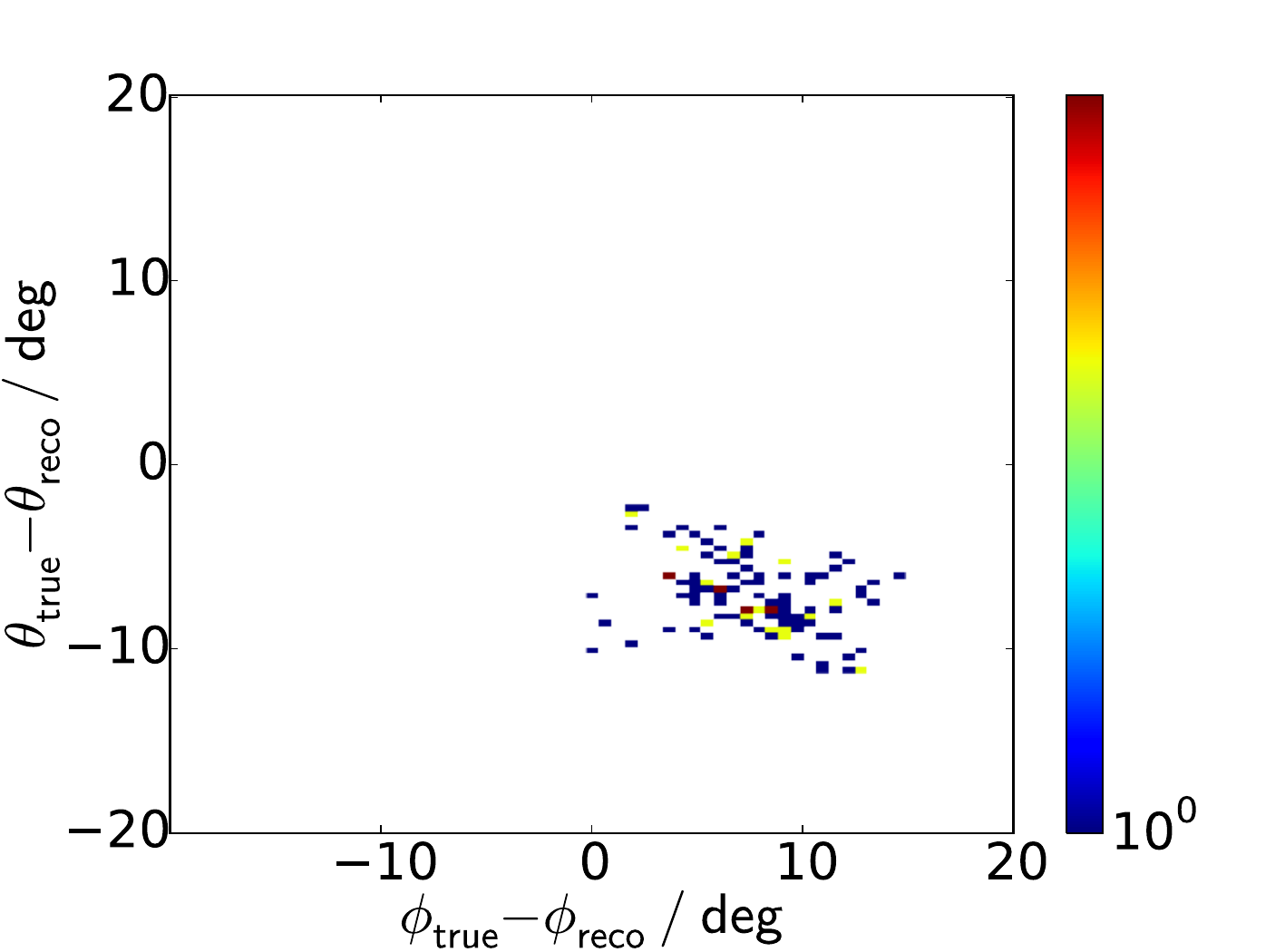}
\label{ARA03impA} } \centering \subfigure[]{
\includegraphics[width=0.9\columnwidth]{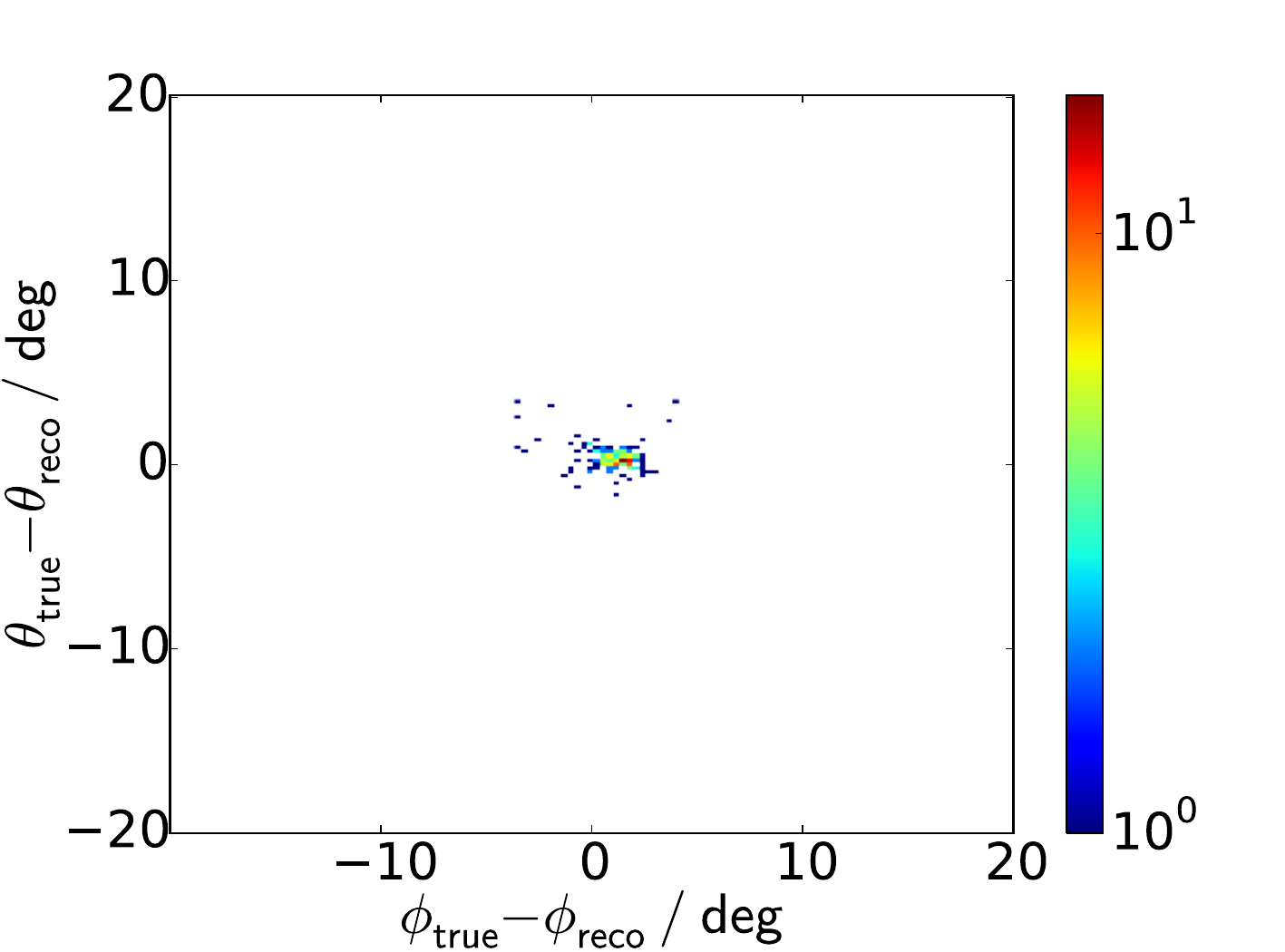}
\label{ARA03impB} } \caption{The directional rooftop pulser reconstruction with A3 (a) before and (b) after the geometrical calibration. The axes show the difference between the reconstructed and the true azimuth (X-axis) and zenith (Y-axis) angle. Reconstruction quality criteria are only applied loosely. It should be noted that the data from the rooftop pulser are not part of the calibration data sample.} 
\label{ARA03RoofPComparisionThetaPhi} 
\end{figure}

The result of this calibration is checked via the reconstruction of an independent pulsing antenna, mounted at a distance of about $\unit{4}{\kilo\meter}$ from both ARA stations on the rooftop of the IceCube Laboratory. The reconstruction algorithm described in Section \ref{sc_reconstruction} is used for this cross check. The figures of merit are the stability of the reconstruction as well as the residual which indicates the internal consistency of the station geometry. Signals from the rooftop pulser are not tied to a GPS clock and have to be filtered out of all recorded data by other means. The residual is plotted before and after calibration for all data recorded during a rooftop pulser run in Figure \ref{ARA03RoofPComparisionRes}. Two peaks are visible in this distribution: one for noise waveforms with high residual around $10^{-1.5}$ and one for signal which shifts to significantly lower values after the calibration. This indicates that the assumed geometry is more consistent with the measured timing after the calibration has been applied. Figure \ref{ARA03RoofPComparisionThetaPhi} shows the actual result of the reconstruction compared to the expected value. After calibration, the reconstruction is much more self-consistent than before. Currently, in the two stations all operating channels but two in A2 are calibrated in position and timing in this way. The two omitted channels in station A2, D3BV and D3BH, show a puzzling timing offset of several nanoseconds which is not corrected in the presented calibration. The source of this offset is unknown to date, but more measurements have been taken to further improve the precision of this calibration.


\subsection{Signal chain calibration}    \label{ch_scCalibration}

In the amplitude calibration of the ARA signal chain, we try to determine the noise temperature of the environment, the noise figure of the signal chain and the directional gain of the antenna.

\subsubsection{Determination of ambient noise} \label{ch_signalChain_ambNoise}
The ARA antennas are exposed to various sources of radio noise. The main contributors are the ice surrounding the antenna, with a depth dependent temperature between $\unit{220}{\kelvin}$ and $\unit{270}{\kelvin}$ \cite{Javaid2012} and sky sources like the atmosphere and the Galactic center. The contributions of the Sun, moon and cosmic background radiation are negligible for the ARA antennas. Also the contribution from the bedrock under the ice, with a temperature of roughly $\unit{273}{\kelvin}$, could be determined to be negligible due to the radio attenuation in ice and the disfavored incoming angle in the directional gain pattern of the antennas.

The radiation from the atmosphere and galactic noise both approach the ice from the top and enter the ice according to Snell's law at the boundary of a medium. The minimum elevation angle, as viewed from the in-ice antennas, above which sources emanating from above the ice can be observed by those in-ice antennas is approximately $\unit{55}{\degree}$. Atmospheric and galactic temperature profiles have been extracted from \cite{itu372.2013,Allison2012457}. The attenuation in the atmosphere is assumed to be negligible and the attenuation through the top ice is normally low. However, the contribution to the noise of the antennas is relatively small, with an equivalent temperature of $T_{\mathrm{sky}} = \unit{18.3}{\kelvin}$ at $\unit{300}{\mega\hertz}$ for Vpol antennas. This is mainly due to the steep incoming angle which is highly disfavored by the directional gain pattern of the antennas.

To calculate the power spectrum received from the ice one has to consider it as divided into semi-transparent volume-elements. The brightness $B'$ of each element is the brightness of a black body B, reduced by the limited emissivity $\varepsilon$ of the ice as:
\begin{eqnarray}
B'(\nu) = \varepsilon \cdot B(\nu) = \frac{2}{\alpha} \cdot B(\nu),
\end{eqnarray}
with $\alpha$ being the attenuation length at $\unit{300}{\mega\hertz}$ at the given depth. This equality is valid under the assumption of thermodynamic equilibrium in the ice \cite{Gary2014}. The total noise power at the antenna is the integral over all volume elements in a given range $R$ around the antenna
\begin{eqnarray}
P_{\mathrm{ice}}(\nu) = \frac{\frac{1}{2} 2\pi\lambda^2 \int\limits_{-\frac{\pi}{2}}^{\frac{\pi}{2}}\int\limits_{0}^{R} B'(\nu) \eta(r, \theta) G(\nu, \theta) \cos{\theta} d\theta dr }{ 2\pi \int\limits_{-\frac{\pi}{2}}^{\frac{\pi}{2}} G(\nu, \theta) \cdot \cos{\theta} d\theta }, \qquad   \label{eq_iceInt}
\end{eqnarray}
with the wavelength $\lambda$, the directional antenna gain G and the reduction factor $\eta$ due to the finite attenuation length for radio waves in ice. The factor $1/2$ is applied to account for the single polarization of the antenna. In the integral of Equation \ref{eq_iceInt} the contribution of radiation reflected at the surface has to be taken into account, which accounts for roughly $T_{\mathrm{refl}} = \unit{53}{\kelvin}$ of noise power.

The total noise temperature seen by an ARA antenna can be calculated to be 
\begin{eqnarray}
T_{\mathrm{ant}} = T_{\mathrm{ice}} + T_{\mathrm{refl}} + T_{\mathrm{sky}}.
\end{eqnarray}
This results in $\unit{247 \pm 13}{\kelvin}$ for Vpol and $\unit{249 \pm 13}{\kelvin}$ for Hpol antennas at $\unit{300}{\mega\hertz}$. The main contributions to the errors come from uncertainties in the ice temperature measurement, the critical angle for surface reflection and the antenna gain pattern. The temperature value currently used in the ARA simulation is calculated for one single frequency of $\unit{300}{\mega\hertz}$. Since this temperature changes over the ARA frequency range by up to $\unit{30}{\kelvin}$, the latter will be assumed as a systematic error for further calculations rather than the smaller calculated error of $\unit{13}{\kelvin}$.

\subsubsection{Signal chain calibration}
The calibration of the signal chain is particularly challenging since the deployment of fixed strings in $\unit{200}{\meter}$ deep holes allows neither a detailed investigation of the directional gain pattern of the antenna nor noise figure measurements \emph{in-situ}.

\begin{figure}[t] 
\includegraphics[width=\columnwidth]{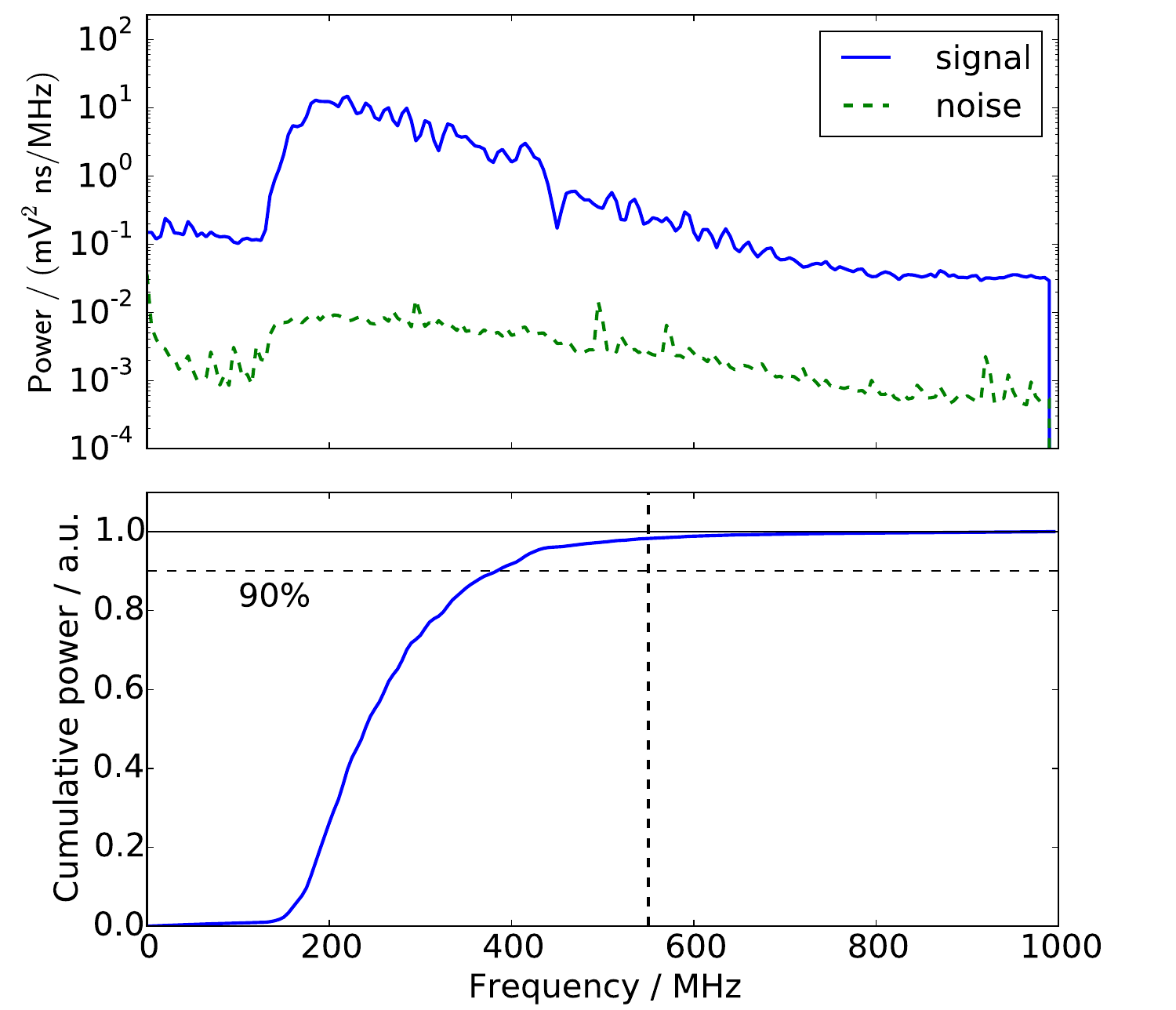}
\caption{Power plotted versus frequency for a recorded Vpol signal from an external source (blue solid line), with a flat input spectrum between 150 and \unit{1000}{\mega\hertz}. The cumulative power distribution shows that most of the signal power is recorded below \unit{500}{\mega\hertz}. The green dashed line shows the recorded spectrum for the natural ambient noise without additional signal for comparison.}
 \label{fig_cumSignalPower}
\end{figure}

The noise factor $F$ of the signal chain has been measured for each channel at the surface at room temperature prior to deployment. This noise factor is used to calculate the noise temperature of the signal chain. The change due to the signal chain being lowered into the ice with a local environmental temperature of roughly $\unit{220}{\kelvin}$ at a depth of $\unit{180}{\metre}$ is estimated as a linear change with ambient temperature. The total noise contribution depends on the transmission coefficient $t$ of the antenna. For perfect coupling it can be calculated to be (e.g., at $\unit{300}{\mega\hertz}$):
\begin{eqnarray}
T_{\mathrm{tot}} &=& t \cdot T_{\mathrm{ant}} + T(\unit{180}{\meter}) \cdot (F - 1)  \nonumber \\
&=& \unit{247}{\kelvin} + \unit{220}{\kelvin} \cdot (1.6 - 1) = \unit{379}{\kelvin},
\end{eqnarray}
where $t=1$ for perfect coupling.

With knowledge of this noise floor, the directional gain can be determined using the external noise sources at each station. These sources are noise diodes, emitting a flat power spectrum in the relevant frequency range, which can be attenuated by up to $\unit{30}{\deci\bel}$. They are connected to the pulser antennas, installed in the vicinity of the ARA stations. Figure \ref{fig_cumSignalPower} shows the recorded power spectrum for a typical receiver antenna and a pulser with a non-attenuated input source. This figure further illustrates the sensitivity of the signal chain in different parts of the frequency spectrum via a cumulative power distribution. Most signal power is recorded below $\unit{500}{\mega\hertz}$ in this measurement.

\begin{figure}[b] \centering
\includegraphics[width=0.9\columnwidth]{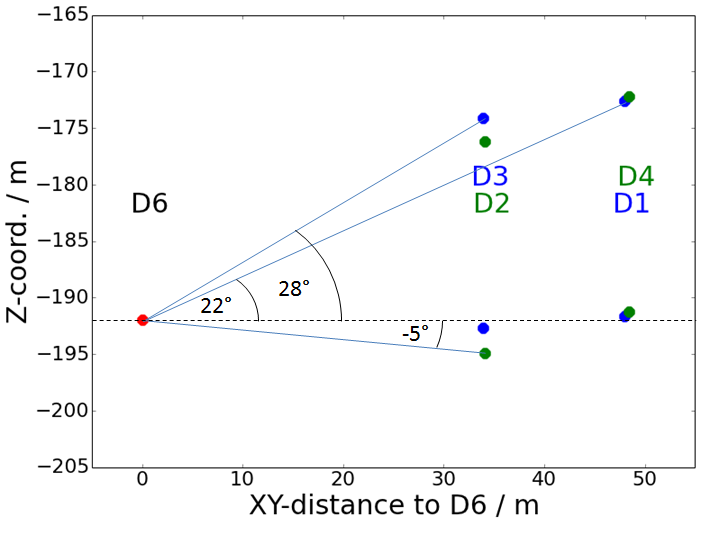}
\caption{The relative geometry of the Vpol antennas to the D6 pulser in A3.} \label{fig_geometryARA03D6} 
\end{figure}

Under the assumption that all installed antennas of the same polarization have the same angular gain pattern, the geometric relation between the pulsers and all 8 measurement antennas of one polarization in both stations can be used to measure the gain pattern for different angles. For this measurement, we assume further that the directionality in azimuth is isotropic. With the available antennas we can establish 24 different relations in elevation which are clustered at distinct angles, as shown by the example of A3 in Figure \ref{fig_geometryARA03D6}. In fact, for station A3 all bottom antennas are at the same depth as the two pulsers (D5 not shown), while the top antennas are elevated by an angle of roughly $\unit{25}{\degree}$. In station A2, the one fully operational pulser is mounted at a similar depth to the top antennas, while the bottom antennas receive the pulser signal at an inclination of $\unit{-25}{\degree}$. The antenna directivity $D$ at a given angle is then calculated via the Friis transmission equation \cite{Shaw2013} and the known noise factor from
\begin{eqnarray}
\sqrt{D_r D_t} &=& \frac{4 \pi R}{\lambda} \sqrt{ \frac{ P_{\mathrm{out}} - N_{\mathrm{out}} }{t_r t_t P_t N_{\mathrm{out}}} } \nonumber \\
& & \cdot \sqrt{ t_r \cdot N_{\mathrm{ant}} + \unit{220}{\kelvin}\cdot (F - 1) },
\end{eqnarray}
with $N_{\mathrm{ant}}$ being the power corresponding to $T_{\mathrm{ant}}$, $t_{r/t}$ the transmission coefficient between a given antenna and the signal chain ($r$ for receiver, $t$ for transmitter), $P_t$ the input power to the noise source antenna and $R$ the distance between transmitter and receiver antennas. $N_{\mathrm{out}}$ is the measured output power without any applied signal and $P_{\mathrm{out}}$ denotes the total recorded power when the noise sources are operating. These are the experimentally measured values. The transmission coefficient $t$ is taken from XFDTD simulations \cite{XFDTD2015} of the deployed Vpol antennas in ice and from NEC2 simulations of the Hpol antennas. For the directivity, we assume for now $D_r = D_t = D$, for a transmitter and receiver of the same polarization.

The antenna gain can then be calculated as
\begin{eqnarray}
G = D \cdot t.
\end{eqnarray}
\begin{figure}[t] 
\centering \subfigure[A3 Vpol]{
\includegraphics[width=1.0\columnwidth]{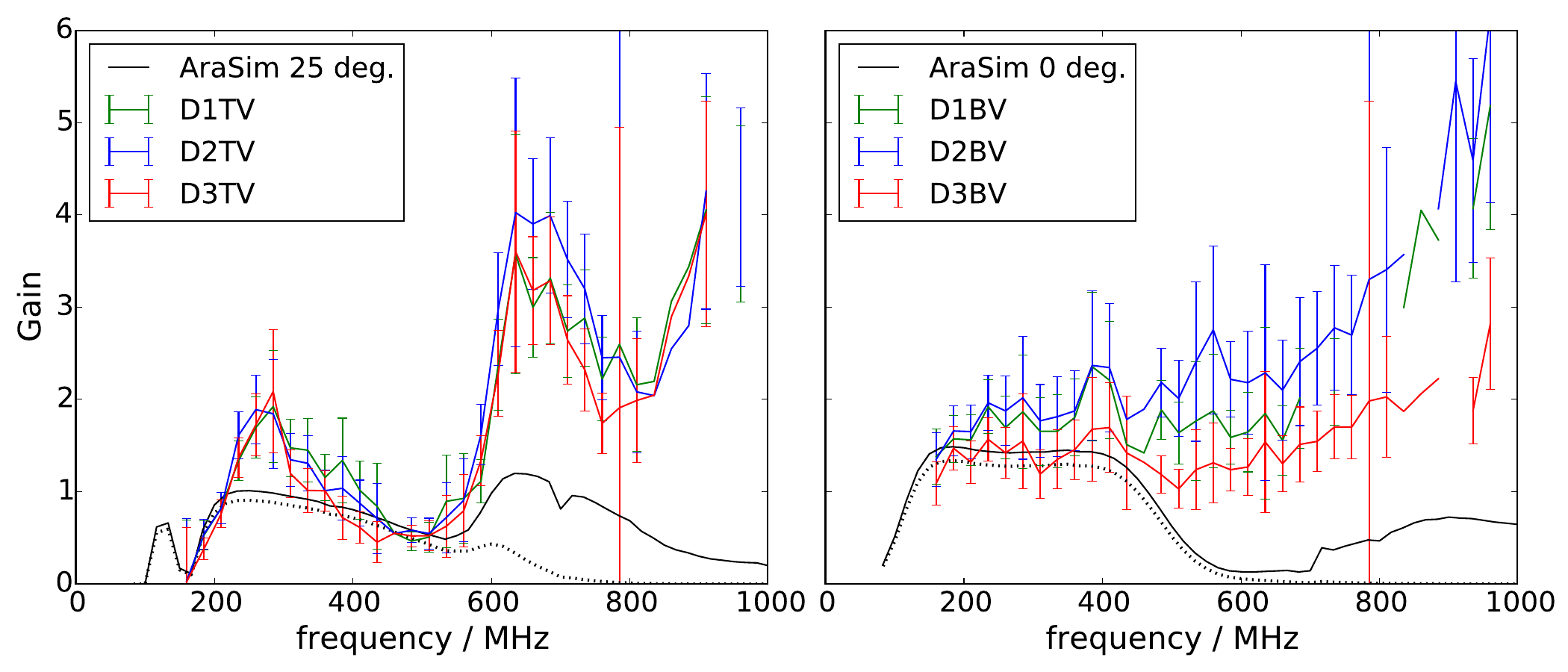}
}
\centering \subfigure[A2 Hpol]{
\includegraphics[width=1.0\columnwidth]{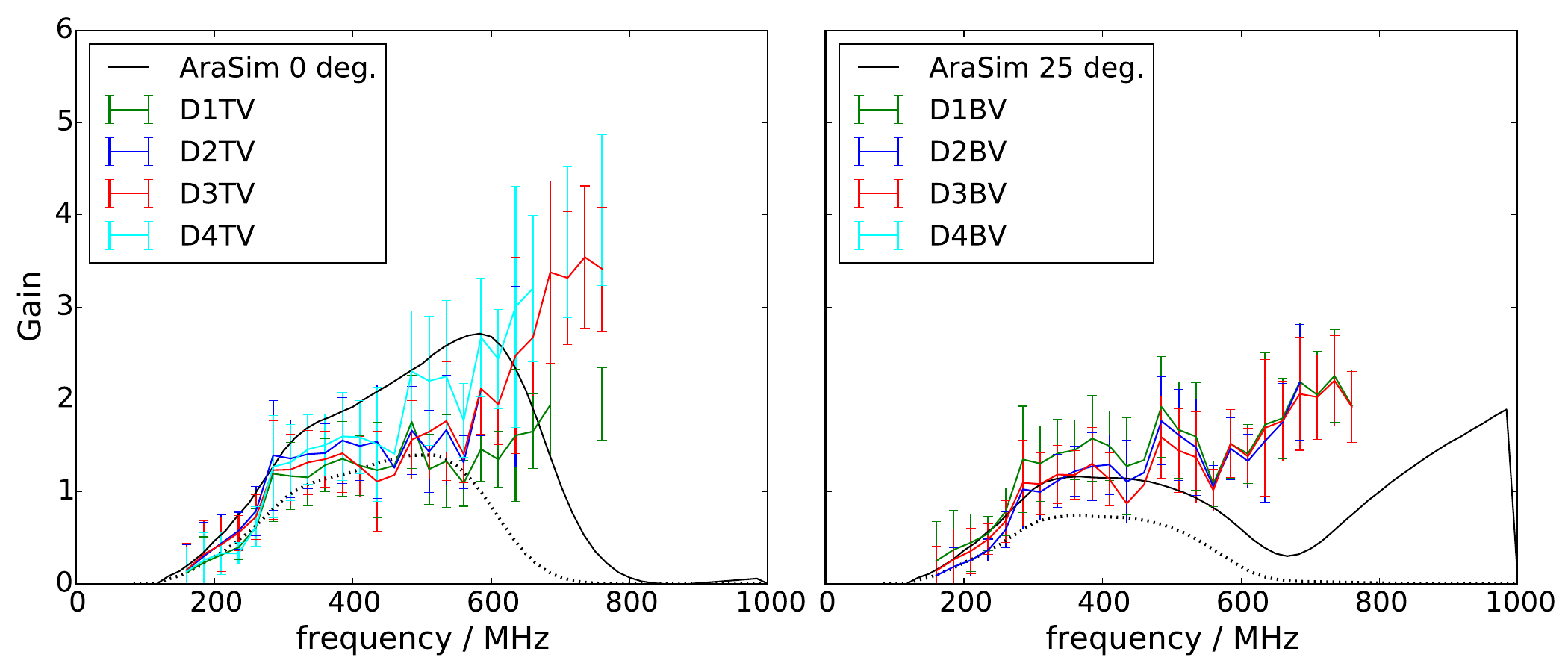}
}
\caption{Sample measurements of the antenna's directional gain at the available angles (see Figure \ref{fig_geometryARA03D6}) for the two ARA stations and polarizations: (a) A3 Vpol, (b) A2 Hpol. The data are plotted on a linear scale versus frequency. For comparison the current status of the simulation is shown as a black solid line. String D4 on A3 was not operating during the time of the measurement. The black dashed line represents the lower limit on the signal gain used to derive the systematic error on the detector sensitivity.} \label{fig_gainARA03D6VP}
\end{figure}
Figure \ref{fig_gainARA03D6VP} shows the gain measured for the A3 Vpol antennas and the A2 Hpol antennas, both plotted with systematic errors, which are dominant over statistical uncertainties in this measurement. They are derived from uncertainties in the input spectrum, the ambient noise power and the used noise figures. The comparison to simulations shows that differences, especially beyond $\unit{500}{\mega\hertz}$, are not covered by our current understanding of these uncertainties. This may be due to an imperfect simulation of the antennas in ice, which is indeed very challenging, or due to unaccounted error sources in the measurement. The signal strength at frequencies beyond $\unit{500}{\mega\hertz}$ is relatively low (see Figure \ref{fig_cumSignalPower}), which is expected due to the lower sensitivity of the signal chain at high frequencies. Such behavior allows for a stronger influence of non-linear effects in this region which are difficult to quantify. One possible source for such non-linearities is the digitizer chip and its calibration. In Appendix \ref{ch_digiCal}, it is shown that non-linearities are observed in the ADC-to-voltage conversion gain. In addition, the sample timing can have errors of $\mathcal{O}(\unit{100}{\pico\second})$. Such imperfections may add artificial power to the Fourier spectrum depending on the level of disturbance, as illustrated in Figure \ref{fig_calDisturbance}. This could explain the excess in the antenna gain measurement at frequencies above $\unit{500}{\mega\hertz}$. The influence of non-linearities on the frequency range below $\unit{500}{\mega\hertz}$, given sufficient signal strength, is however expected to be very small. The visible difference between top and bottom antennas is due to cable feed-through. In the ARA stations, an antenna contains a vertical cable feed-through for each antenna which is mounted below. This has a significant influence on the angular gain pattern, which has not yet been fully simulated.

\begin{figure}[b] 
\centering
\includegraphics[width=1.0\columnwidth]{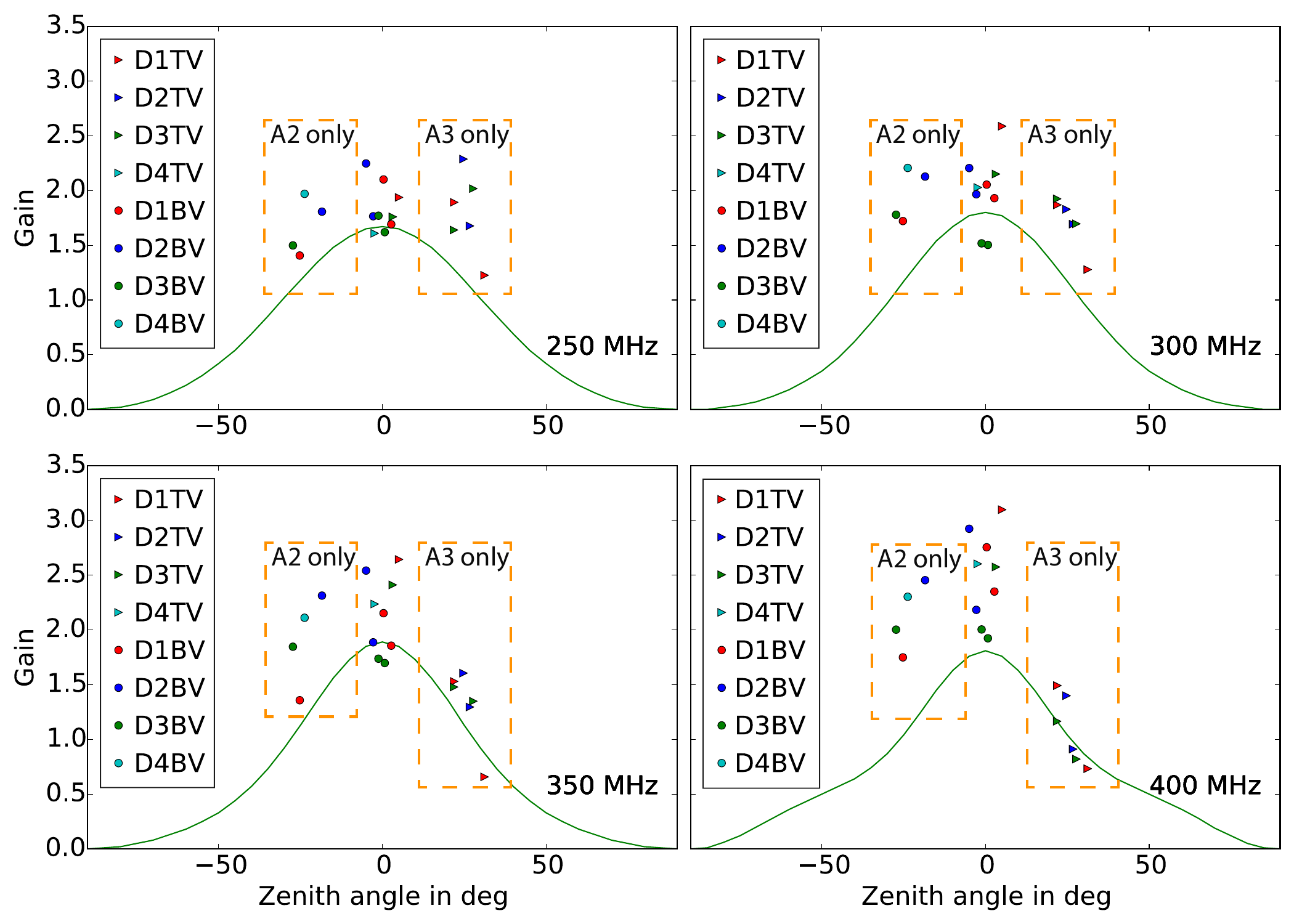}
\caption{The directional gain results versus reception angle from A2 and A3 for all Vpol antennas compared to the current simulation of the bottom antennas (green line) at different frequencies. All data are normalized to an isotropic directionality pattern. The three data points per antenna originate from the three available calibration sources used in this calibration. Gains around $\unit{-25}{\degree}$ angles can only be measured in the A2 geometry, gains around $\unit{+25}{\degree}$ only in A3.} 
\label{fig_allPulserGain}
\end{figure}

Figure \ref{fig_allPulserGain} shows the gain measured by all possible Vpol antennas versus the reception angle. As is evident, the antennas are not distributed equally over the angular range, but concentrated at a few points. Therefore, it is difficult to make a prediction for the full angular response. The gain pattern currently used for the bottom antennas in the ARA simulation (Section \ref{ch_simulation}) is included in Figure \ref{fig_allPulserGain} for comparison. This simulation has been derived for a Vpol antenna in ice based on an adapted NEC2 simulation.

The visible asymmetry in zenith angle, strongest at $350$ and $\unit{400}{\mega\hertz}$, is expected to be due to the connecting cable which is fed vertically through the antennas to a connector in the center. This effect has been observed in calibration measurements in air. Its strength in ice could not be quantified in the presented measurement since antennas of one type (bottom/top) have not been measured at positive and negative angles. Furthermore, a difference in the source could influence the current picture since certain angular combinations only appear for a given source.

The wide clustering of data points is most likely caused by the change of input source and the difference between top and bottom antennas. Furthermore, the shape of the hole walls around the antennas may have a significant influence which still needs to be quantified. 

\subsubsection{Discussion}	\label{ch_signalChain_discussion}
The calibration presented above is an initial step to understand the behavior of the ARA signal chain and the ambient noise. Several discrepancies between the measurement and the currently used simulation have been pointed out which exceed the shown systematic errors on the measurement. These discrepancies need to be resolved, to obtain a comprehensible frequency spectrum of recorded signals. Plans to improve the shown measurements are currently under development. One possibility to achieve a better understanding of the antenna gain is to move pulsers vertically in the hole and to take measurements at different depths.

From the signal power distribution in this measurement we conclude that the calibration values are only reliable in a frequency range of up to roughly $\unit{500}{\mega\hertz}$. Above that frequency we consider our signal chain to be understood more poorly.

In the lower part of the spectrum, up to roughly $\unit{500}{\mega\hertz}$, simulations appear to underestimate the Vpol antenna gain at the measured angles and to overestimate Hpol antennas. An average underestimation for Vpols of $15\%$ and an overestimation for Hpol antennas of $30\%$ can be obtained between simulation and measurement. Given this, we choose the negative systematic error on the antenna gain to be 0 for the Vpol antennas and $-30\%$ for the Hpol antennas.

To account for all errors determined in this signal chain calibration we combine them into an uncertainty on the signal to noise ratio (SNR), which is used to determine the error on the detector sensitivity (see Section \ref{ch_sysError}). The signal to noise ratio in power can be calculated to be
\begin{eqnarray}
SNR = \frac{ t \cdot D \cdot P_{\mathrm{sig}} }{ t \cdot N_{\mathrm{ant}} + N_{\mathrm{sc}}},
\end{eqnarray}
with the transmission coefficient $t$, the directivity $D$, the incoming neutrino signal $P_{\mathrm{sig}}$, the ambient noise $N_{\mathrm{ant}}$ and the signal chain noise $N_{\mathrm{sc}}$.
The uncertainty on the signal chain noise figure has been measured to be $10\%$ on average, and the general uncertainty on the transmission coefficient is assumed to be $10\%$. As mentioned in section \ref{ch_signalChain_ambNoise} the error on the ambient noise temperature is taken to be $\unit{30}{\kelvin}$. All error values are summarized in Table \ref{tab_sysErrorsSNR}. Given those values the resulting relative error on the SNR in power is $-32\%$ for Hpols and $-10\%$ for Vpols.
\begin{table}[h!]
\caption{The estimated errors from various sources included in the lower systematic error on the SNR in power.}    
\centering
\begin{tabular}{c c c}
\hline 
\hline \noalign{\smallskip}
Source & estimated error \\  \noalign{\smallskip}
\hline
Directivity $D$ & $30\%$ (Hpol), $0\%$ (Vpol)\\
Transmission coefficient $t$ & $10 \%$\\
Signal Chain $N_{\mathrm{sc}}$ & $\approx10\%$ (frequency dependent)\\
Ambient noise temp $N_{\mathrm{ant}}$ & $\unit{30}{\kelvin}$\\  \noalign{\smallskip}
\hline \noalign{\smallskip}
Total & $ $32\%$ (Hpol), $10\%$ (Vpol)$\\
\hline
\hline \noalign{\smallskip}
\end{tabular}
\label{tab_sysErrorsSNR}
\end{table}

Since we do not have conclusive calibration results for the upper part of the spectrum, we assume a worst case scenario for the lower systematic error and apply a low pass filter on the signal in this area. To quote an upper systematic error, the error sources in the upper frequency range need greater understanding than has been achieved at this point. Therefore, no upper systematic error is presented for now. This, however, has no impact on the flux limits presented later in this document.

The relative lower limit on the SNR is represented by the dashed black line in Figure \ref{fig_gainARA03D6VP}. For the calculation of the systematic error on the detector sensitivity, this lower limit gain is applied to simulated neutrino signals to estimate its influence on the effective area (see Section \ref{ch_sysError}).

\section{Simulations}   \label{ch_simulation}

The simulation of neutrino vertices for ARA is performed with the AraSim code, which is described in detail in \cite{Hong2013,TestBed2014}. In this section a short summary of the simulation is presented.

In AraSim, forced neutrino interactions are generated uniformly over a cylindrical volume. This volume is centered on the simulated ARA station with incoming neutrino directions uniform in $\cos{\theta}$. It is bounded by the bedrock under the ice in depth and by an energy dependent radius, chosen to include all possibly triggering events. For each interaction, a weight is calculated based on the probability that the neutrino would interact at the given point after having passed earthbound material along its trajectory. This probability depends on the energy dependent cross-section for a neutrino interaction \cite{Connolly2011} and the summed number of radiation lengths along the neutrino path. The primary neutrino energy spectrum can be chosen freely.
\begin{figure}[b]
\centering
\includegraphics[width=\columnwidth]{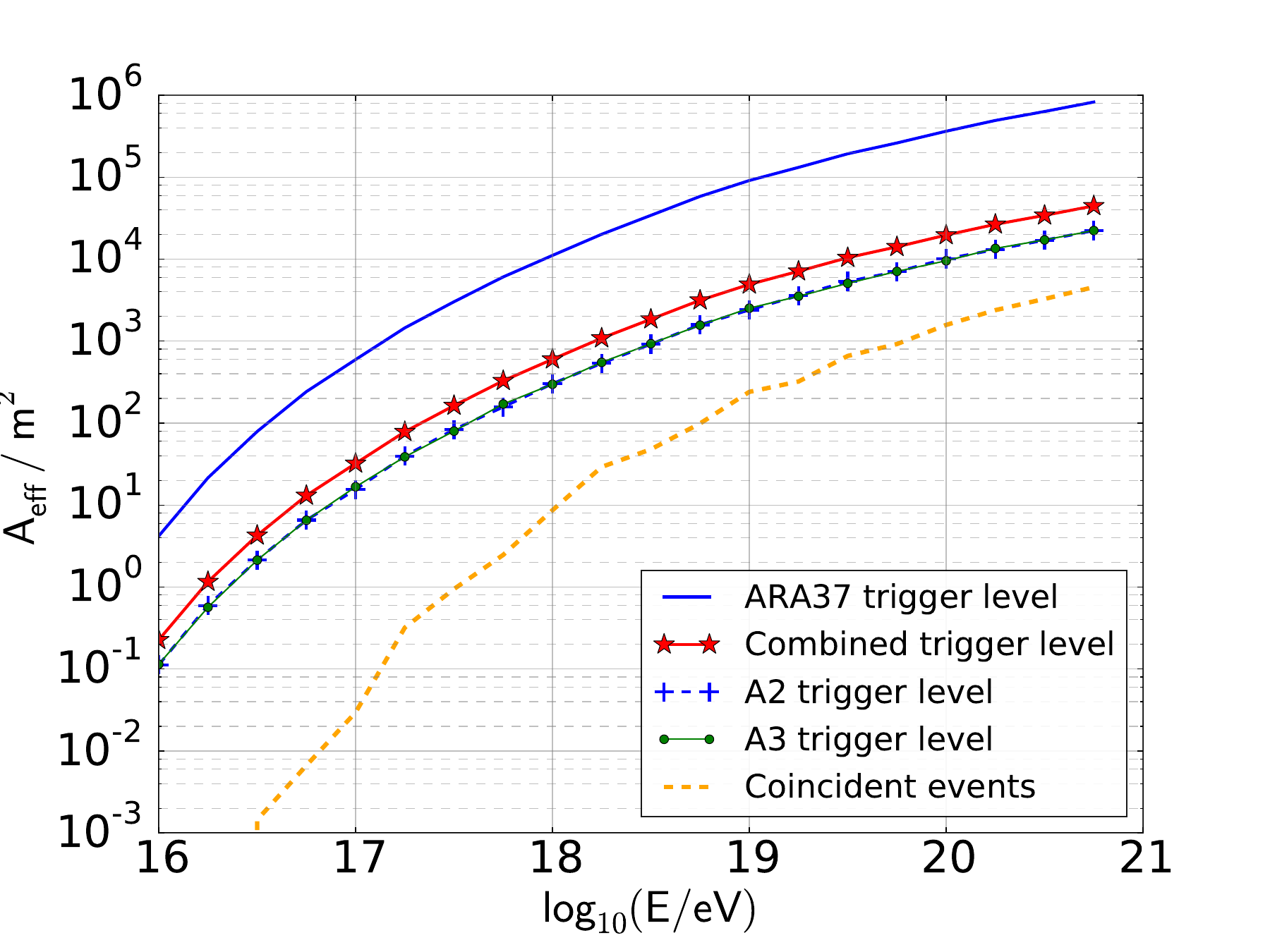}
\caption{The effective area of the two ARA stations as a function of neutrino energy.}
\label{fig_AeffTriggerLevel}
\end{figure}

For each simulated neutrino interaction, a cascade and its radio frequency emission are modeled from theoretical approximations. The results obtained in the following analysis are based on modeling of the Askaryan emission in the frequency domain according to \cite{AlvarezMuniz1997}, which has since been updated to a semi-analytical method of simulating the emission for each event based on \cite{AlvarezTD2010}.

The trajectory of the RF signal to the antenna is calculated using fitted models for the index of refraction \cite{Kravchenko2004} and the final signal strength is derived taking into account the depth dependent attenuation length \cite{Allison2012457}.

Finally, the signal chain and the trigger system are modeled after calibration measurements of their components in the laboratory. For the antenna response, a NEC2 simulation is used. The trigger logic is modeled in AraSim as it is currently set up in the ARA stations: whenever 3 out of 8 antennas of the same polarization cross a given threshold, an event is recorded.

Thermal background noise is modeled using the average frequency spectrum from unbiased forced trigger events recorded throughout 2013, measured with each antenna in station A3. The production of a significant sample of thermal noise data is difficult due to the $6 \sigma$ power-threshold which is currently used in the trigger system to limit the event rate to $\unit{5}{\hertz}$. Therefore, simulated noise is exclusively used to develop and initially test algorithms while final checks and the estimation of background are performed on recorded data.

From simulations, the effective area of the two ARA stations can be calculated at the trigger level to be:
\begin{eqnarray}
A_\mathrm{{eff}}(E) = \frac{V_{\mathrm{gen}}(E) }{N_{\mathrm{gen}}(E)} \frac{1}{L_{\mathrm{int}}(E)} \cdot \sum_{i,\mathrm{trig}}{\omega_i},     \label{eq_Aeff}
\end{eqnarray}
where $V_{\mathrm{gen}}$ is the cylindrical volume over which events are generated, $N_{\mathrm{gen}}$ is the number of events which have been generated, $\omega_i$ is the weight of each event $i$ which triggered the detector and $L_{\mathrm{int}}$ is the interaction length at the given energy. The effective area for the combined A2 and A3 detector is plotted at the trigger level in Figure \ref{fig_AeffTriggerLevel}. In this plot, the single effective areas per station are also shown as well as the effective area for events which are coincident to both stations. These coincidences amount to roughly $5\%$ of all events at an energy of $\unit{10^{18}}{\electronvolt}$.

The simulated dataset used in the present analysis contains equal numbers of neutrinos in quarter-decade energy bins between $\unit{10^{16}}{\electronvolt}$ and  $\unit{10^{21}}{\electronvolt}$. All figures, showing simulated neutrino events are based on this sample.

\section{Data analysis}

The analysis of the first data from A2 and A3, recorded in 2013, has been optimized for sensitivity to neutrino interactions at a fixed rejection of thermal and anthropogenic backgrounds.

The ARA detector records events at a rate of roughly $\unit{5}{\hertz}$. These events are mostly thermal noise and to a lesser degree, backgrounds of anthropogenic origin. In this analysis these backgrounds are reduced in two steps: with a thermal noise filter and by application of angular cuts to reconstructed vertices. All algorithms have been developed and tested on a $10\%$ subset of the full recorded data, in the following called ``burn-sample'', to avoid a bias in the analysis. Cuts are developed to reduce the expected background to approximately a factor $10$ beneath the level of expected neutrino events. After the cuts are finalized, the analysis is applied to the full recorded data set for the year 2013.

\subsection{Thermal noise filtering}

Thermal noise filtering is performed with the time sequence filter, developed for the close-to-cubical ARA station geometry. Further details about a first version of the method, described below, can be found in \cite{Meures2013}.
\begin{figure}[t]
\subfigure[]{
\includegraphics[width=\columnwidth]{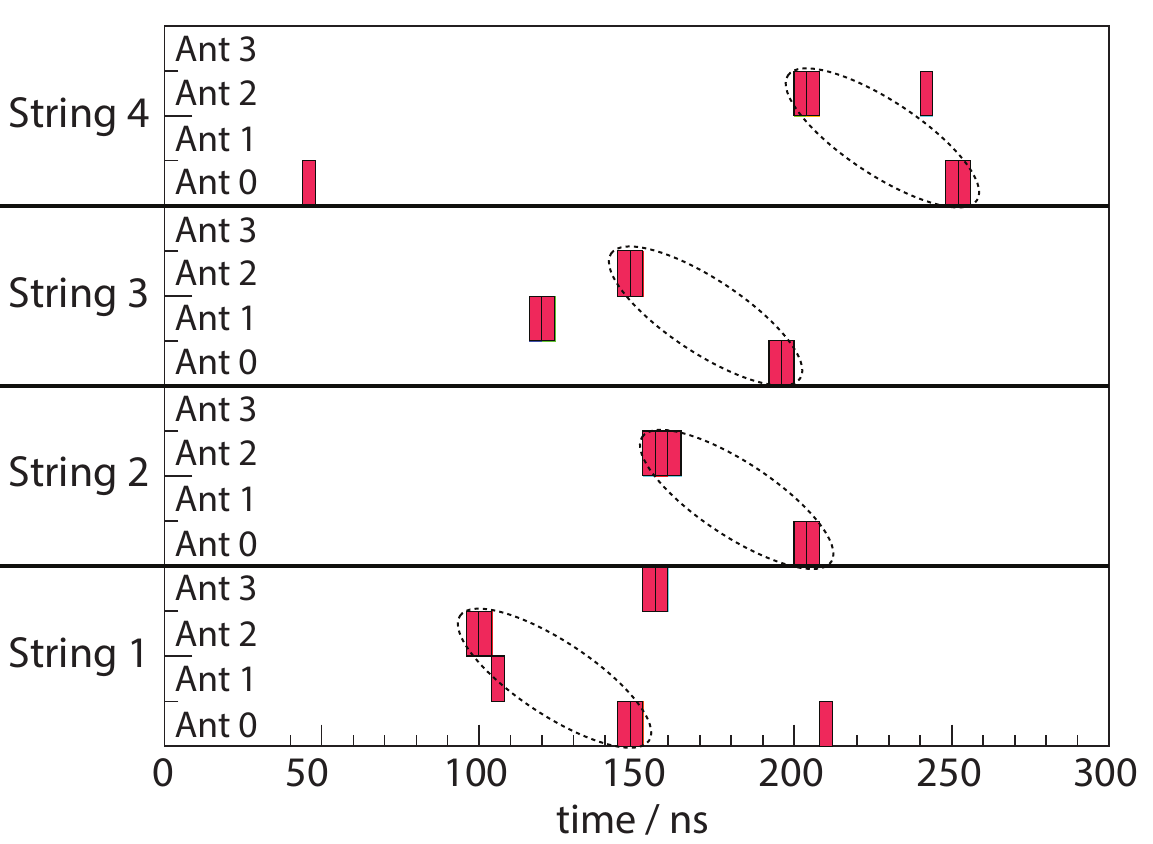}
\label{fig_signalPattern}
}
\subfigure[]{
\includegraphics[width=\columnwidth]{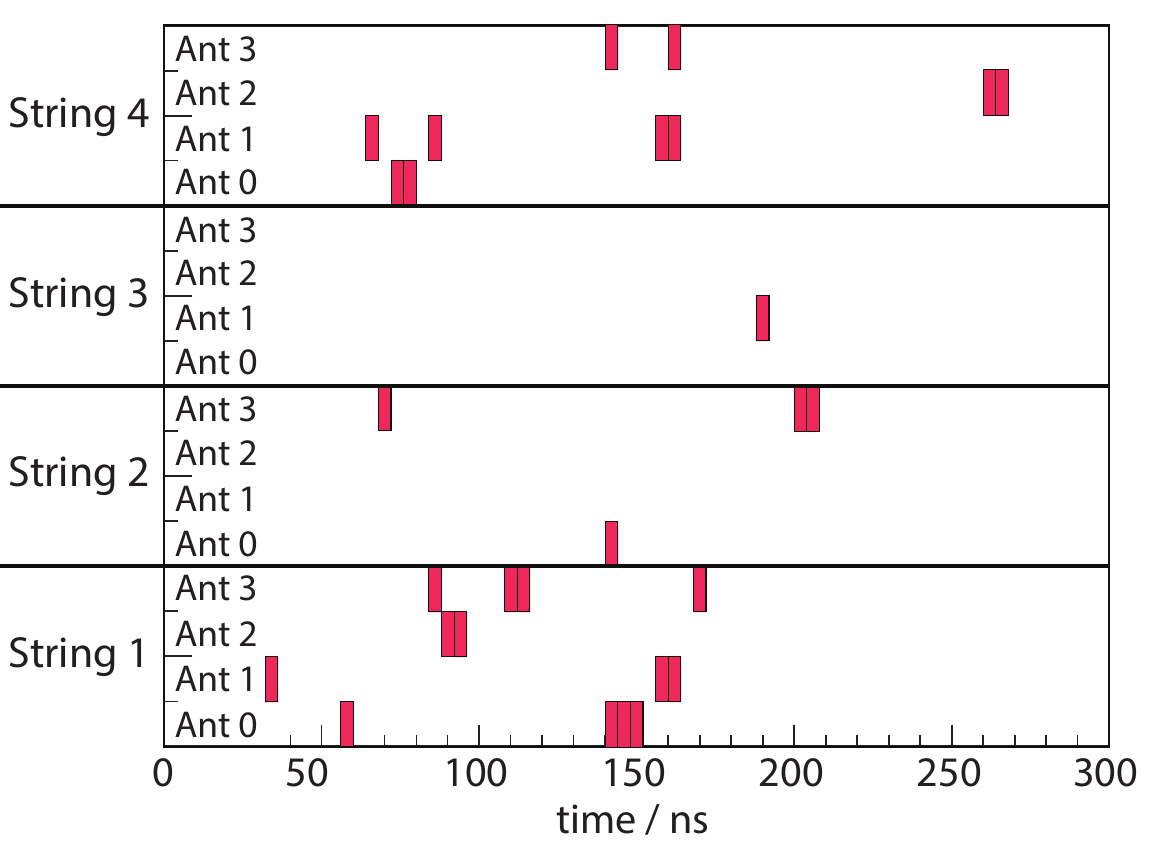}
\label{fig_noisePattern}
}
\caption{(a) The hit pattern for a simulated event containing a signal waveform. The hits are indicated by the colored squares. Hits which correspond to the signal wavefront are marked by black dashed ellipses to underline their visible regular pattern. (b) A typical hit pattern for pure thermal noise.}
\label{fig_hitPatterns}
\end{figure}

The algorithm works in three steps. First, a so-called energy envelope is calculated for each recorded waveform and a dynamic signal threshold is set. Then, for any signal with energy above this given threshold, a hit is recorded for the given antenna at that threshold-crossing time. In this way, hit patterns are generated for each event which, in the third step, are checked for consistency with incoming planar radio waves.

The ``energy'' envelope is in principle calculated as the RMS of a sliding \unit{5}{\nano\second} time window of the voltage data to enhance the signal-to-noise ratio. For each event and antenna channel a threshold is defined as $\mu_E + 4 \cdot \sigma_E$, with $\mu_E$ being the average and $\sigma_E$ the RMS of the full energy envelope of a waveform. Whenever an envelope crosses the threshold, a hit is recorded for the given channel with a coarse timing precision of $\unit{5}{\nano\second}$. The hits of all channels taken together form a hit pattern (Figure \ref{fig_hitPatterns}).
\begin{figure}[t]
\subfigure[]{
\includegraphics[width=0.91\columnwidth]{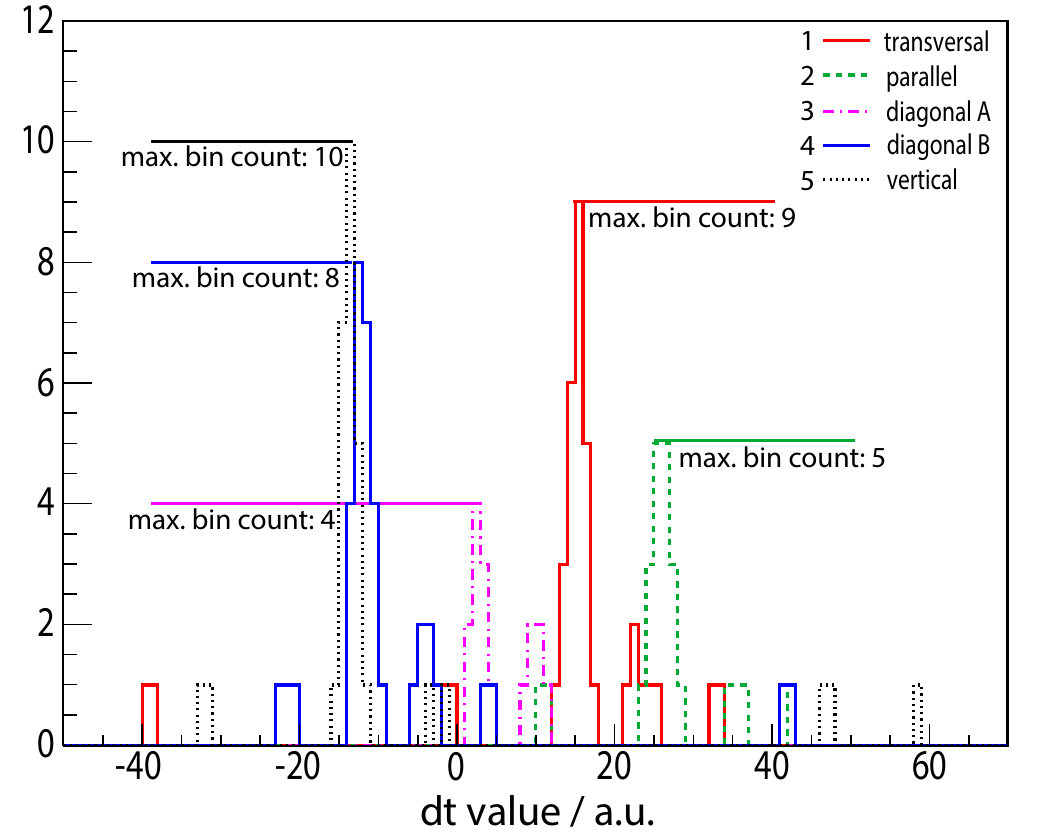}
\label{fig_signalHisto}
}
\subfigure[]{
\includegraphics[width=0.91\columnwidth]{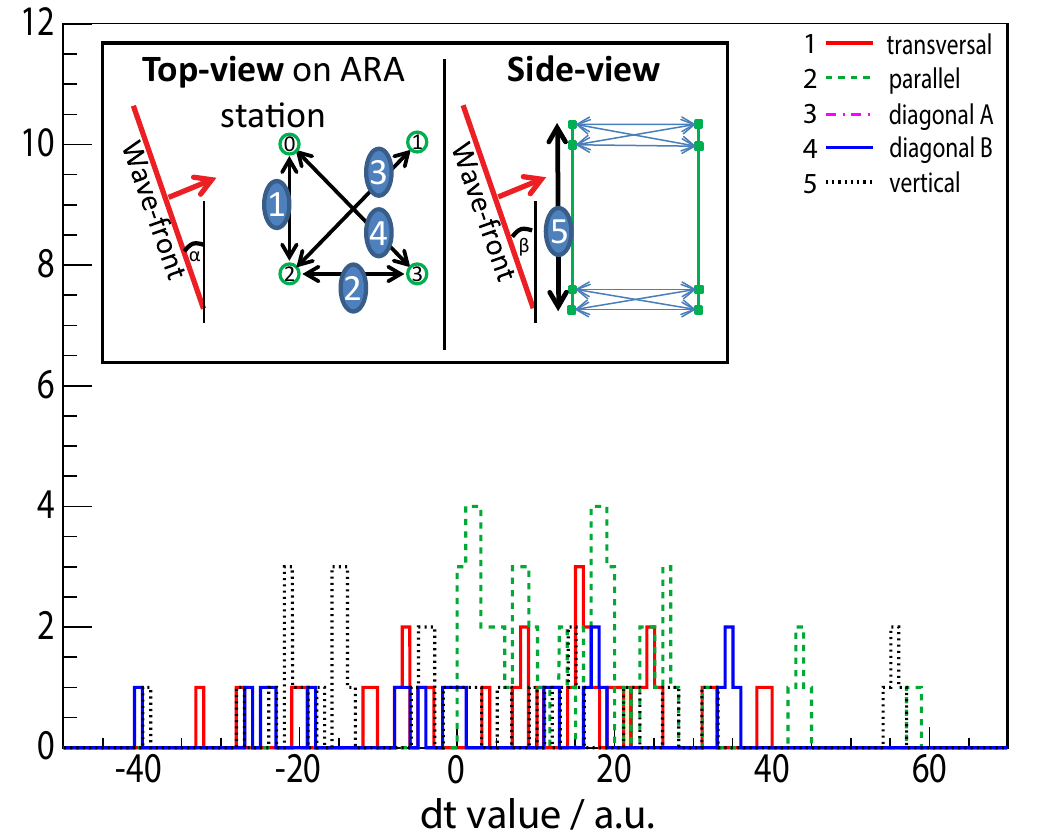}
\label{fig_noiseHisto}
}
\caption{(a) The hit time differences in histograms for each of the five geometrical groups for the event shown in Figure \ref{fig_signalPattern}. The pair groups of the histograms are schematically shown as inlay in the bottom plot. The quality parameter here $(QP)=1.6$. (b) The hit time difference histograms for the noise event shown in Figure \ref{fig_noisePattern}. In this second case the quality parameter equals $0.5$.}
\label{fig_QPHistos}
\end{figure}

In the next step, pairs are formed from antennas at roughly the same depth (horizontal pairs) and antennas on the same string (vertical pairs). For pairs with the same geometrical orientation, the time difference, divided by the antenna distance, is filled into a common histogram as shown in Figure \ref{fig_QPHistos}. This figure illustrates how radio signal patterns and thermal noise patterns are separated in the presented algorithm. In total there are five groups of pairs with the same geometrical orientation. For an incoming plane wave these histograms are expected to show a strong peak while they should be flat for thermal noise. The normalized sum of the maximum bin counts from each histogram is used as the time sequence quality parameter (QP), to distinguish incoming wavefronts from thermal noise.

\begin{figure}[tb]
\centering
\includegraphics[width=0.9\columnwidth]{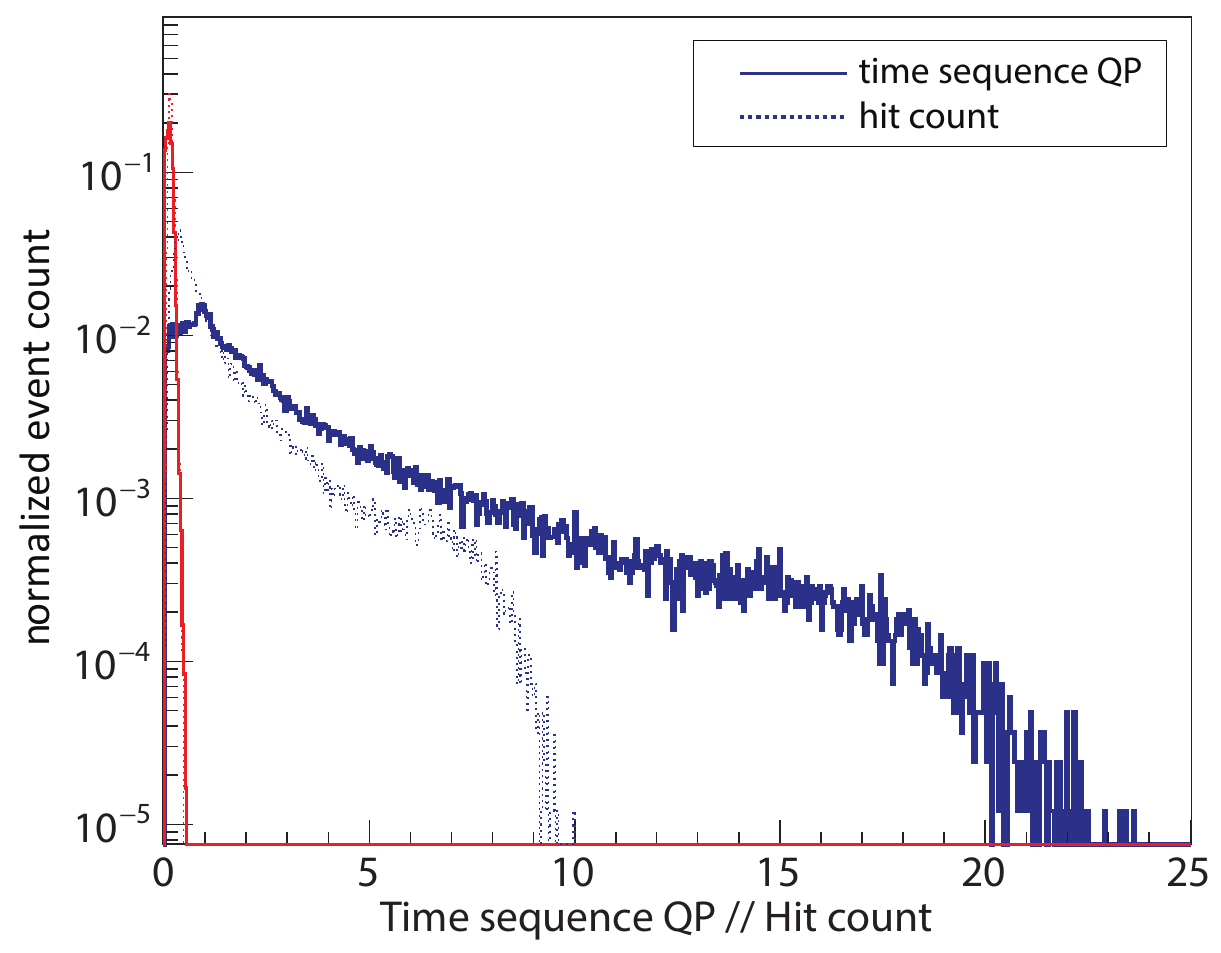}
\caption{The quality parameter $QP$ (solid line) compared to a simple count of all hits in a pattern as they appear in the example in Figure \ref{fig_hitPatterns} (dashed line) for simulated neutrinos with energies between $\unit{10^{16}}{\electronvolt}$ and $\unit{10^{21}}{\electronvolt}$  (blue) and thermal noise events (red). All values are scaled to cumulatively reach $99\%$ in both noise distributions at the same X value. The distributions are normalized to the total event count.}
\label{fig_QPdistribution}
\end{figure}

The noise rejection power of this filter is shown in Figure \ref{fig_QPdistribution} for simulated neutrinos between $\unit{10^{16}}{\electronvolt}$ and $\unit{10^{21}}{\electronvolt}$. In the range between $\unit{10^{18}}{\electronvolt}$ and $\unit{10^{19}}{\electronvolt}$, $92\%$ of neutrino signals are kept at $99.9\%$ noise rejection. The actual cut has to be tightened to provide adequate thermal noise rejection for the full data sample. However, as will be shown, the signal efficiency remains high.

\subsection{Vertex reconstruction} \label{sc_reconstruction}

\begin{figure}
\centering
\subfigure[]{
\centering
\includegraphics[width=0.9\columnwidth]{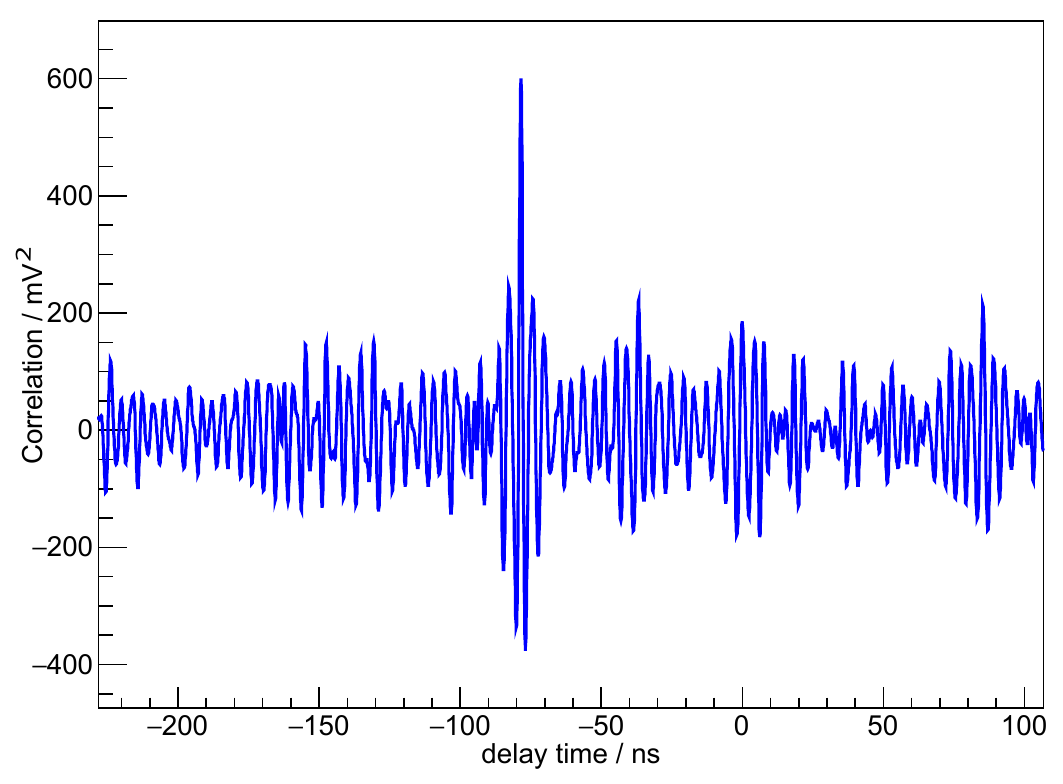}
}
\subfigure[]{
\centering
\includegraphics[width=0.9\columnwidth]{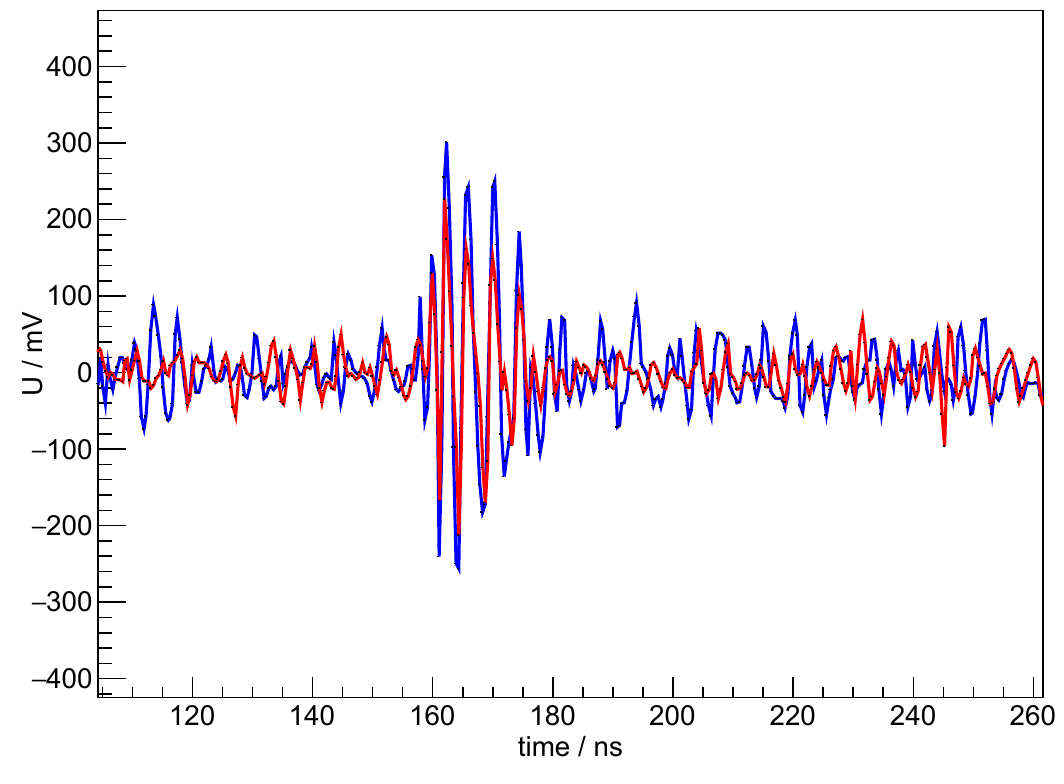}
}
\caption{(a) The cross-correlation graph for a calibration pulser signal using waveforms measured in two different antennas. (b) The two recorded waveforms after a shift by the time of maximum correlation $dt=\unit{-78.4}{\nano\second}$.}
\label{fig_correlationExample}
\end{figure}

After separating impulsive radio signals from thermal noise, we use a directional vertex reconstruction to distinguish neutrino-induced emission from anthropogenic noise. Here, we use the fact that man-made signals will reconstruct to the surface while only neutrino signals originate from within the ice itself. Furthermore, signals generated above the ice undergo refraction at the ice air boundary. Under the assumption of an index of refraction of $1.755$ at the antennas, the critical angle at this interface can be calculated to be $\unit{55}{\degree}$ and events produced in air are thus limited to a zenith angle between $\unit{90}{\degree}$ and $\unit{55}{\degree}$ as viewed in-ice from the ARA stations. Events which are generated directly at the ice boundary, for example by driving vehicles, will arrive with a minimal angle of roughly $\unit{40}{\degree}$. However, this is not a concern for winter running, when on-ice activities are minimized.

Background radio pulses from cosmic ray air showers can mainly be emitted from above the ice or as transition radiation from particle bunches at the ice-air boundary \cite{deVries2015}. For the ARA stations they will appear in the same angular region as the anthropogenic background described above. In the background estimation they will be treated as surface events. Only penetrating high energy muon bundles might be able to generate dense enough cascades in the deep ice via catastrophic energy loss to produce detectable radio signals. Studies of UHE neutrinos predict neutrino fluxes to be roughly $100$ times smaller than the cosmic ray flux at $\unit{10^{18}}{\electronvolt}$ \cite{Ahlers2010}. However, a study of the energy loss of muon bundles in South Pole ice below $\unit{1450}{\meter}$ indicates that at such a primary cosmic ray energy the probability for an energy loss of $\unit{10^{16}}{\electronvolt}$ over $\unit{5}{\meter}$, the lower end of the ARA sensitivity, is less than $10^{-6}$ \cite{Bai2009}. This gives us confidence that the background expected from muon bundles is small compared to neutrino signals. The actual amount of background, produced by such muon bundles in the ARA detector is currently under more detailed investigation.

The reconstruction algorithm developed for this analysis is based on a system of linear equations, formed from the signal arrival times on the different antennas.  In the ARA stations relative arrival time differences can be measured using the cross-correlation $g$ between two antenna waveforms $f_1$ and $f_2$ (Figure \ref{fig_correlationExample}):
\begin{eqnarray}
g(t) = f_1 \star f_2 =  \mathcal{F}^{-1}\left( (\mathcal{F}(f_1))^* \cdot \mathcal{F}(f_2) \right),
\end{eqnarray}
where $\mathcal{F}$ stands for the Fourier transform of a given function. The maximum of the correlation graph should occur at the delay time between the two input signals. With this method, a timing precision of $\unit{100}{\pico\second}$ on average can currently be achieved in ARA (see Appendix \ref{ch_digiCal}).

In principle, the time differences from all possible pairs can be used in the reconstruction. However, to exclude antennas that did not register a signal waveform from the algorithm, the maximum correlation amplitude in the correlation graph is used as a selection criterion for good antenna pairs. For a pair to be selected, the squared correlation amplitude has to cross a dynamic threshold adapted to the overall signal amplitude in an event, but at least a fixed lower limit above the product of the integrated power of the two correlated waveforms. All pairs passing the threshold are initially used in the reconstruction prior to a more refined channel pair selection performed in further steps, which is based on quality criteria applied to the outcome of the reconstruction. With the time differences found for good antenna pairs, a system of equations is set up, using the equality between the distance to the signal source and the measured travel time:
\begin{eqnarray}
c^2 ( t_{v} - t_{i} )^2 &=& ( x_v - x_i )^2 \\
& &\qquad +\ ( y_v - y_i )^2 \nonumber \\
& &\qquad \qquad +\ ( z_v - z_i )^2,\nonumber 
\end{eqnarray}
where $t_v$ is the time of emission at the vertex, $t_i$ the time of reception by the antenna $i$ and $x, y$ and $z$ the respective spatial coordinates. The speed of light $c$ is assumed to be constant and equal to the average speed at the station depth $c = \unit{0.3/1.755}{\meter \per\nano\second}$ (from \cite{Kravchenko2004}). The changing index of refraction with depth is not taken into account in this reconstruction, which has an influence on the zenith reconstruction precision. This, as will be shown, does not have a significant effect on the efficiency of the analysis.
\begin{figure}
\centering
\includegraphics[width=\columnwidth]{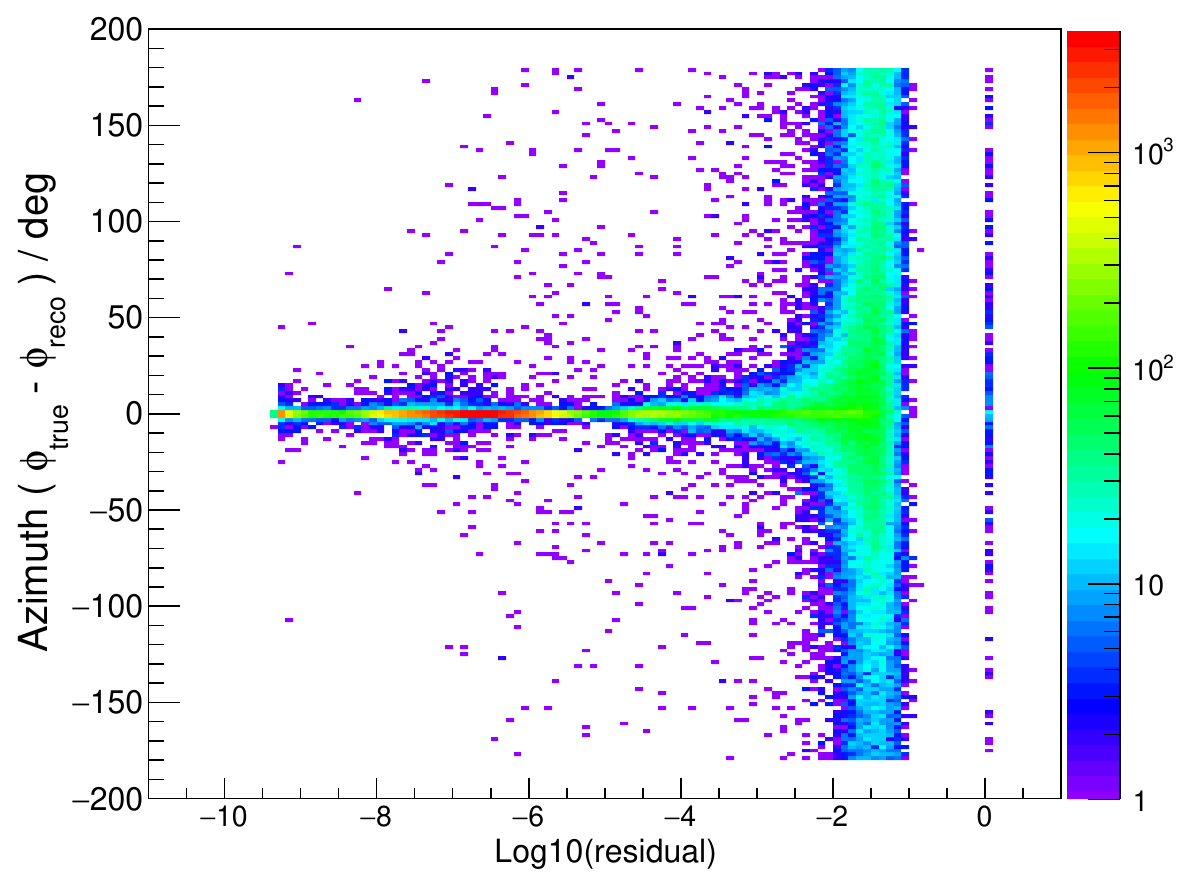}
\caption{The dependence of the azimuthal reconstruction on the residual (Equation \ref{eq_res}). The Y-axis indicates the difference between the reconstructed angle and the true angle for 165000 simulated neutrino events with energies between $\unit{10^{16}}{\electronvolt}$ and $\unit{10^{21}}{\electronvolt}$. Events with a high residual triggered the detector but do not contain strong enough signal to be properly reconstructed and are likely thermal in origin.}
\label{fig_phiVsResidual}
\end{figure}

When subtracting this relation for pairs of antennas, and after some reordering, one can obtain for each pair of antennas $i$ and $j$:
\begin{eqnarray}
\boldsymbol{x_v} \cdot 2 x_{{ij}} + \boldsymbol{y_v} \cdot 2 y_{{ij}} + \boldsymbol{z_v} \cdot 2 z_{{ij}} - \boldsymbol{t_{v,\mathrm{ref}}} \cdot 2 c^2 t_{{ij}} \nonumber \\
		 = r_i^2 - r_j^2  - c^2 (t_{{i,\mathrm{ref}}}^2 - t_{{j,\mathrm{ref}}}^2).     \label{eq_linEqBase}
\end{eqnarray}
Here, the index $ij$ indicates the difference between the values of a certain parameter for antennas $i$ and $j$, the parameter $r$ denotes the distance to the center of the coordinate system and the index ``ref'' indicates a reference antenna for which the signal arrival time is set to be $t_0=0$. This relation is used to set up a system of equations, linear in the vertex coordinates and emission time, represented by the matrix equation
\begin{eqnarray}
 \textbf{A} \vec{v} = \vec{b},      \label{eq_matrixEq}
\end{eqnarray}
with $\vec{v}$ containing the vertex coordinates and emission time and the matrix $\textbf{A}$ and the vector $\vec{b}$ offsets and arrival time differences. It should be noted that this approach is similar to Bancroft's solution of GPS equations, described in \cite{Bancroft1985}. The solution of this equation can be obtained using matrix decomposition tools \cite{NR2007,Eigen2014}. It is thus not seed-dependent and very fast, which allows us to perform several thousand reconstructions per second. Their precision and stability depend strongly on the precise knowledge of the relative antenna positions and possible time offsets between their recorded signals, which can be caused by cables or other electronic components. A calibration of the station geometry and systematic time delays between antenna waveforms has been presented in Section \ref{ch_geomCal}. For each reconstruction, a residual is calculated as
\begin{eqnarray}
res = \left| \frac{\vec{b}}{| \vec{b} |} - \frac{\textbf{A} \cdot \vec{v} }{|\textbf{A} \cdot \vec{v}|} \right|^2  \cdot \frac{1}{N_{\mathrm{chp}} },   \label{eq_res}
\end{eqnarray}
with $N_{\mathrm{chp}}$ the number of involved channel pairs. This indicates how well the reconstructed values fit the measured arrival time differences. Based on the residual, the channel pair selection is refined to further exclude noise antenna pairs if they have not been identified in the first step of channel selection.
\begin{figure}
\centering
\subfigure[]{
\includegraphics[width=0.97\columnwidth]{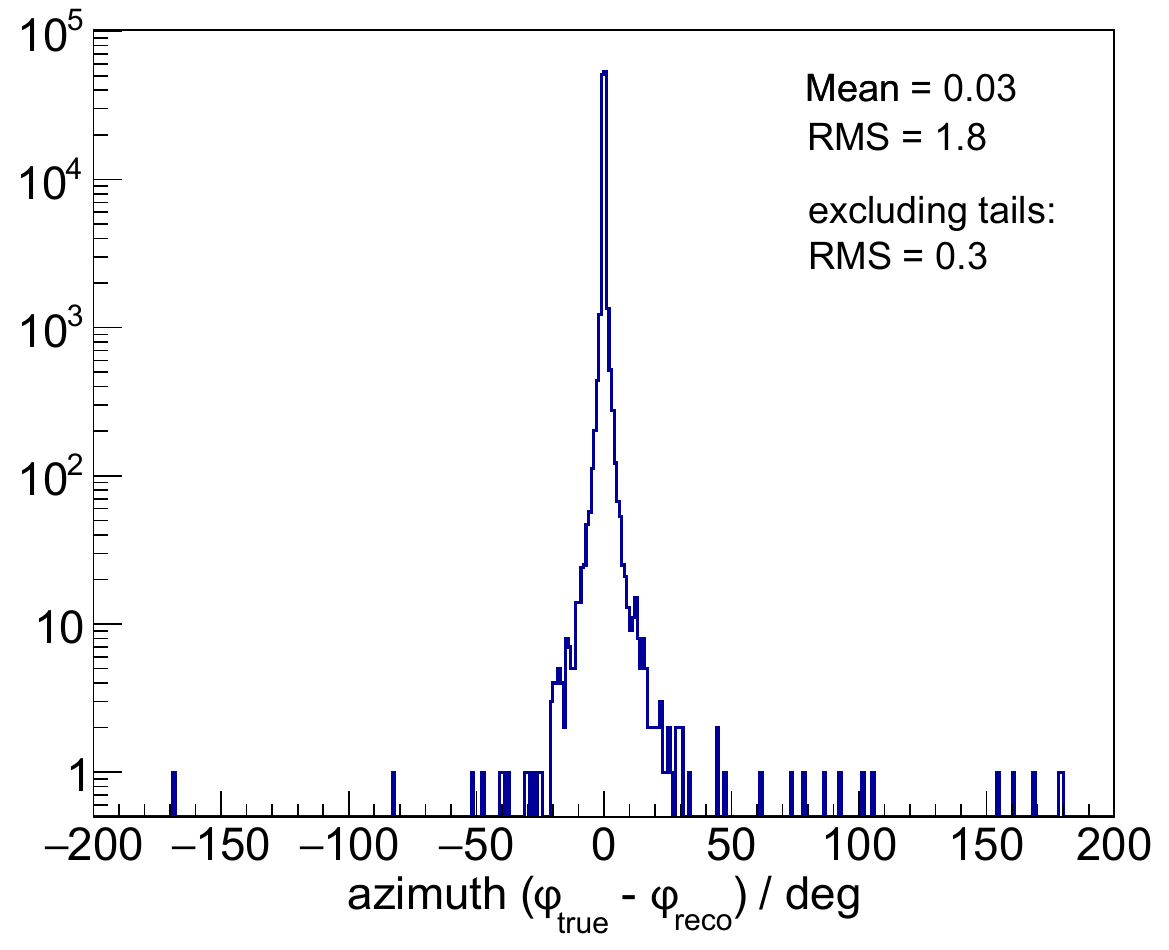}
}
\subfigure[]{
\includegraphics[width=0.97\columnwidth]{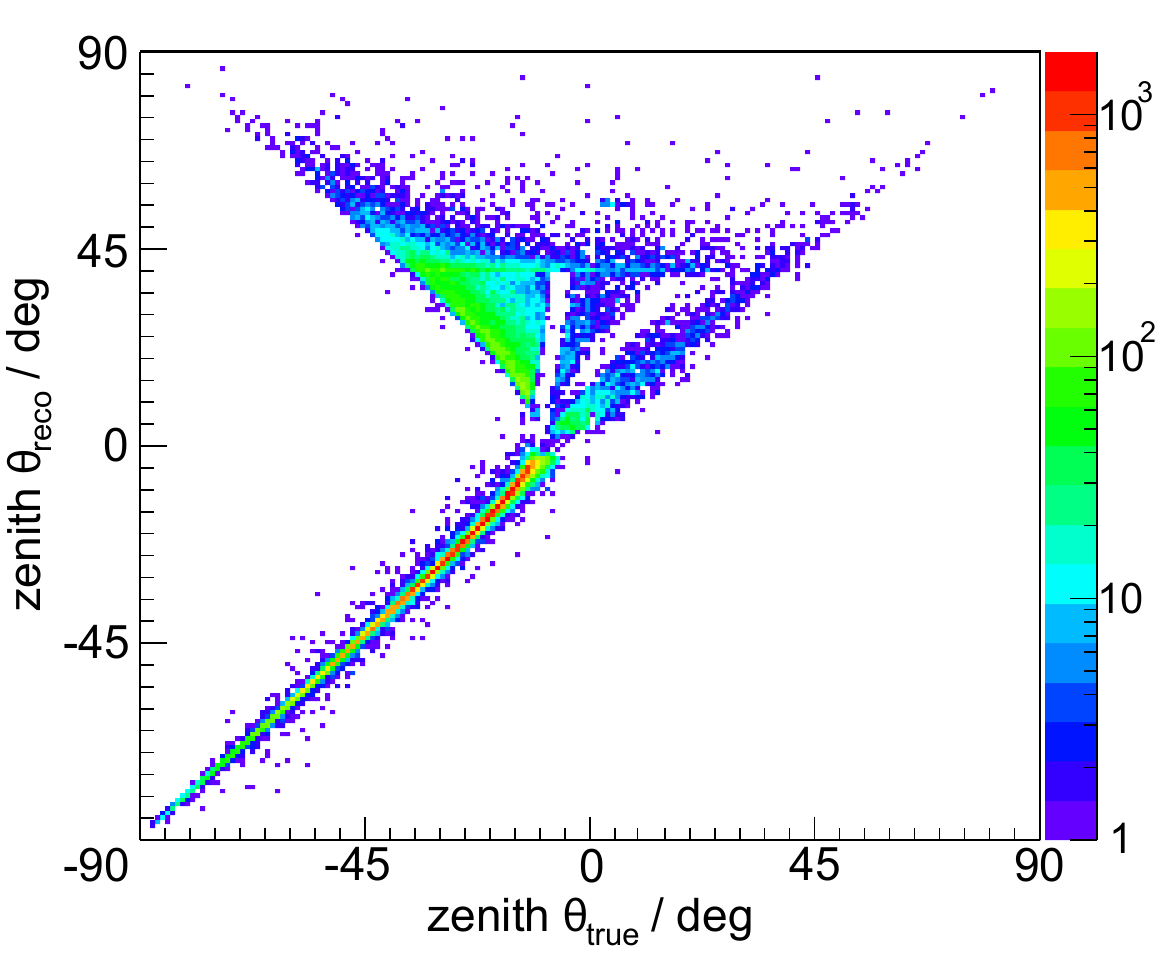}
}
\caption{
Simulated neutrino vertices reconstructed in azimuth and zenith angle with all quality criteria applied. (a) The difference between reconstructed and true azimuth. (b) The reconstructed zenith angle plotted as a function of the true zenith angle of each event.} \label{fig_anglePrecisionSim}
\end{figure}

The residual, alongside with some minor quality parameters, indicates whether a reconstruction is considered trustworthy. Figure \ref{fig_phiVsResidual} shows the dependence of the azimuthal reconstruction on the residual. For low residuals, a high quality reconstruction can be obtained, while reconstructions with a residual above $10^{-2.5}$ appear to point to locations that are broadly distributed in azimuth with respect to the true value. The residual also rejects more thermal noise events in favor of signal events although that is not its main purpose.

The result of reconstructions from a set of simulated neutrinos after application of all reconstruction quality criteria is shown in Figure \ref{fig_anglePrecisionSim}. While for the azimuth reconstruction a precision of better than $\unit{2}{\degree}$ ($\unit{0.3}{\degree}$ when excluding the tails) can be achieved, the zenith reconstruction is significantly degraded by surface reflections and ray-tracing effects. About $30\%$ of the events show behavior which causes the reconstruction to miss by several degrees in zenith angle. This causes an efficiency loss of roughly $6\%$ due to application of our angular cuts on radio waves coming from the surface, which is not dramatic.
\begin{figure}[!t]
\centering
\subfigure[]{
\includegraphics[width=\columnwidth]{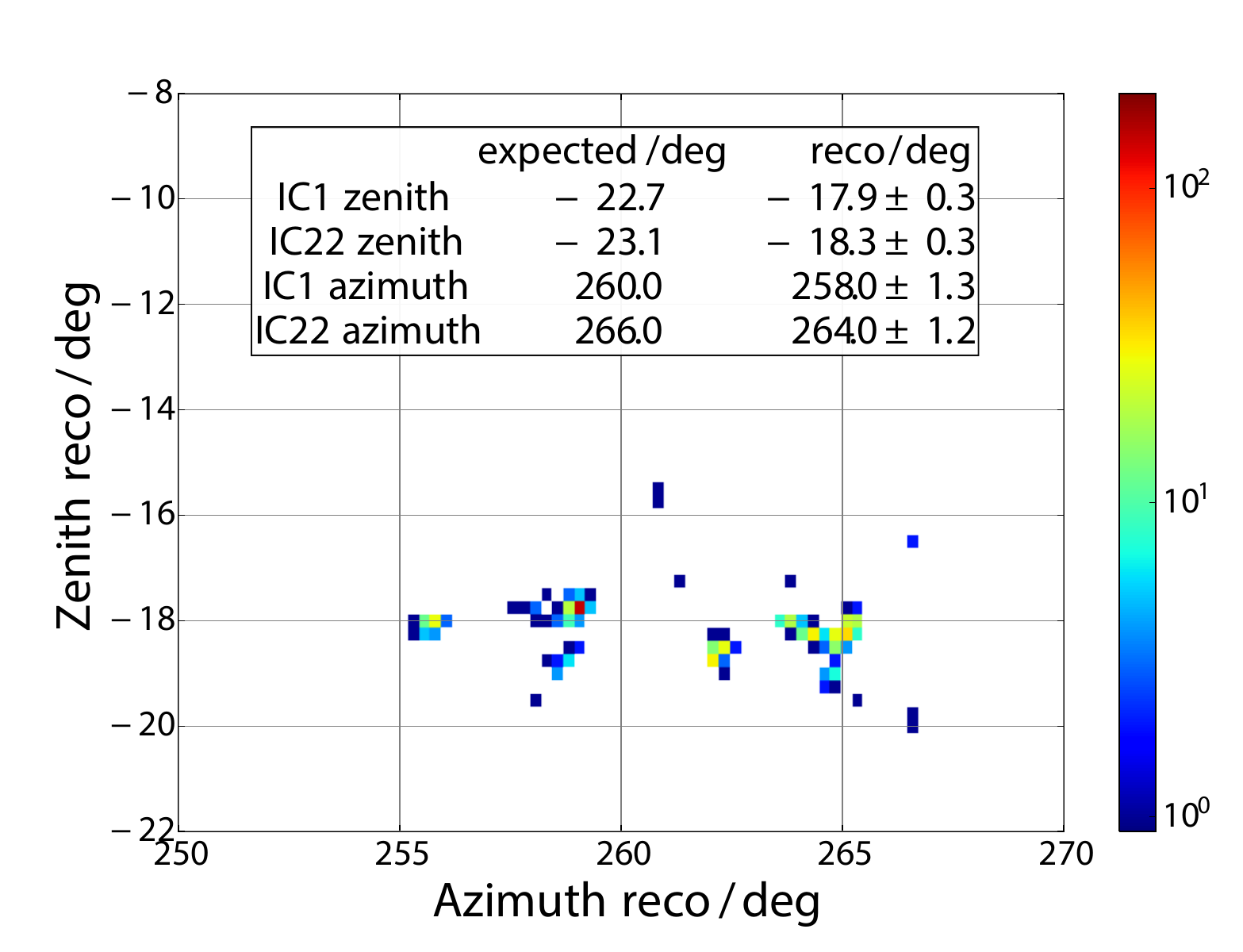}
}
\subfigure[]{
\includegraphics[width=\columnwidth]{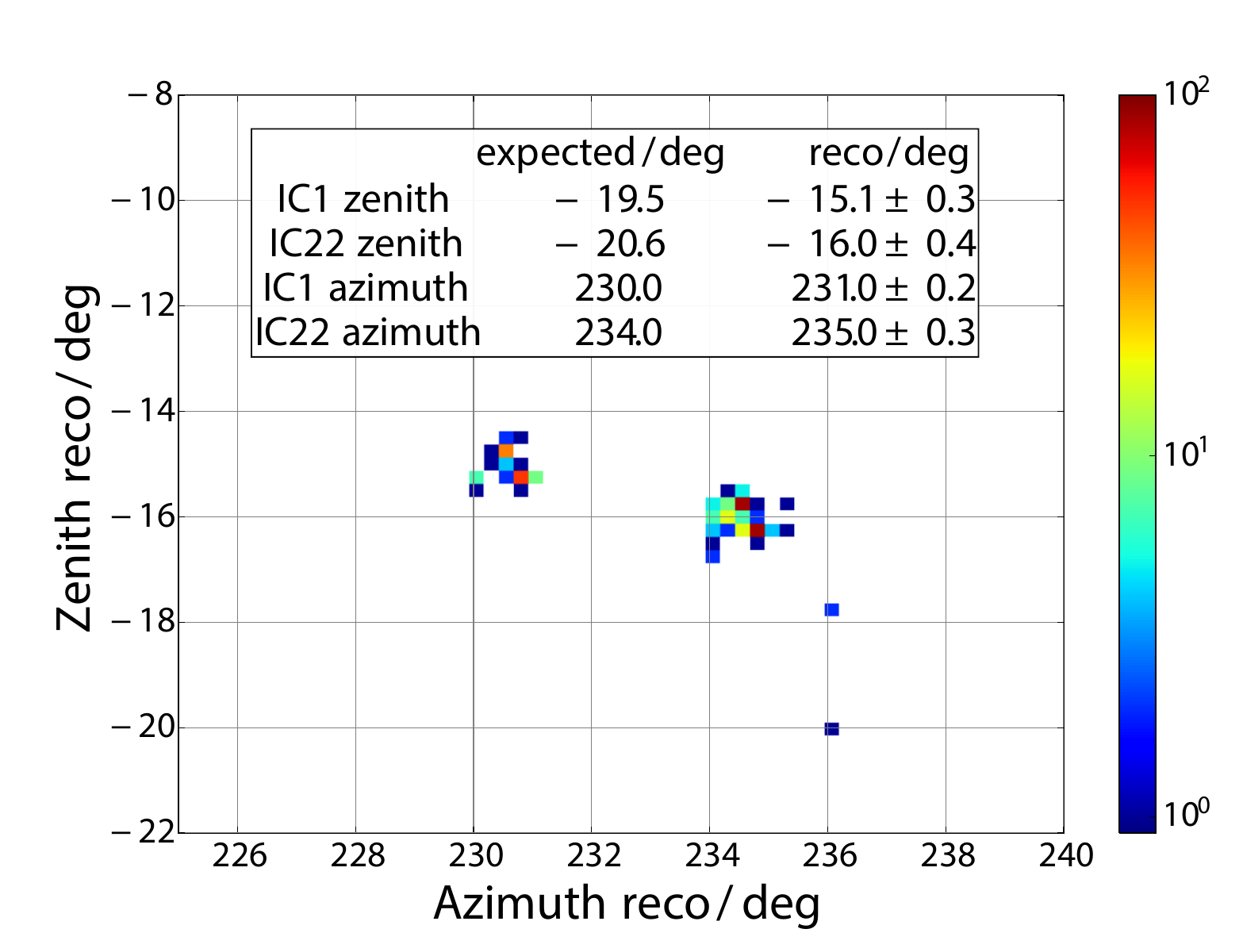}
}
\caption{
Reconstruction of pulsers deployed in IceCube holes in the deep ice (a) with A2, (b) with A3. The plots include data for the expected positions and the reconstructed positions with their standard deviation. The influence of systematic differences of a few degrees between the true and reconstructed angles on neutrino identification is negligible.}
\label{fig_deepPulserReco}
\end{figure}

Figure \ref{fig_deepPulserReco} shows the reconstruction of two pulsers deployed at a depth of $\unit{1450}{\meter}$ in the ice on IceCube strings 1 and 22 deployed in the final season of IceCube construction. Their distance to the ARA stations is roughly $\unit{4000}{\meter}$. These pulsers are therefore our most neutrino-like calibration tool. The plots show that both pulsers can be reconstructed with good precision by both stations. In addition to the roof pulser, this is another external source which confirms that the reconstruction algorithms should work properly for neutrinos.

\subsection{Cuts and background estimation}		\label{ch_dataCuts}
\begin{figure}[b]
\centering
\subfigure[]{
\includegraphics[width=0.895\columnwidth]{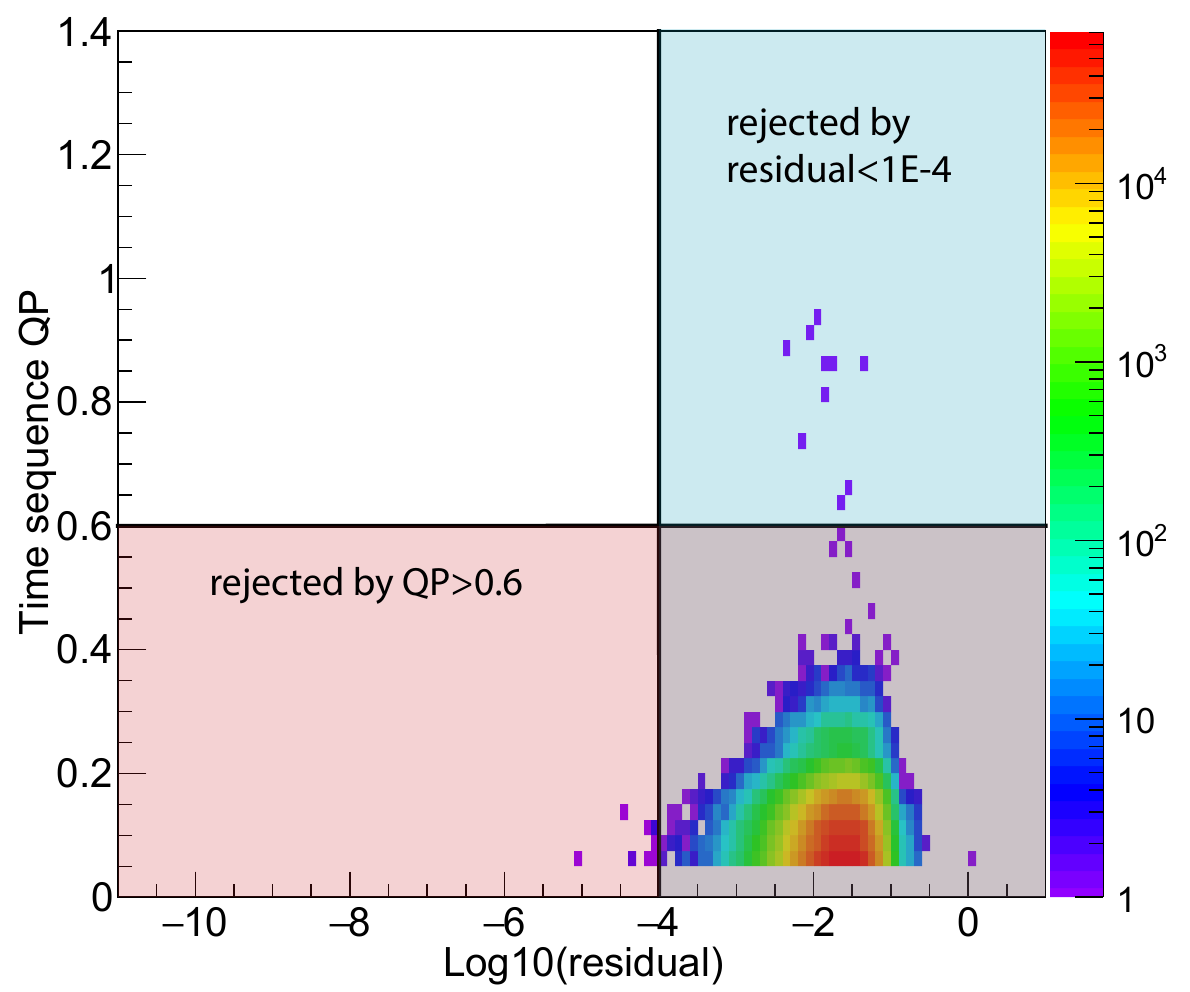}
}
\\
\subfigure[]{
\includegraphics[width=0.895\columnwidth]{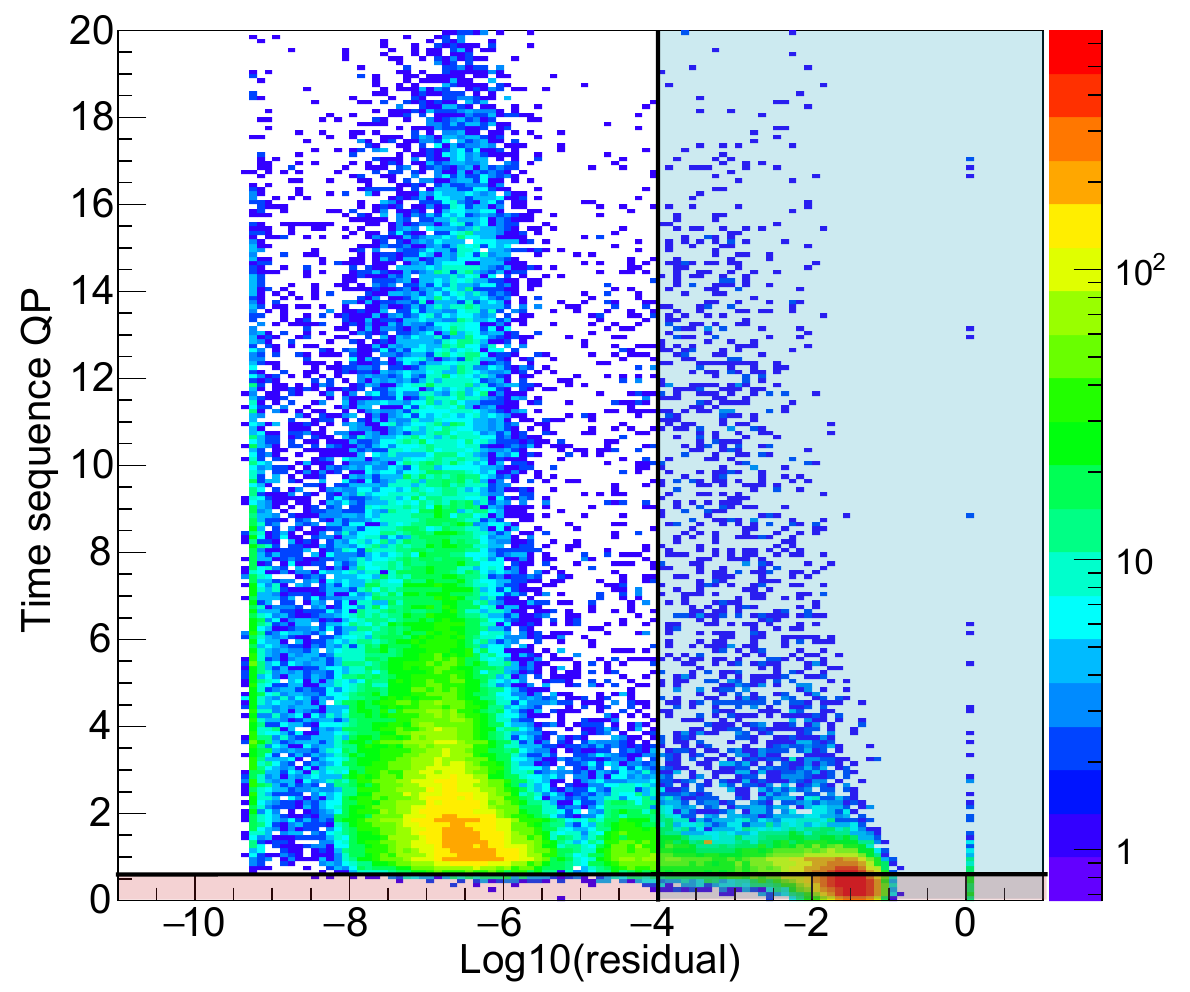}
}
\caption{(a) The distribution of recorded data in the two main cut parameters. All data correlated to known radio source locations (calibration pulser, surface) by reconstruction are removed. Hence, the remaining events can be thermal noise, mis-reconstructed radio events or neutrinos. (b) The distribution of simulated signal (see Section \ref{ch_simulation}) in the two main noise parameters.}
\label{fig_QPRes}
\end{figure}

Based on the presented algorithms, three cuts are used to distinguish neutrino signals from thermal and anthropogenic background. Thermal noise must be reduced by a factor $10^{-10}$ to reach the goal of ten times less events than the expected number of neutrinos. This can be achieved mainly by requiring a time sequence quality parameter of at least $0.6$ to select an event. Furthermore, to select high quality reconstructions and reject the remaining thermal noise, the residual is required to be less than $10^{-4}$. In this way only well-reconstructed impulsive radio signals are kept, which can further be reduced by angular cuts. As the cut values for the time sequence parameter and the reconstruction residual have not been optimized in a strictly systematic way, we note that there might be room for improvement in a subsequent analysis employing these algorithms. Figure \ref{fig_QPRes} illustrates the steep decline of the noise distribution towards the cut boundaries in the two main cut parameters. As a representative thermal noise sample, all data from the $10\%$ burn-sample which cannot be correlated to a known signal source (pulsers, surface activities) are used. Simulated signal events from the dataset described in Section \ref{ch_simulation} are distributed broadly compared to that noise sample.

Angular cuts are placed around the known locations of calibration pulsers inside the ice and are specific to a given station. In addition, a surface cut is applied, rejecting all events reconstructed to a zenith angle of $\theta > \unit{35}{\degree}$ for A2 and $\theta > \unit{40}{\degree}$ for A3. This cut can be a bit looser for A3 since the reconstruction errors are smaller for this detector (see Figure \ref{fig_deepPulserReco}). The reason for this difference in precision is due to the different number of channels in both stations which are available for reconstruction. Whereas all 16 channels are used in station A3, in A2 only 13 channels are included in vertex reconstruction. One channel, D4BH, is broken, while two other channels show a puzzling timing offset which could not be removed in the geometrical calibration. The cut values are chosen for each angular requirement separately, such that each allows less than $0.01$ background events to enter the signal sample in the full data set. The number of background events expected to pass a given cut is estimated from the $10\%$ data subset by fitting an adequate Gaussian or exponential function to the tail of an event distribution close to each cut. The best fit parameters and the position of the cut are used to obtain the number of background events expected to leak from the angular region being excluded. The uncertainty on the number of background events is derived from the fit errors. Note that calibration pulser events are normally tagged by the DAQ as calibration events and excluded from the analyzed data sample. However, due to possible mis-tagging, pulser events may leak into the final sample. Therefore, all pulser events are taken into account in the background estimation, even if they are not part of the $10\%$ burn-sample. This is a very conservative estimate but strengthens the analysis against mis-tagged pulser events.

After these cuts, the background expectation for the full data recorded in the year 2013 is $0.009 \pm 0.010$ for A2 and $0.011 \pm 0.015$ for A3. The number of neutrinos expected to be observed by the combined two-station detector from the flux prediction in \cite{Ahlers2010} for a crossover energy from Galactic to extragalactic cosmic ray sources of $E_{min} = \unit{10^{18.5}}{\electronvolt}$ amounts to $0.10 \pm 0.002 (stat)$ events.

\section{Results and cross checks}
\begin{figure}[t!]
\centering
\subfigure[]{
\includegraphics[width=\columnwidth]{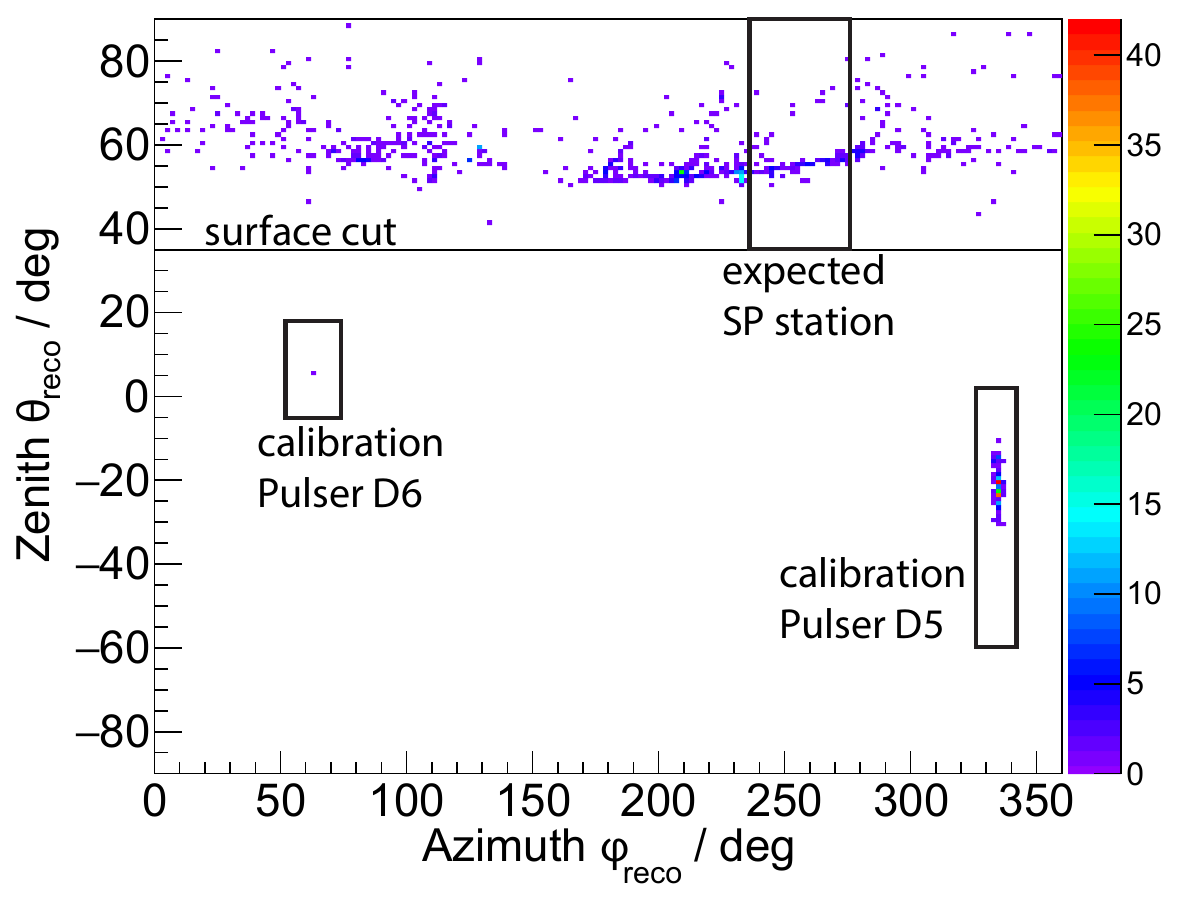}
\label{fig_finalEventSelectionARA02}
}
\subfigure[]{
   \includegraphics[width=\columnwidth]{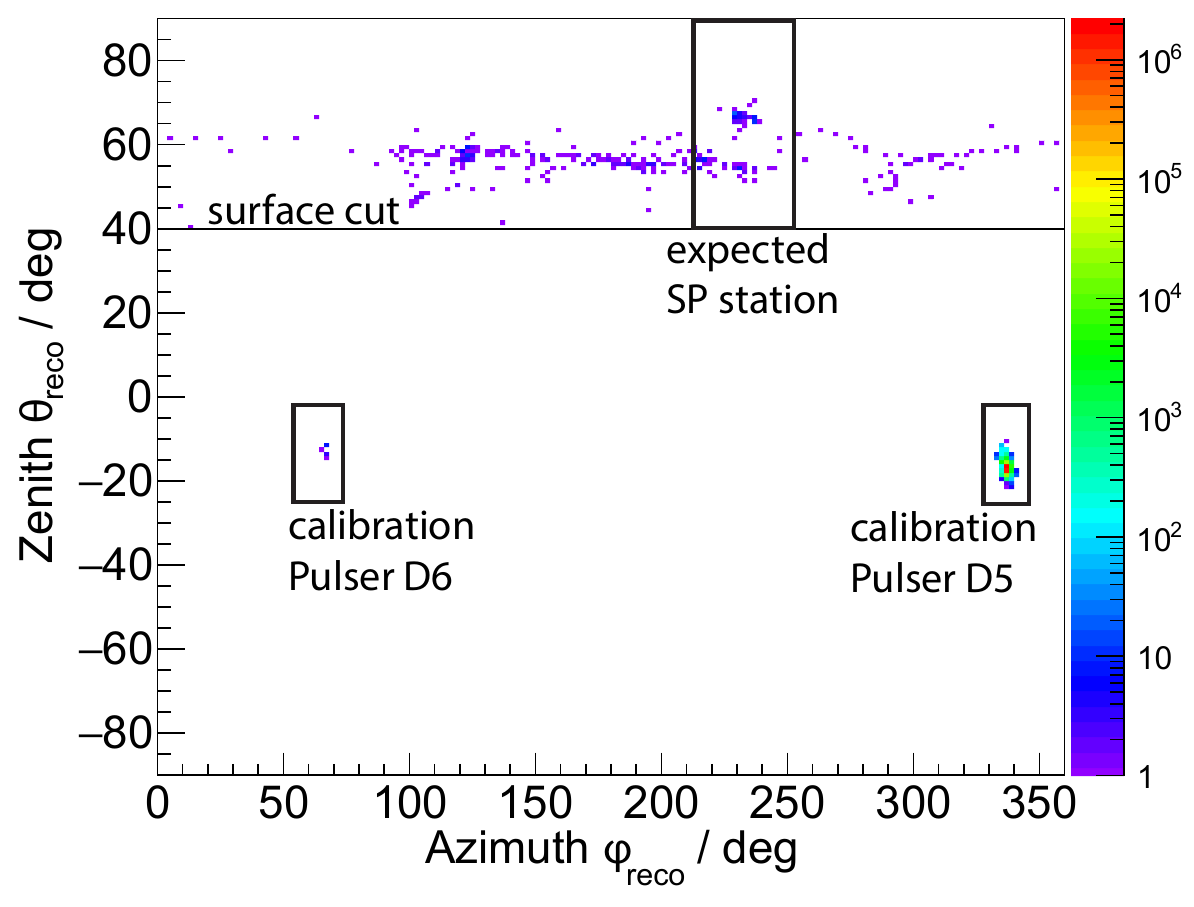}
\label{fig_finalEventSelectionARA03}
}
\caption{Reconstructed events that passed the thermal noise and reconstruction quality cuts for (a) A2 and (b) A3. The black boxes indicate the angular cut regions around the calibration pulser positions and the black line indicates the surface cut. Events inside the squares and above the surface line are rejected.}
\label{fig_finalEventSelection}
\end{figure}
\subsection{Results}

\begin{figure}
\centering
\subfigure[]{
\includegraphics[width=0.9\columnwidth]{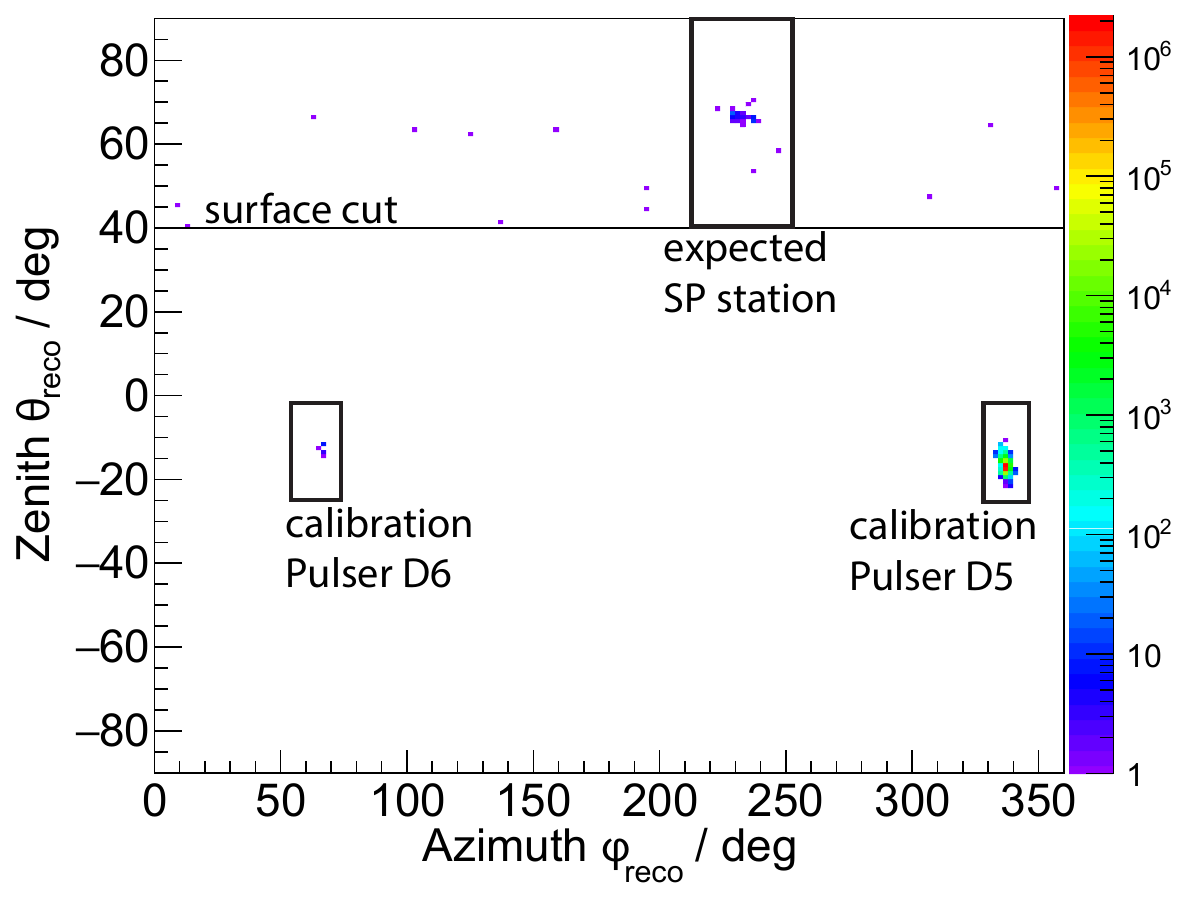}
\label{fig_winterEventsARA03}
}
\subfigure[]{
   \includegraphics[width=0.9\columnwidth]{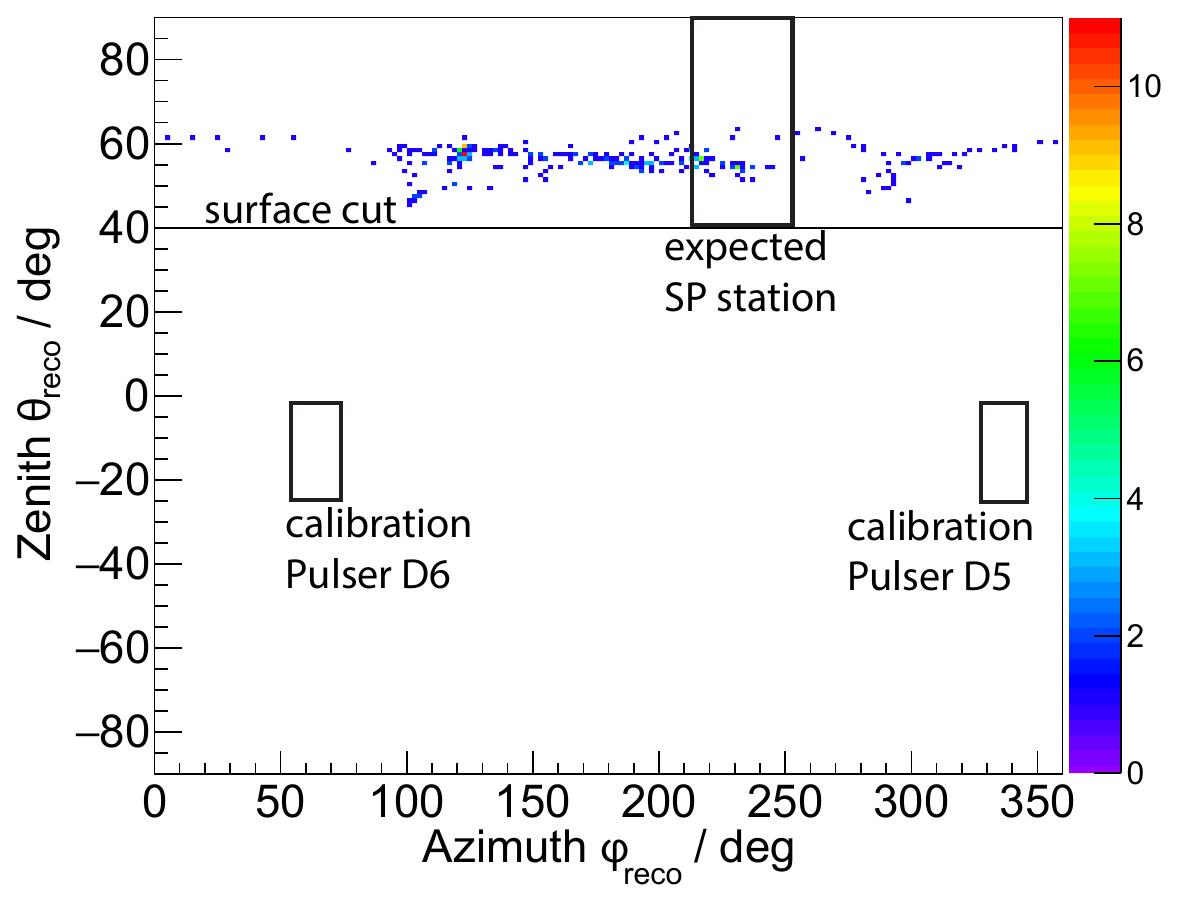}
\label{fig_summerEventsARA03}
}
\caption{Reconstructed events that passed the thermal noise and reconstruction quality cuts for A3 (a) in the austral winter and (b) in the austral summer when station activity is maximal. Note that due to improper tagging of calibration pulser events in the first months of winter, many such events entered the final data sample. This is however accounted for in the final angular cuts of the analysis.}
\label{fig_WinterSummerEvents}
\end{figure}

The results of the above described analysis are summarized in the two sky maps in Figure \ref{fig_finalEventSelection}. No events are found outside the angular cut regions which implies that no neutrino candidates have been observed. This agrees with the expectation of $0.1$ signal and roughly $0.02$ background events in the two stations. The difference between summer and winter source locations at South Pole is presented in Figure \ref{fig_WinterSummerEvents}, showing the impact of human activities during summer which is limited to surface events.

\begin{figure}[t]
\centering
\includegraphics[width=\columnwidth]{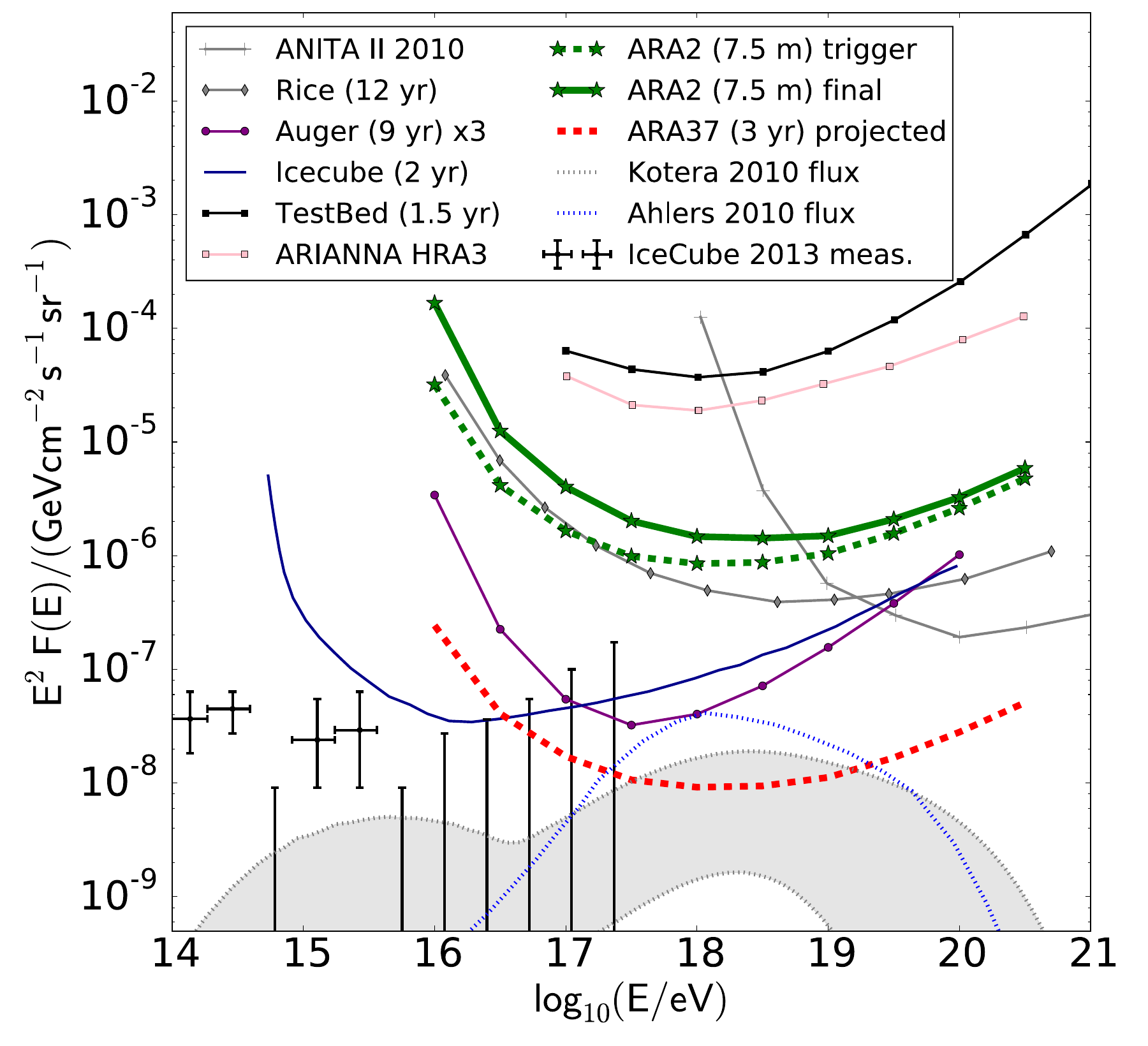}
\caption{Neutrino limits and sensitivities from various experiments including the 7.5 months data analysis of the two ARA stations described herein. Systematic errors, as derived in Section \ref{ch_sysError}, have been accounted for in the ARA2 limit. The ARA37 (3 yr) sensitivity is projected from the ARA2 (7.5 m) trigger level without accounting for systematic errors ($K(E)=2.44$). Data for other experiments are taken from \cite{TestBed2014,Aartsen:2013dsm,Gorham2010,RICE2011,Aab:2015kma,Aartsen:2014gkd}. The neutrino fluxes are derived in \cite{Ahlers2010} and \cite{Kotera2011}. See Appendix \ref{app_Emin1} for $E F(E)$ scaling on the y-axis.}
\label{fig_SensAnaLevel}
\end{figure}

With this result from two ARA stations (designated as ``ARA2'') we can calculate a differential limit on the neutrino flux in the sensitive energy region as shown in Figure \ref{fig_SensAnaLevel}. For the neutrino energy range between $\unit{10^{18}}{\electronvolt}$ and $\unit{10^{19}}{\electronvolt}$, the energy where most neutrinos are expected to be observed, the limit is $E^2 F_{\rm{up}}(E) =\unit{1.5\times10^{-6}}{\giga\electronvolt/\centi\meter^{2}/\second/\steradian}$.
\\
 This limit is calculated as
\begin{eqnarray}
E^2 F_{\mathrm{up}}(E) &=&  E^2 S(E) \cdot \frac{K(E)}{dE} \nonumber \\
 &=&  E S(E) \cdot \frac{K(E)}{\ln(10)},
\end{eqnarray}
where the factor $K(E)$ is derived with the construction described in \cite{Feldman1998} as the $90\%$ Poisson confidence limit for no observed events under the expectation of zero background. Systematic errors on the signal efficiency, as described in Section \ref{ch_sysError}, are accounted for in this factor following the method presented in \cite{Conrad2002} with the improvement proposed in \cite{Hill2003}. The error caused by uncertainties on the cross section is not taken into account. $S(E)$ denotes the sensitivity of the detector which is calculated from the effective areas $A_{\mathrm{eff}}$ and live times $T$ of each detector as
\begin{eqnarray}
S(E) = \frac{1}{4 \pi \cdot ( A_{\mathrm{eff,2}} \cdot T_2 + A_{\mathrm{eff,3}} \cdot T_3) }.      \label{eq_sensitivity}
\end{eqnarray}
Furthermore, the limit is presented as a half decade interpolation for a logarithmic energy scale with a resolution of $\mathrm{dLog_{10}}(E) = 1$ logarithmic bins. Therefore we obtain:
\begin{eqnarray}
dE = E \cdot \ln(10) \mathrm{dLog_{10}}(E) = E \cdot \ln(10).
\end{eqnarray}
The resulting limit for two ARA stations 10 months after deployment is not yet competitive with the current best limits from the IceCube detector. In spite of this they show, when projected to the full size of ARA37, that the completed detector is expected to be sensitive to mainstream models for neutrinos from the GZK process.

\subsection{Cross checks}	\label{ch_crossChecks}

Although the limit obtained with the currently deployed ARA stations is not competitive yet, this first data analysis proves the capabilities of the full ARA detector. Cross checks have been performed with events that are observed coincidentally in both stations to demonstrate that the employed algorithms select impulsive radio signals and that the directional reconstruction works.  Events that have passed both the thermal noise and reconstruction quality cuts are considered coincident if they trigger both stations within a time window of $\unit{12}{\micro\second}$. This time corresponds to the maximal in-ice travel time of signals between the two stations given their separation of $\unit{2}{\kilo\meter}$. Sequences of such events, appearing in short time frames of roughly $\unit{60}{\second}$ have been found which show evidence of originating from airplane communication transmitters.
\begin{figure}[t]
\centering
	\includegraphics[width=0.95\columnwidth]{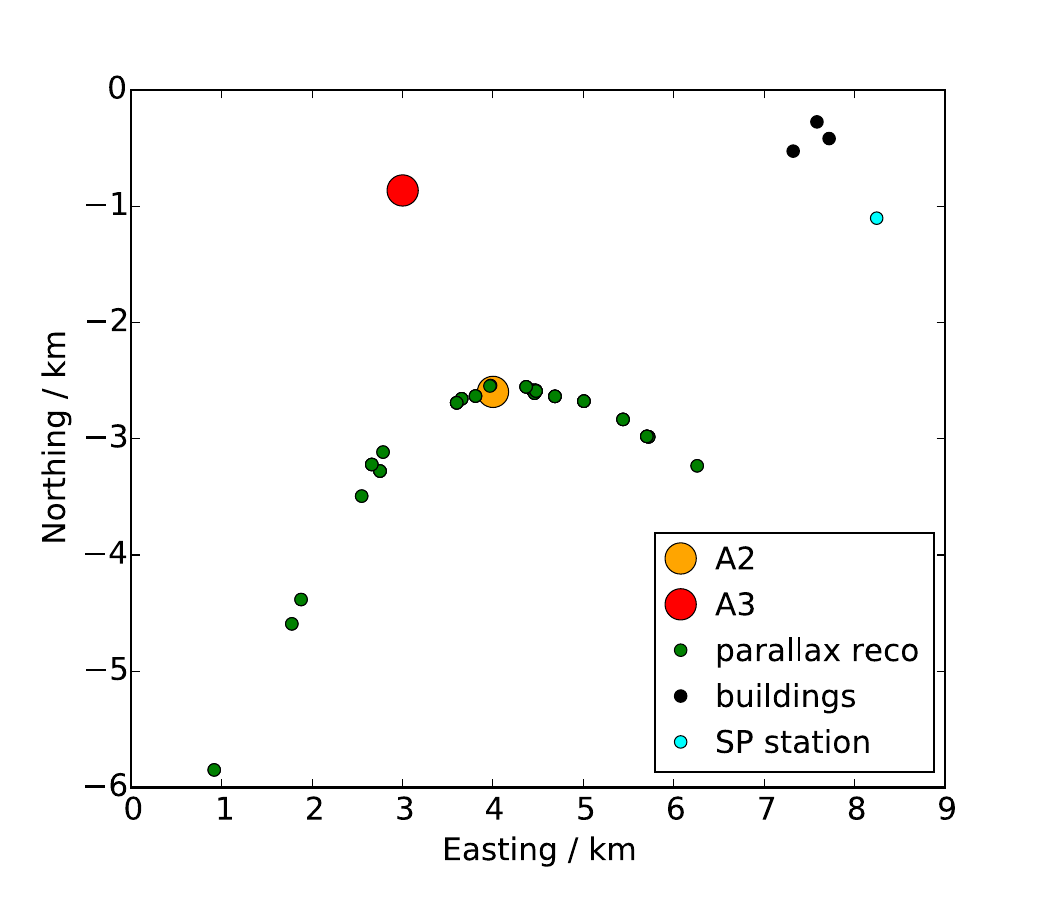}
\caption{A South Pole map with the two ARA stations and the XY reconstruction of an A2/A3 coincident event series via the parallax method (green dots). The events in this series are distributed over a time of $\unit{50}{\second}$ starting from the rightmost point in the map.}.
\label{fig_parallaxReco}
\end{figure}

\begin{figure}[t]
\centering
\includegraphics[width=0.95\columnwidth]{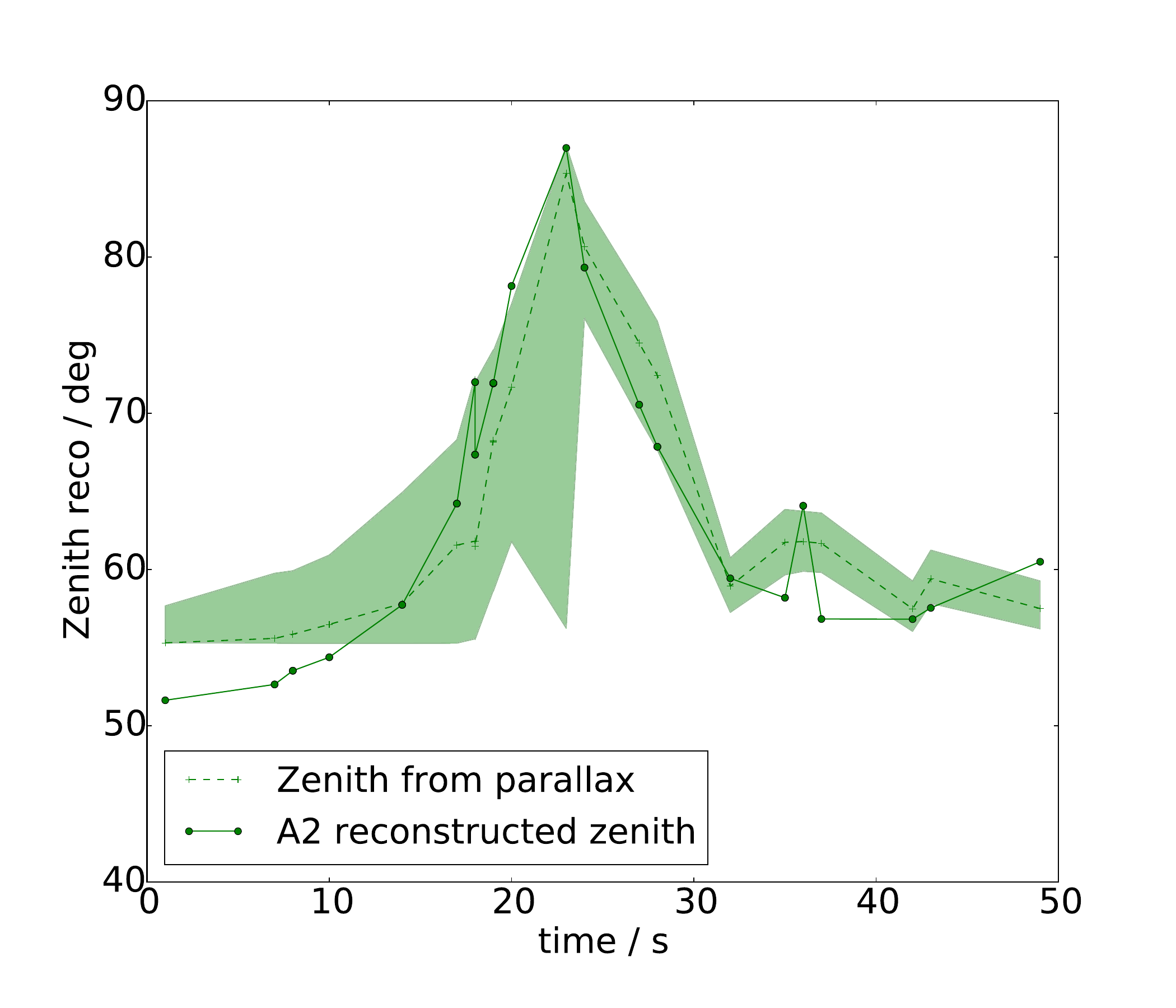}
\label{fig_zenithCrossCheck}
\caption{Comparison between the expected zenith from the A2/A3 XY parallax reconstruction (green dashed line) and the reconstructed zenith angle by A2 only (green solid line). Errors from the XY reconstruction, calculated from the errors on the reconstruction of a surface pulser with known position, are shown as a green band.
}
\label{fig_coincCrossChecks}
\end{figure}

With the azimuthal reconstruction from each station, the XY position of the source can be determined via trigonometric calculations. The crossing point of the two beams pointing to the azimuthal reconstruction of both stations is used as its XY position. This method is in the following referred to as the ``parallax reconstruction''. One particularly interesting event sequence is shown in Figure \ref{fig_parallaxReco}. The positions of the events within this hit series form a smooth track, indicating an emitting object that moves at a speed of several hundred $\unit{}{\kilo\meter \per \hour}$ at an increasing height of order $\unit{500}{\meter}$ above the ice surface. Waveforms recorded for events at different points of the sequence are presented in Appendix \ref{ch_ARA02_coincEvents}. This track is very useful as a cross check since it passes on top of station A2 which should be evident in the zenith reconstruction. In Figure \ref{fig_coincCrossChecks}, the expected zenith angle from the XY position is compared to the zenith reconstruction of station A2, showing good agreement within the error bars.

The track confirms that the detector is capable of observing radio sources. Furthermore, the agreement between different positioning methods and the smoothness of the reconstructed track are evidence that the used analysis tools work properly to identify such sources and reconstruct their position.

\section{Systematic uncertainties}     \label{ch_sysError}

The systematic uncertainties of the presented analysis result from errors on the theoretical models of neutrino interactions and radio wave propagation, as well as on the calibration of the detector. The error estimation is performed in a similar way as has been described in \cite{TestBed2014}.\\

\begin{figure}[t]
\centering
	\includegraphics[width=0.95\columnwidth]{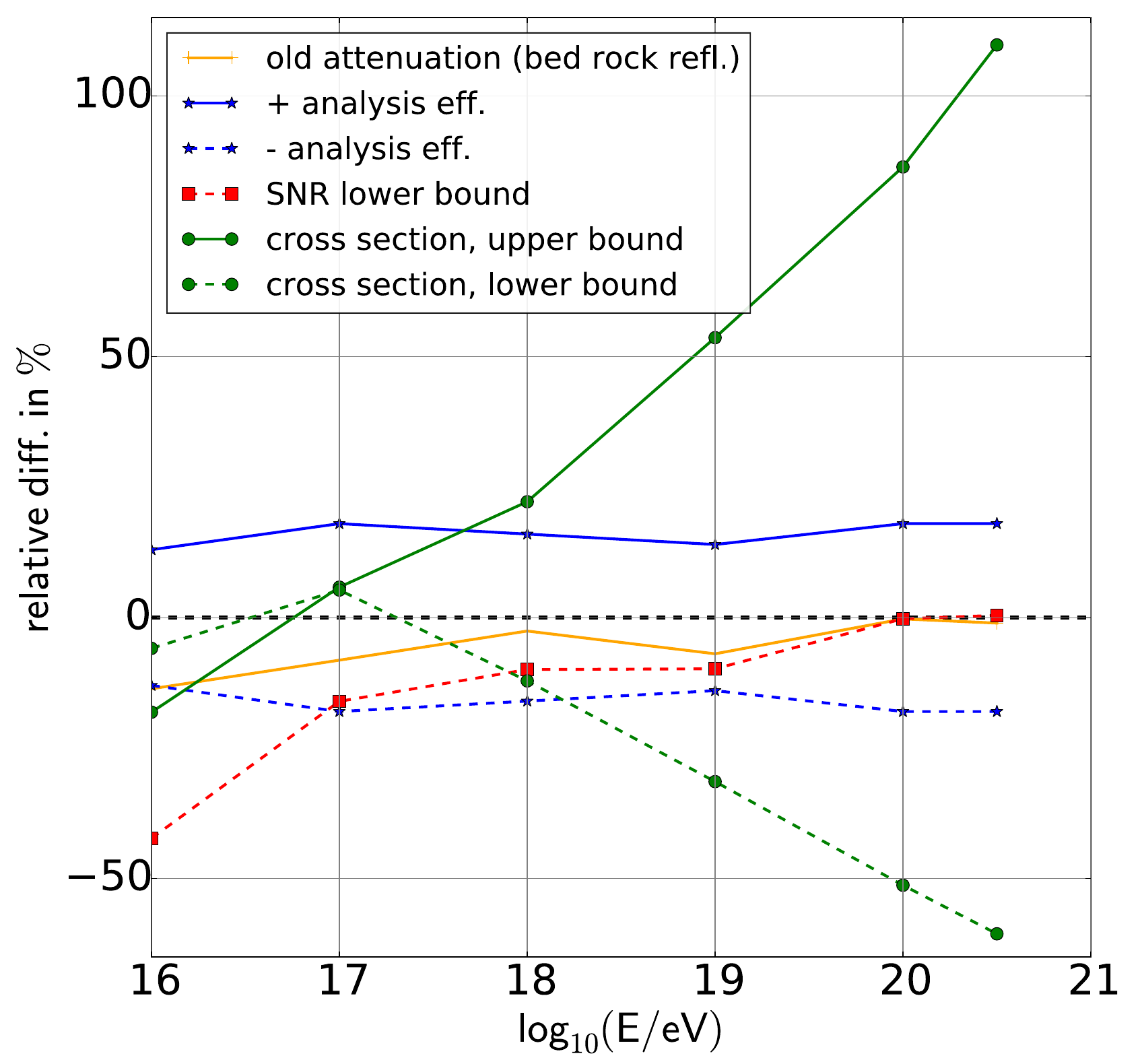}
\caption{The relative difference in effective area at analysis level, caused by various systematic error sources.}
\label{fig_errorsAeff}
\end{figure}

The neutrino interaction cross section at the energies of interest above $\unit{10^{16}}{\electronvolt}$ is calculated based on measurements at much lower energies and is thus subject to large uncertainties. To check the influence of this uncertainty on the effective area of the ARA detector, simulations are run using the upper and lower limits of the cross section estimates from \cite{Connolly2011}. The effect on the final analysis result can be seen in Figure \ref{fig_errorsAeff}. Especially for the highest energies, this is the dominant uncertainty in the analysis.\\

A further uncertainty results from the error on the radio attenuation length measurement in the South Pole ice sheet. This measurement has been performed using calibrated pulsers, deployed at different depths with the last IceCube strings \cite{Allison2012457}. With the obtained data, the local attenuation length at a given depth can be inferred with knowledge of the temperature and density profile of the ice. The difference between this result and an earlier measurement, using the bed rock under the ice as a reflector for radio waves emitted and received at the surface \cite{Barwick2005}, provides a measure of the uncertainty in the attenuation length. The error on the effective area is again obtained by comparing simulations with different sets of parameters. As visible in Figure \ref{fig_errorsAeff}, it contributes only slightly to the final error.

One should note that the uncertainty on the changing index of refraction inside the ice is not a major concern for ARA due to the deep deployment of the stations at $\unit{180}{\meter}$ below the ice surface. Below this depth the index of refraction does not change appreciably and approaches the value of $1.78$ for the deep ice.\\
\begin{figure}[t]
\centering
\subfigure[]{
\includegraphics[width=0.9\columnwidth]{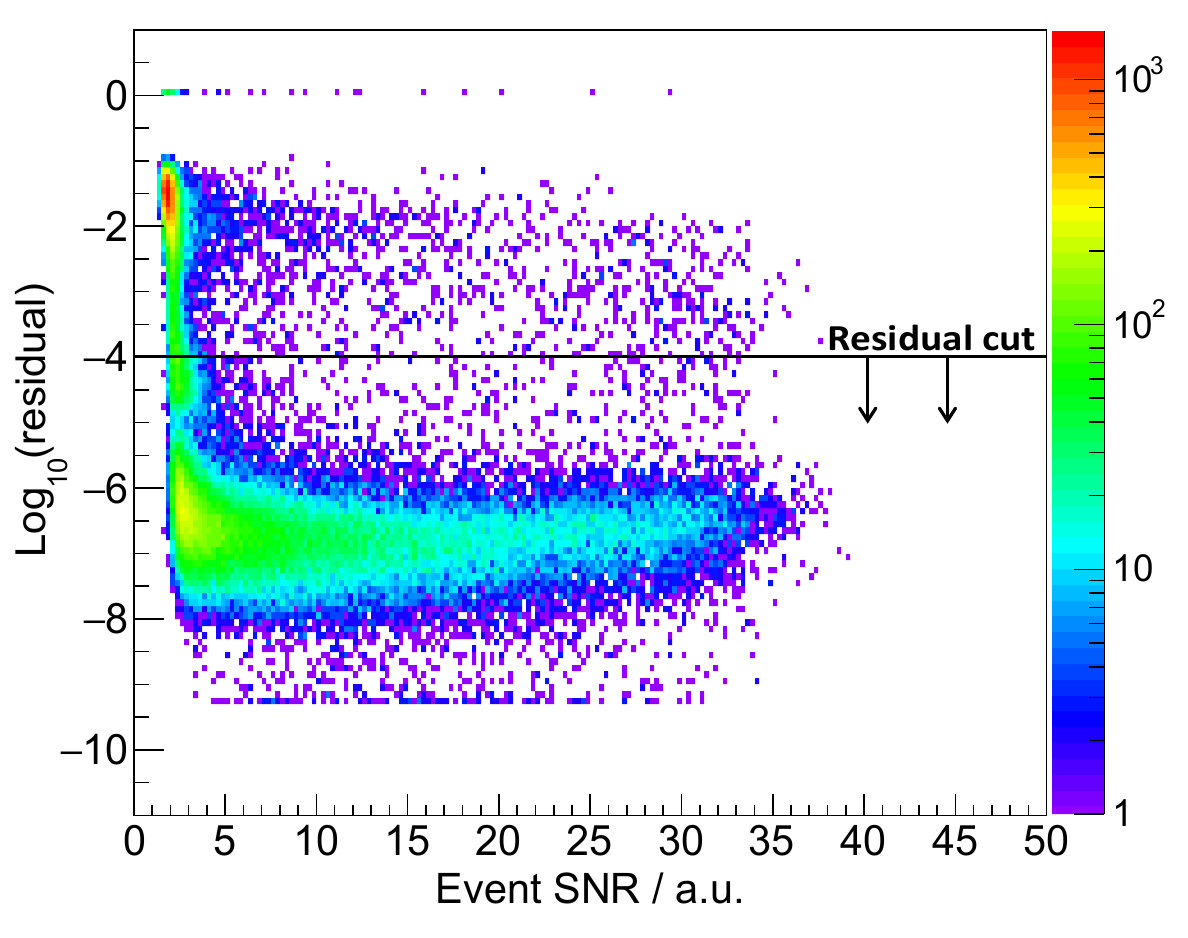}
}
\subfigure[]{
\includegraphics[width=0.9\columnwidth]{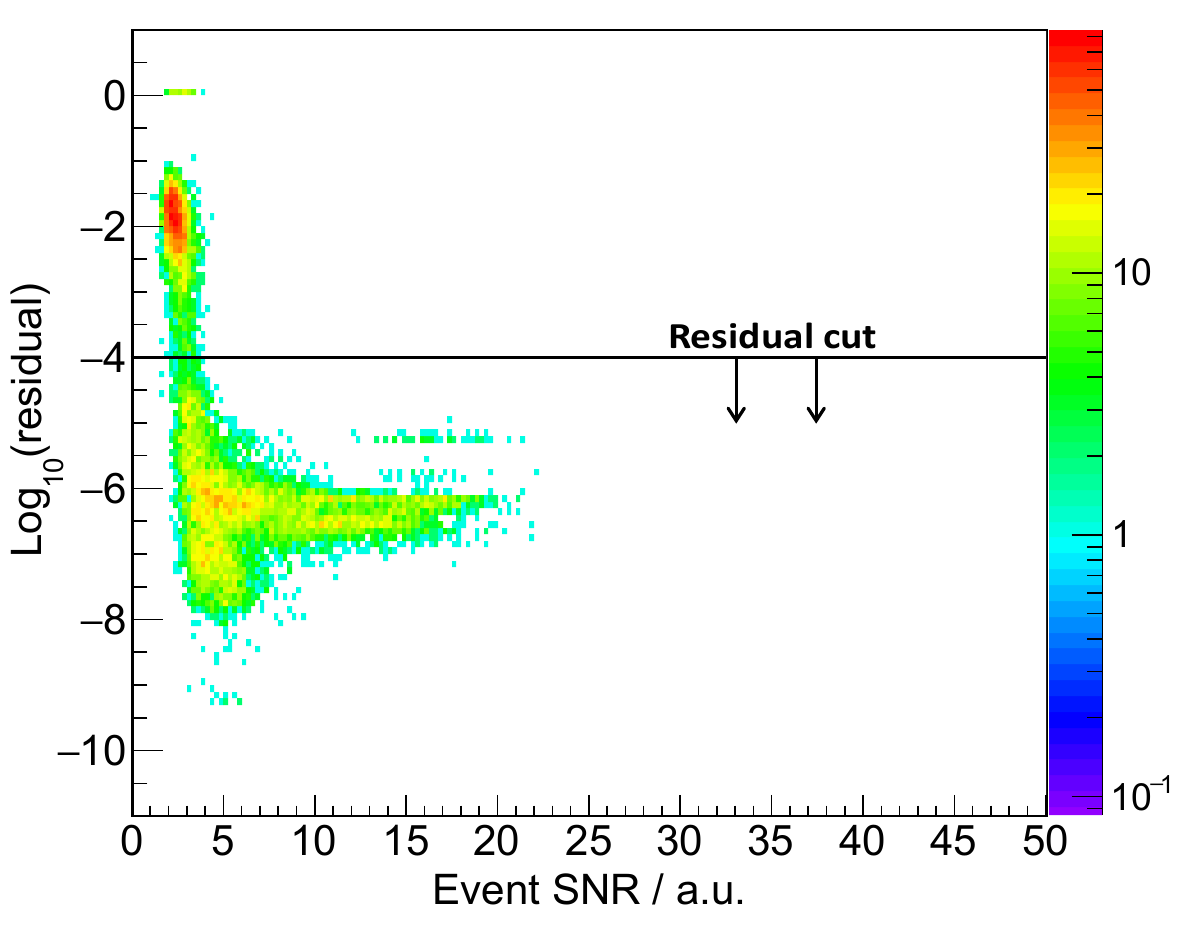}
}
\caption{The residual of reconstructed events as function of the $SNR$ for examples of (a) neutrino simulations in A3 and (b) calibration pulser signals recorded by A3.}
\label{fig_ARA03SNRComparison}
\end{figure}

One more important uncertainty relates to the final signal to noise ratio recorded by the signal chain. This depends on the assumed ambient noise, the antenna directivity, the transmission coefficient and the assumed noise figure as explained is Section \ref{ch_signalChain_discussion}.
To estimate the resulting systematic error on the effective area, the overall amplitude of the incoming radio signal is reduced according to the results from Section \ref{ch_signalChain_discussion}. Under these conditions, sets of neutrino events are simulated and the analysis is re-run. The presented analysis is nearly exclusively based on coarse envelopes of time domain waveforms or time differences, derived from cross-correlation. Therefore, precise knowledge of the frequency response is of secondary importance. Figure \ref{fig_errorsAeff} shows that the uncertainty on the signal chain calibration has the greatest impact at low energies, when most of the incoming signals are weak and the signal to noise ratio is low. At higher energies, this error loses importance compared to the error on the cross section. Due to the limited knowledge about the upper systematic errors in that measurement, as explained in Section \ref{ch_signalChain_discussion}, such a limit is currently not quoted. Only the lower error is used in the determination of the neutrino limit presented in Figure \ref{fig_SensAnaLevel}.\\

The last estimated uncertainty is the difference in the analysis efficiency obtained from simulated neutrinos and recorded data. The shape of noise and signal waveforms will not perfectly match between simulations and real data. Therefore, the used analysis algorithms are compared between real data from calibration sources and simulated neutrino signals. For this comparison, the signal to noise ratio (SNR) is chosen as a simple parameter, independent of the shape of the waveform. In Figure \ref{fig_ARA03SNRComparison}, calibration data from station A3 are compared to simulated neutrino events with energies between $\unit{10^{16}}{\electronvolt}$ and $\unit{10^{21}}{\electronvolt}$ for one of the two main cut parameters, the reconstruction residual. Generally, the distribution for simulations aligns well with the recorded data. The difference around the cut value seems especially small. 
\begin{figure}[t]
\centering
	\includegraphics[width=0.47\textwidth]{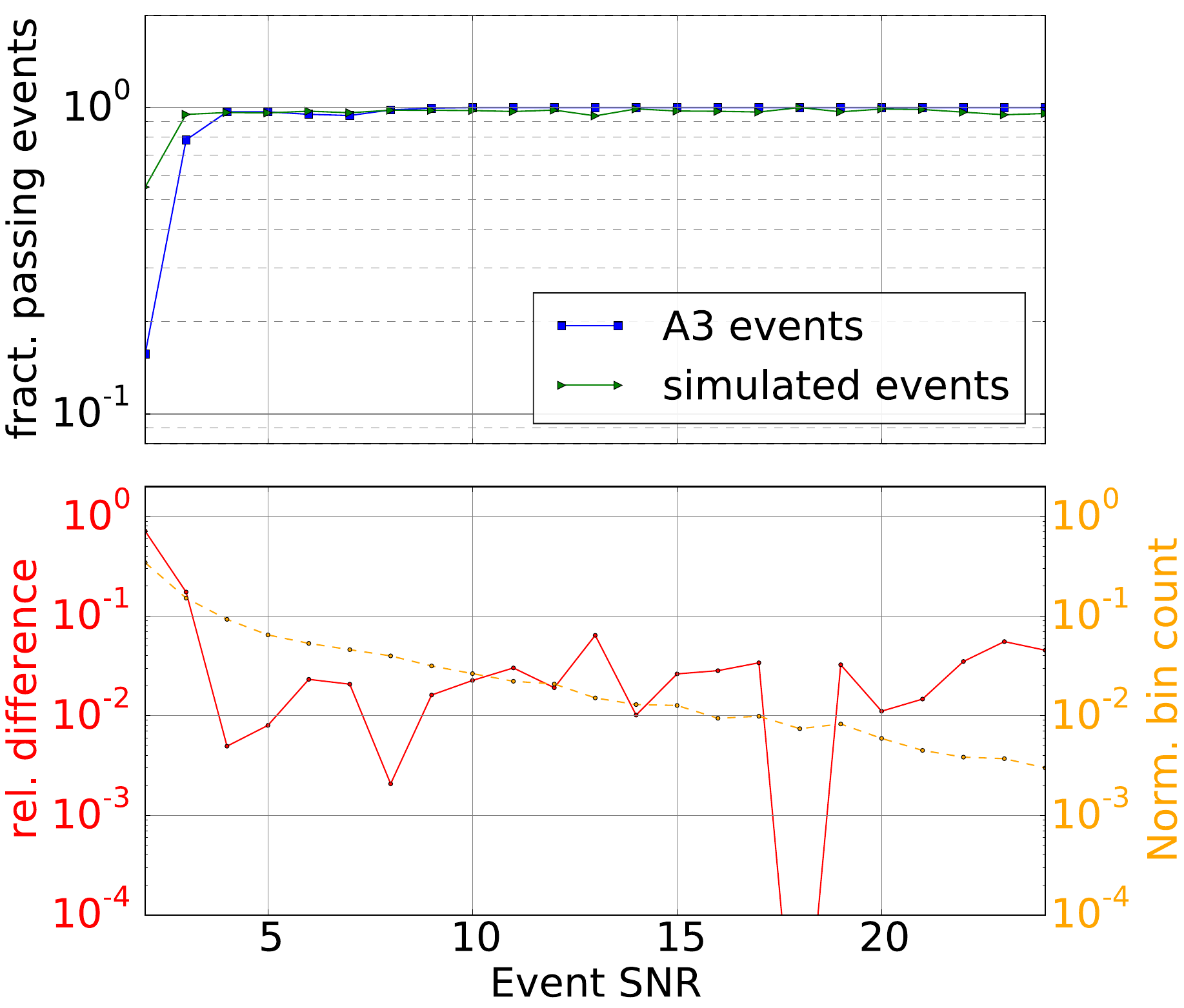}
\caption{Top panel: The analysis efficiency of the ARA stations for simulated events (green) and for calibration pulser events on A3 (blue). Bottom panel: The relative difference in efficiency between simulation and pulser data (red) and the normalized event count $N_i$ per SNR bin (orange). The shown simulated events are produced with a primary energy of $\unit{10^{18}}{\electronvolt}$.}
\label{fig_SNREfficiencyComp}
\end{figure}

For the systematic uncertainties it is interesting to quantify the analysis efficiency, i. e. the fraction of events that pass the applied cuts, for real data and simulations and to calculate their relative difference. To do so, one can compare its dependence on SNR after separating the simulated neutrinos into single decade energy bins. Figure \ref{fig_SNREfficiencyComp} shows this comparison for a neutrino energy of $\unit{10^{18}}{\electronvolt}$. The difference $|\Delta_{\mathrm{eff},i}|$ is plotted relative to the efficiency in simulation. For each energy an average difference in efficiency, weighted by the number of events in each SNR bin, is calculated as
\begin{eqnarray}
\sigma_{\mathrm{sys}} = \frac{\sum \left( |\Delta_{\mathrm{eff},i}|\cdot N_i \right) }{\sum N_i},
\end{eqnarray}

with $N_i$ the normalized number of events per SNR bin. This difference is propagated as a systematic error on the analysis efficiency into the limit calculation. As visible in Figure \ref{fig_errorsAeff}, this is a non-negligible contribution to the systematic error. However, with a better knowledge of the detector, simulations will become more precise and this uncertainty can be reduced.

The systematic uncertainties calculated for the number of recorded neutrinos are summarized in Table \ref{tab_sysErrors}.

\begin{table}[h!]
\caption{The systematic error from various sources on our signal expectation of 0.10 neutrino events for a flux prediction from \cite{Ahlers2010}.}    
\centering
\begin{tabular}{c c c}
\hline 
\hline \noalign{\smallskip}
Source & positive error & negative error \\  \noalign{\smallskip}
\hline
Cross section & $0.035 (34\%)$ & $0.020 (19\%)$ \\
Attenuation & $-$ & $0.005 (5\%)$\\
Signal Chain & $-$ & $0.011 (10\%)$\\	
Analysis efficiency & $0.018 (17\%)$     & $0.018 (17\%)$   \\  \noalign{\smallskip}
\hline   \noalign{\smallskip}
TOTAL & $0.040 (38\%)$ & $ 0.030 (29\%)$ \\  \noalign{\smallskip}
\hline
\hline \noalign{\smallskip}
\end{tabular}
\label{tab_sysErrors}
\end{table}

\section{Summary and outlook}


We have demonstrated the power of ARA as an UHE neutrino detector through an analysis of data from the two deep stations currently in operation. Through calibrations, good timing precision and geometrical understanding of the detector have been achieved, which are key factors in the detection of radio vertices. Furthermore, initial analysis algorithms have been presented. These show a good efficiency for retaining signal ($60\%$ efficiency at $\unit{10^{18}}{\electronvolt}$) with background rejection by $10$ orders of magnitude, leaving   0.02 events, at the current trigger settings. Thermal noise can be rejected by simple algorithms to a high level and radio vertices can be reconstructed with an angular precision of a few degrees. This relatively simple reconstruction algorithm is found to have a RMS of  $<\unit{2}{\degree}$ in azimuth reconstruction and appears to be stable for most cases of zenith reconstruction (see Figure \ref{fig_anglePrecisionSim}) without accounting for ray tracing effects. Improvements and alternatives to the algorithm are currently being developed.

In addition, cross checks confirm that the used analysis algorithms select radio signals from background and return sensible directional reconstructions.

Since the probability for a neutrino detection within the given time of operation was very low and of order 0.10 events, the analysis algorithms have not been optimized for reconstruction of neutrino four-vectors. To determine the incoming direction and energy of a neutrino, one needs to know the polarization of the event and the distance to the vertex. The polarization can be determined by comparing the signal strength in Hpol and Vpol. The distance reconstruction is important for the determination of the neutrino energy. Since the distance reconstruction is not crucial for neutrino identification, it is not essential to the ARA analysis strategy. For now, only a lower limit on the distance can be provided for each event, which results in a lower limit for the neutrino energy. The achievable energy resolution of the ARA detector is thus still unknown and more work in this direction is needed to prepare for the case of a neutrino detection.

\begin{figure}[tbp]
\centering
\includegraphics[width=0.97\columnwidth]{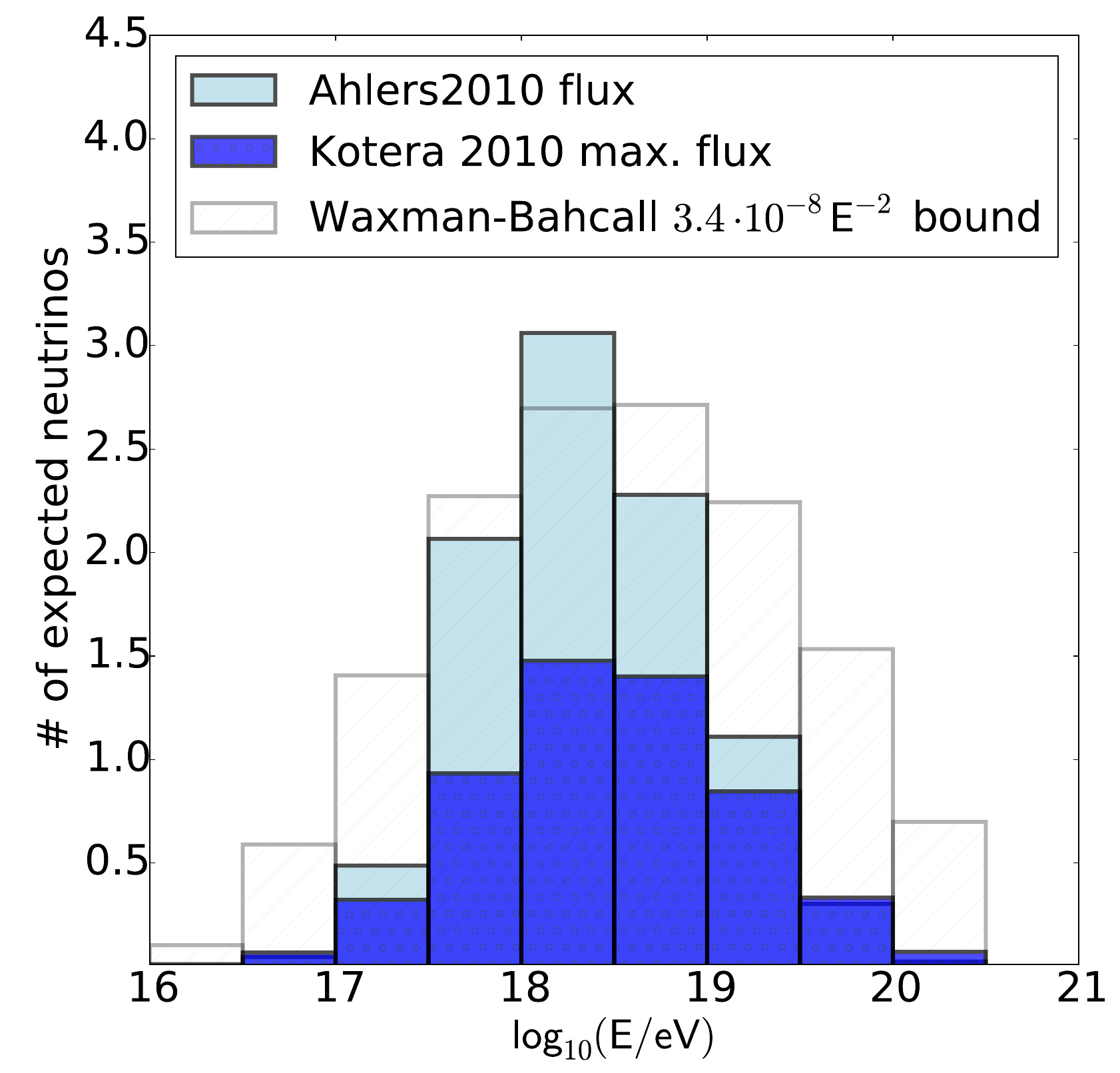}
\caption{Numbers of neutrinos versus energy in half-decade bins, as expected to be seen with the ARA37 detector within 3 years. These numbers have been extrapolated from the presented trigger and analysis efficiency for the first two stations. The numbers are calculated for 3 different flux predictions which are partly shown in Figure \ref{fig_SensAnaLevel} and estimated in \cite{Ahlers2010,Kotera2011,WB2001}. The total expected number of neutrinos for the predictions are 9.4 (Ahlers 2010) and 5.5 (Kotera 2010).
}
\label{fig_neutrinoNumbers}
\end{figure}

The presented limit, resulting from one year of data taking, is not yet significant but raises the expectations for the discovery of UHE neutrinos with the full ARA detector. \\
\\
The results presented in this paper reflect the status of detector operations in 2013 and of currently available analysis tools. Since then, several improvements have been developed, benefiting coming analysis results, of which the most important are the following:
\begin{itemize}
\item The live time per year has been increased by roughly $25\%$, thanks to newly developed monitoring tools which allow for quick debugging, particularly in the case of downtime due to issues in the detector electronics.
\item The trigger and readout windows have been optimized and widened to enhance the detector sensitivity and to render analysis tools, complicated by cutoff signal waveforms, more efficient.
\item A PCI Express bus has been integrated into the DAQ to replace the previously used USB connection. This allows the recording of data at an event rate several times higher than before, and allows lowering the trigger threshold, enhancing the neutrino sensitivity.
\item New reconstruction algorithms are underway which take ray-tracing effects into account and are therefore expected to show a significant improvement in precision and reconstruction efficiency.
\item The detector calibration is under continuous improvement which will help especially to reduce systematic uncertainties on the detector geometry, timing and the signal chain, which consequently improve analysis algorithms like the angular reconstruction.
\end{itemize}

Figure \ref{fig_neutrinoNumbers} shows the numbers of neutrinos which are expected to be seen with the ARA37 detector, given the trigger and analysis efficiency for 2013, within 3 years of operation at different energies and from different fluxes. Additionally, an expectation for a power law flux, normalized to the Waxman-Bahcall bound from \cite{WB2001} is plotted to illustrate the response of the detector over a wide range of energy. This figure shows that the planned full ARA detector is a promising candidate for the detection of ultra-high energy neutrinos.

\section{Acknowledgments}

We thank the National Science Foundation for their generous  support  through  Grant  NSF  OPP-1002483 and  Grant  NSF  OPP-1359535. We further thank the Taiwan National Science Councils Vanguard Program: NSC 102-2628-M-002-010 and the Belgian F.R.S.-FNRS Grant 4.4508.01. We are grateful to the U.S. National Science Foundation-Office of Polar Programs and the U.S. National Science Foundation-Physics Division. We also thank the University of Wisconsin Alumni Research Foundation, the University of Maryland and the Ohio State University for their support. Furthermore, we are grateful to the Raytheon Polar Services Corporation and the Antarctic Support Contractor, Lockheed, for field support. A. Connolly thanks the National Science Foundation for their support through CAREER award 1255557, and also the Ohio Supercomputer Center. K. Hoffman likewise thanks the National Science Foundation for their support  through CAREER award 0847658. A. Connolly, H. Landsman and D. Besson thank the United States-Israel Bi-national Science Foundation for their support through Grant 2012077. A. Connolly, A. Karle and J. Kelley thank the National Science Foundation for the support through BIGDATA Grant 1250720. D. Besson and A. Novikov acknowledge support from National Research Nuclear University MEPhi (Moscow Engineering Physics Institute). R. Nichol thanks the Leverhulme Trust for their support.



\appendix
\section{Calibration of the IRS2 digitizer chip}   \label{ch_digiCal}

\begin{figure}[b]
\centering
\includegraphics[width=\columnwidth]{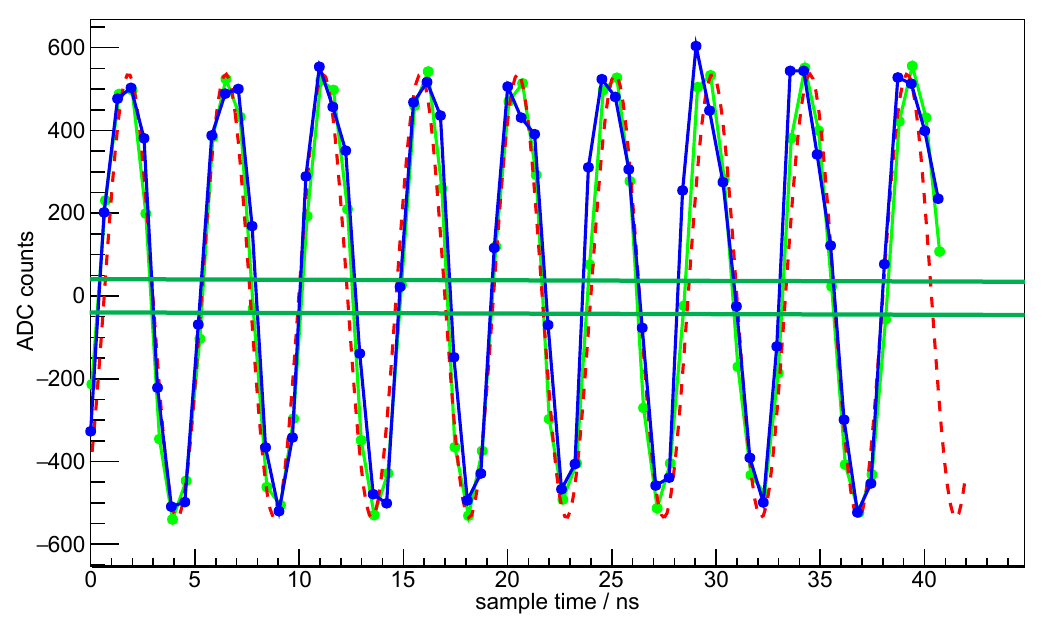} 
\caption{A typical calibration waveform separated by even (blue) and odd (green) samples with a fit waveform (red). The horizontal lines indicate the range for individual sample correction in timing.} \label{fig_timeCal} 
\end{figure}

The IRS2 chip is a custom ASIC for radio frequency applications \cite{IRS_spec}. It is designed to digitize at a speed of several $\unit{}{\giga S \per \second}$ at low power consumption of less than $\unit{20}{\milli\watt}$ per channel. These features merit usage of CMOS technology for the implementation of the sampling and digitization steps, as well as utilization of a large analog buffer, which allows for a slow digitization technique without inducing dead time. In the IRS2 chip the data are sampled via a Switched Capacitor Array (SCA) which utilizes finely tuned delay elements to set up a sampling sequence of the input data. These delay elements can differ from their nominal delay width due to variations during the chip fabrication and have to be calibrated individually. The SCA consists of 128 sampling capacitors per channel, equally divided into even and odd samples on two delay lines and each with a delay element requiring individual calibration in timing. In addition to that the ADC to voltage conversion gain needs to be determined for each of the 32768 buffer elements on each channel to obtain a proper voltage calibration.

\begin{figure}[t]
\subfigure[Even samples]{ 
\centering
\includegraphics[width=\columnwidth]{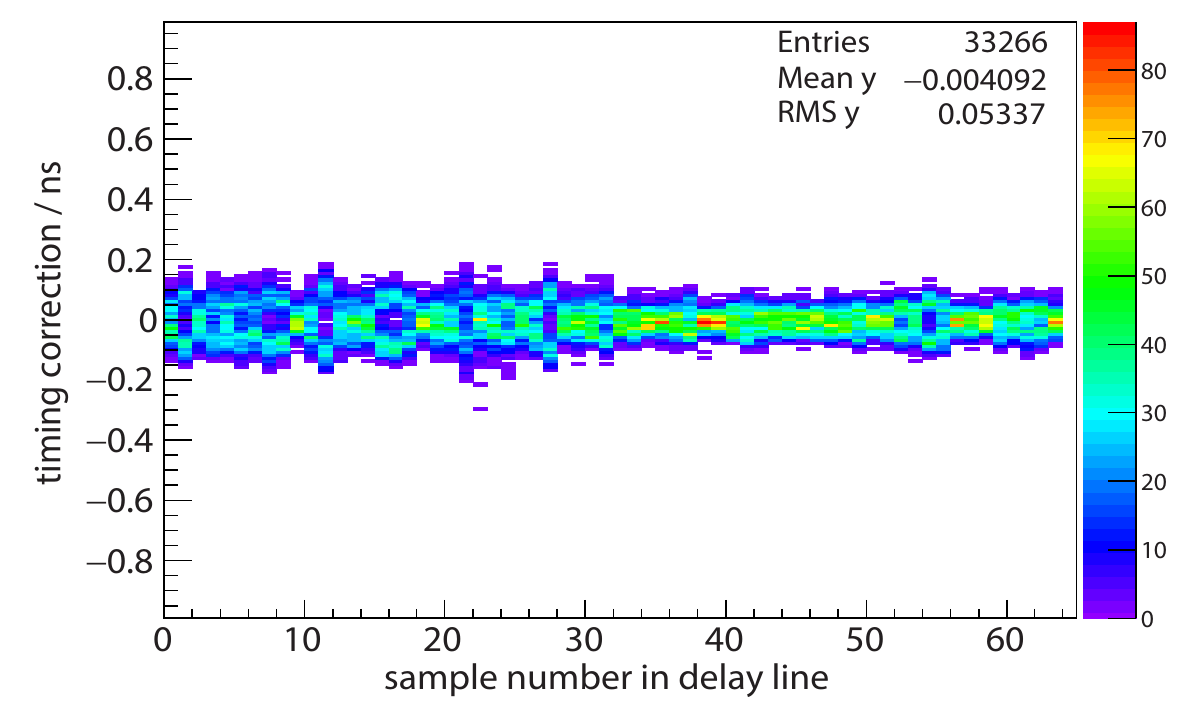} }
\subfigure[Odd samples]{
\centering
\includegraphics[width=\columnwidth]{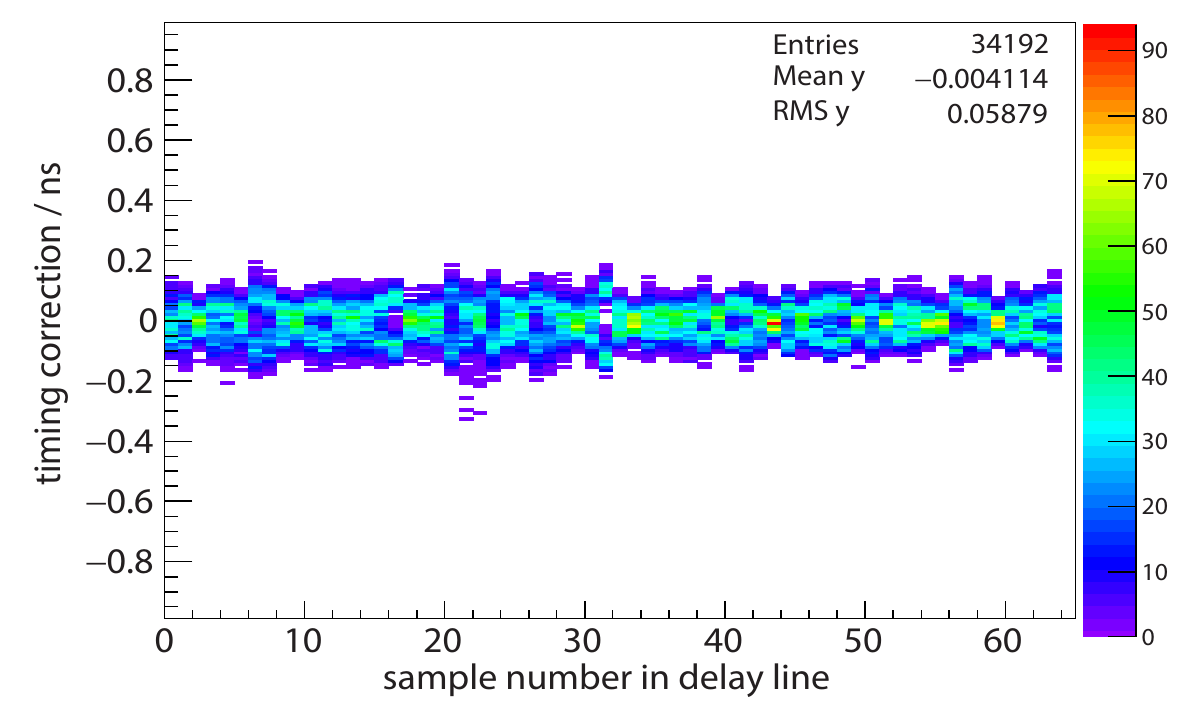} }
\caption{The timing corrections for all delay elements (X-axis) of one channel in station A3, separated by an even (a) and odd (b) delay line after several iterations of calibration.} \label{fig_timeCorrCh16ARA03}
\end{figure}

For both calibrations, timing and ADC gain, sine waves are recorded and fit to ideal waveforms (Figure \ref{fig_timeCal}). The data are recorded in the laboratory with the instrument at the final in-ice temperature of $\unit{-50}{\celsius}$.

For the timing, first the fit frequency is compared to the frequency of the input waveform to calibrate the average sampling speed. Then, each individual sample timing is compared to the fit waveform to obtain a correction factor for the given delay element. This correction is applied if samples have an absolute ADC count below $30$, thus if the derivative of the sine wave is maximal and the influence of voltage errors is small. Corrections are directly applied and the process is repeated for several iterations until the correction factors converge. Figure \ref{fig_timeCorrCh16ARA03} shows the final timing corrections needed for samples of a selected channel after several iterations of calibration. One can observe a statistical fluctuation symmetric around zero, which indicates that all systematic errors have been removed by the calibration. The underlying calibration values are then used to set the correct timing for each delay element. The visible spread of some distributions is connected to a non-linearity in the voltage of the respective sample.

\begin{figure}[t!]
\centering
\includegraphics[width=\columnwidth]{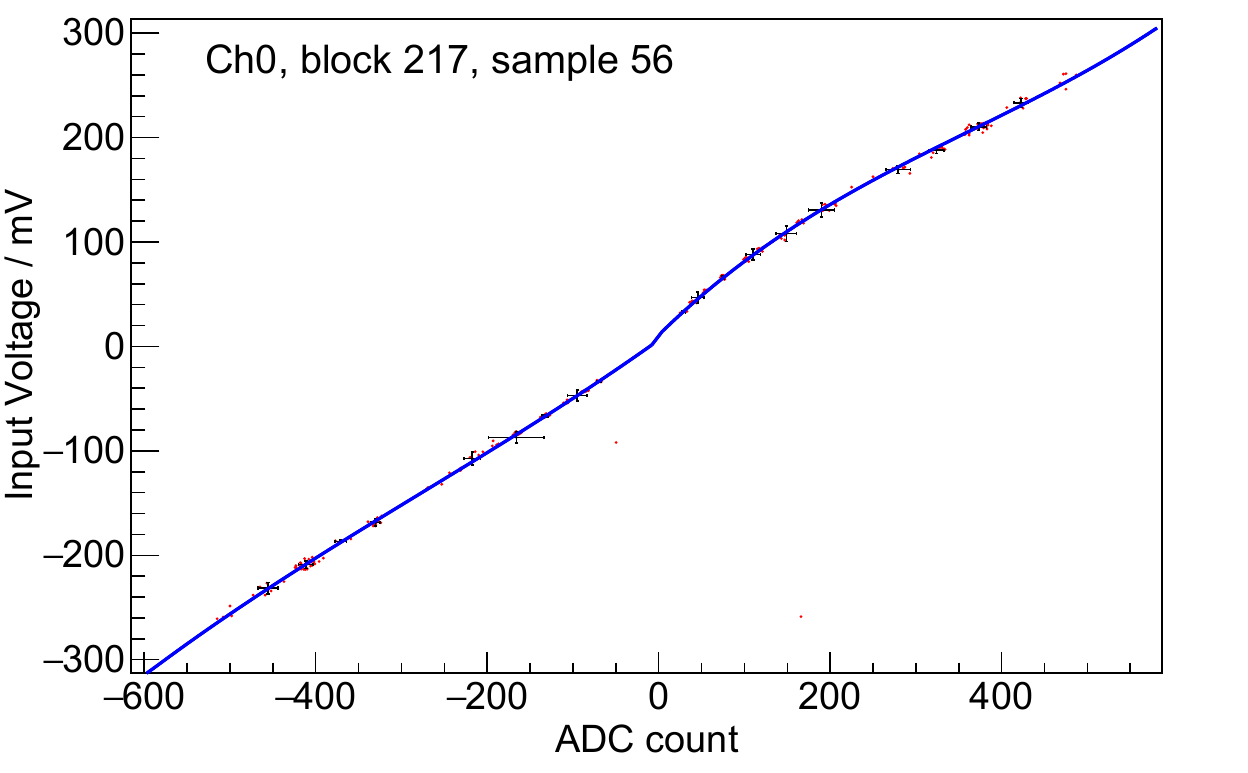}
\caption{Calibration data for the calculation of the ADC-to-voltage conversion for a single storage sample: Collected data (red dots), averaged data (black points) with errors and a broken 3rd order polynomial fit to the average data (blue line).} 
\label{fig_adcVSvolts} 
\end{figure}

For the voltage calibration, timing-corrected waveforms are used as input. Due to the density of the chip structure, a slight non-linearity and asymmetry around 0 is induced in voltage which depends strongly on the channel number. Furthermore, we need calibration data for about $650000$ storage elements per station, which requires a huge data sample. These conditions render classical voltage calibration methods difficult. For the calibration of the ADC to voltage conversion gain of the IRS2, input sine waves of known amplitude and frequency are fit and ADC samples are compared to the fit whenever its derivative is smaller than $45\%$ of the maximal value. In contrast to the timing calibration, a small derivative is required, this time to minimize the influence of timing errors. Following this procedure a statistically significant sample for each storage element can be collected, using input waveforms over a wide range in amplitude (Figure \ref{fig_adcVSvolts}).

Further calibrations which have been performed on the IRS2 chip are a check of 
\begin{itemize} 
\item the frequency response, which could not be determined conclusively with the used data set, 
\item the temperature dependence of timing and voltage, which appears to be negligible in the temperature range of the ARA experiment. 
\end{itemize} 

\begin{figure}[t!]
\centering
\includegraphics[width=0.9\columnwidth]{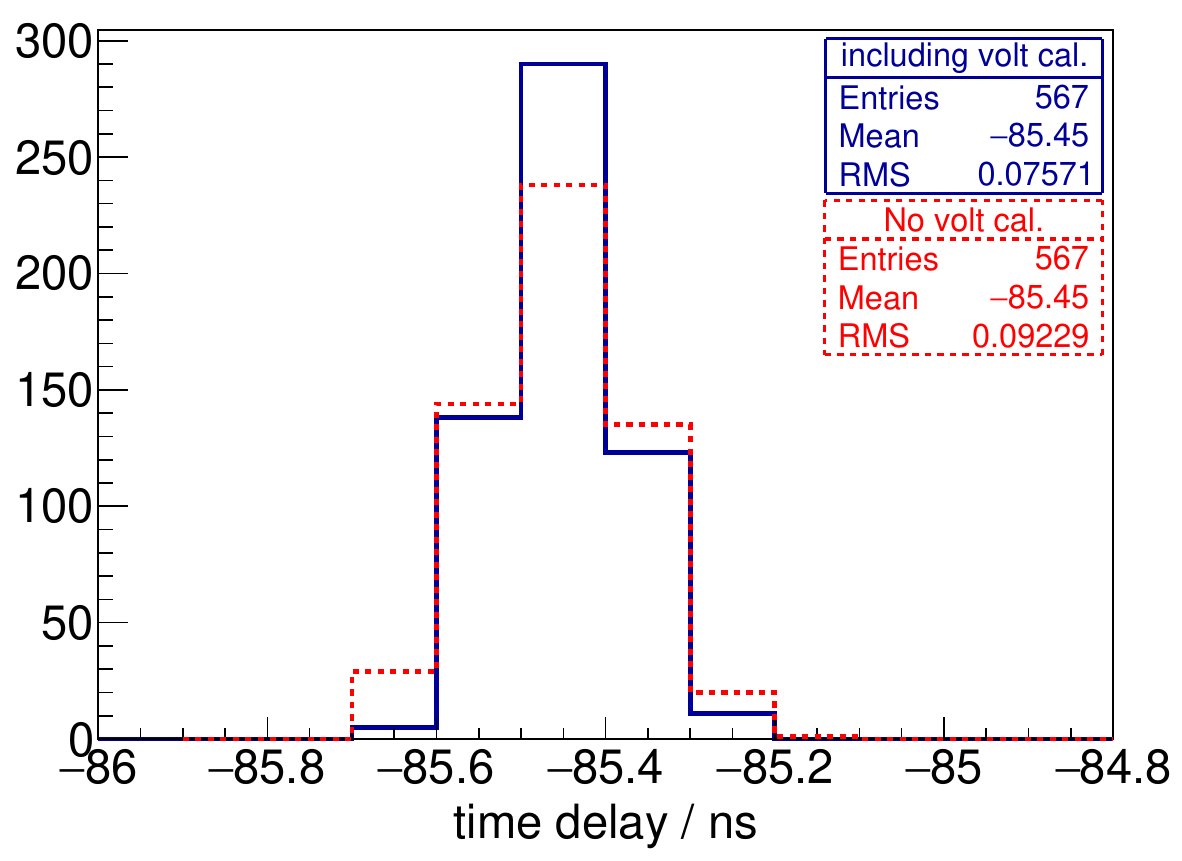}
\caption{Arrival time difference for a calibration pulser signal on two channels of station A2 after timing calibration (red) and the full calibration (blue).} \label{fig_tempCalCorr}
\end{figure}

The main purpose of the calibration is to obtain good correlation timing between incoming waveforms. Therefore, the calibration is cross checked with calibration pulser waveforms, recorded on different channels. On average, a precision of $\unit{100}{\pico\second}$ can be achieved (Figure \ref{fig_tempCalCorr}), which is entirely adequate for good vertex angular reconstructions.  Determination of the radius of curvature of the incoming wavefront is considerably more difficult with stations of limited size; therefore, for sources more than tens of meters from the station, the range to emission is effectively an unknown at the current level of analysis.

The influence of the timing jitter and possible non-linearities in the ADC to voltage conversion gain are tested on a simulated waveform, sampled at $\unit{3.2}{\giga S \per \second}$, with a frequency spectrum similar to what is expected for an ARA signal. Timing calibration errors are modeled as Gaussian distributed random jitter for different standard deviations and added to the sample timing. Non-linearities, left in the calibration of the ADC to voltage conversion gain, are modeled with a simple third order polynomial as
\begin{eqnarray}
g_{\mathrm{dist}} = g + k \cdot (g^2 + g^3),	\label{eq_nonlin}
\end{eqnarray}

\begin{figure}[b]
\subfigure[]{
\centering
\includegraphics[width=\columnwidth]{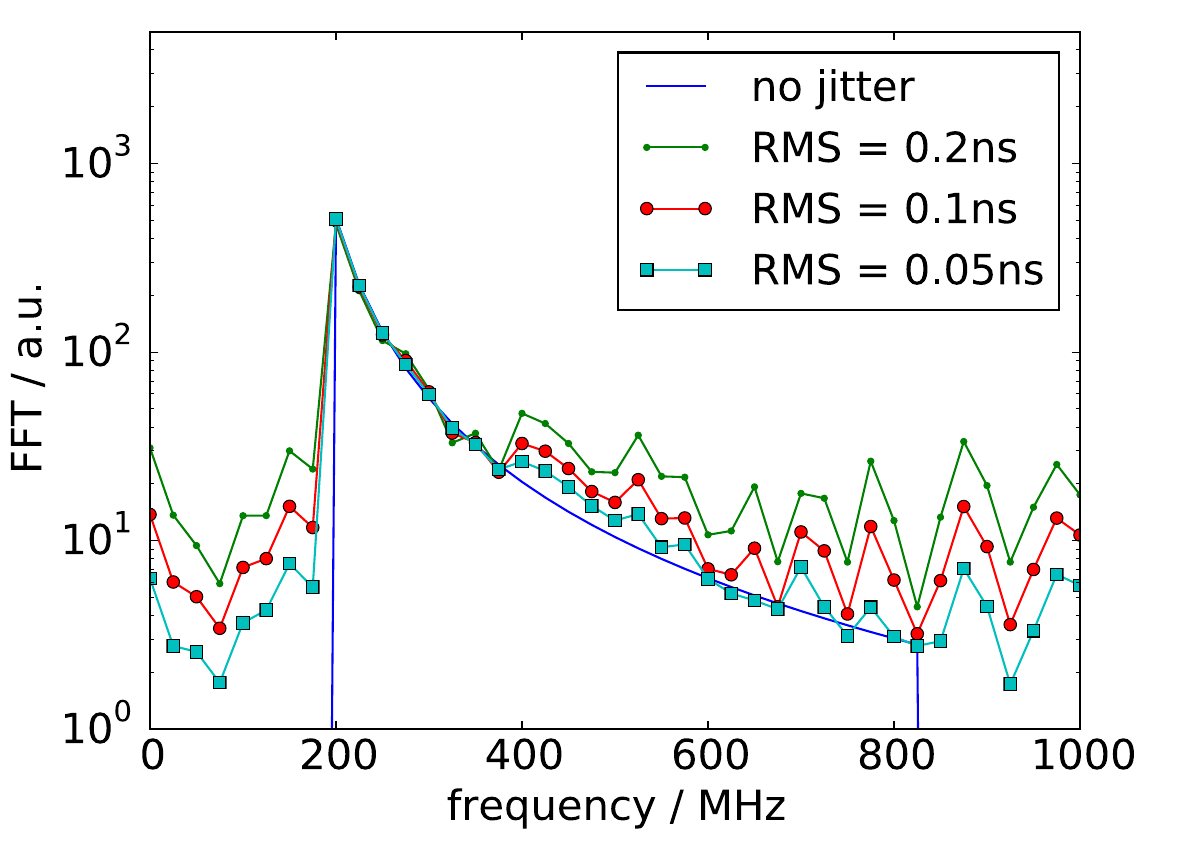}
}
\subfigure[]{
\centering
\includegraphics[width=\columnwidth]{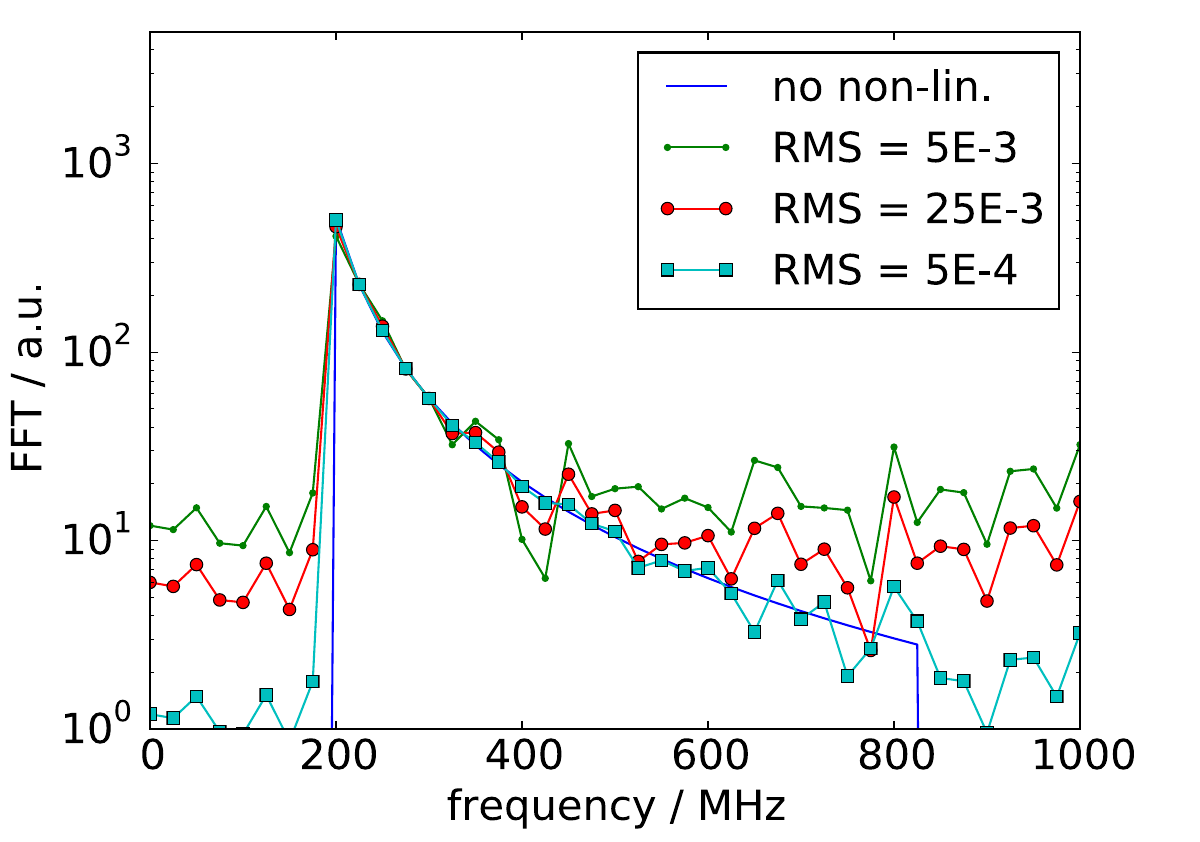}
}
\caption{Comparison of Fourier spectra for a simulated waveform (undisturbed in blue) at various error levels in the digitizer calibration. (a) Different magnitudes of timing jitter applied to each sample. The RMS represents the width of the Gaussian distribution used to generate the random jitter for each sample. (b) Different levels of a non-linear addition in the ADC to voltage conversion gain as presented in Equation \ref{eq_nonlin}. The RMS represents the width of the distribution used to generate $k$.}
\label{fig_calDisturbance}
\end{figure}

with $g$ the linear relation between the ADC count and the input voltage and $k$ the level of a non-linear addition. As for the timing jitter, $k$ is a random number for each sample following a Gaussian distribution of a given width. The resulting Fourier spectra of the original and smeared waveform for different error levels are presented in Figure \ref{fig_calDisturbance}. As is visible, the smearing adds a broadband component to the spectrum which has a small influence on regions of strong signal but adds significantly at low signal levels. The shown levels of smearing have been chosen to be within a reasonable range as observed during the calibration process. Additional investigations are needed to determine the actual amount of jitter and non-linear gain components for each sample to quantify the influence on the Fourier spectrum more precisely.

\section{Neutrino limits with alternative $E F(E)$ scaling}   \label{app_Emin1}

\begin{figure}[]
\centering{
\includegraphics[width=\columnwidth]{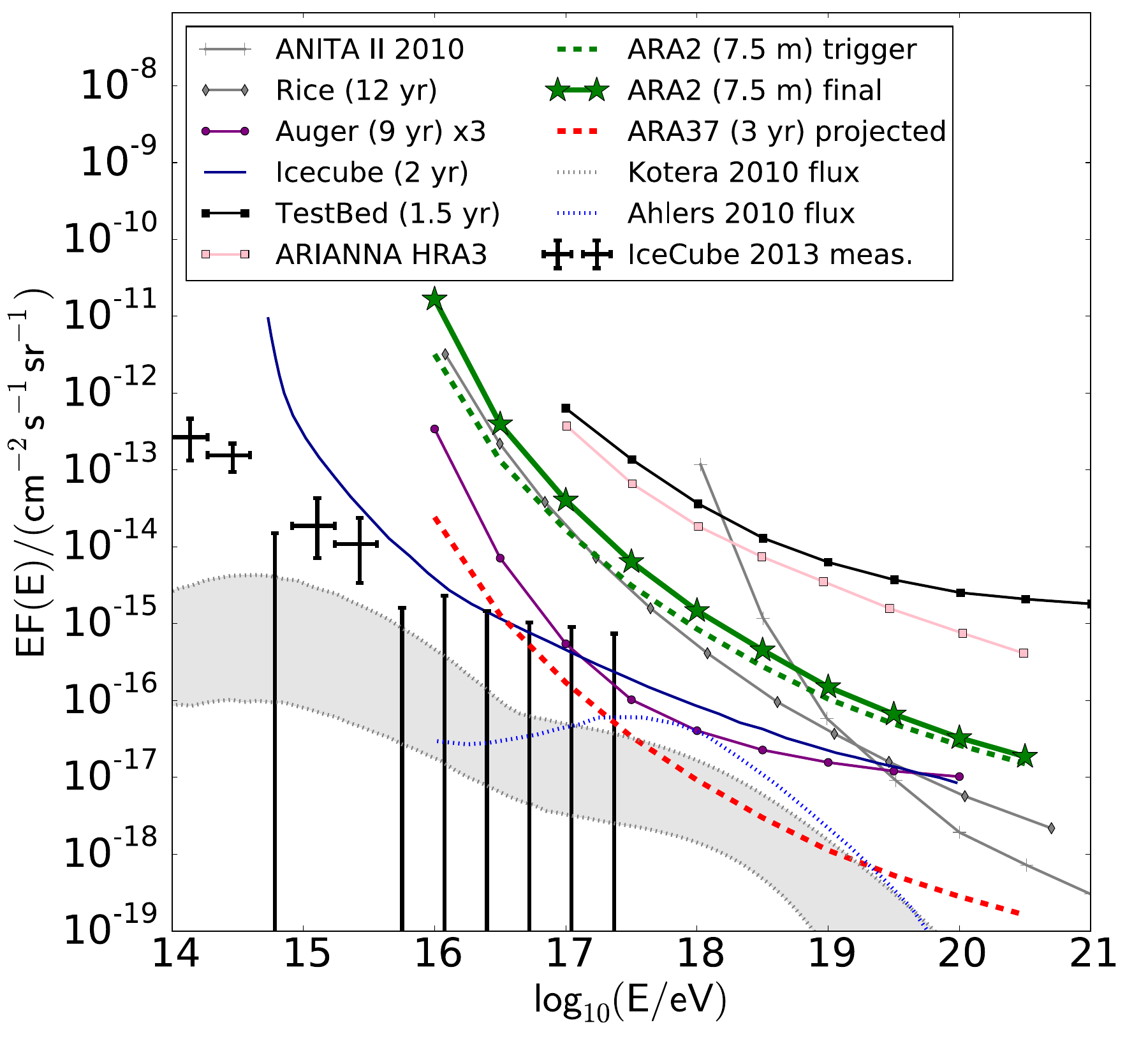}
\caption{The neutrino limits and fluxes of Figure \ref{fig_SensAnaLevel} with an alternative y-axis scaling.}
\label{fig_Emin1ScaledLimit}
\vspace{-50pt}
}
\end{figure}

Figure \ref{fig_Emin1ScaledLimit} shows the same data as Figure \ref{fig_SensAnaLevel} with the Y-axis scaling changed from $E^2 F(E)$ to $E F(E)$. This plot has been added for convenience of the reader. 

\newpage

\begin{widetext}

\section{A2 events reconstructed to different zenith angles}	\label{ch_ARA02_coincEvents}

In the event sequence for which sources are observed coincidently in both stations, as explained in Section \ref{ch_crossChecks}, the zenith angle towards A2 changes drastically over the development of the track. This allows for observation of a shift in polarization in these events. Figures \ref{fig_StartTrackEvent}, \ref{fig_MidTrackEvent} and \ref{fig_EndTrackEvent} show a sequence of three events drawn from three different parts of the track. As expected, the event emitted vertically above station A2 shows a strong Hpol component, which is considerably weaker in those two events observed at shallower incident zenith angles.

\begin{figure*}[h!]
\centering
\includegraphics[width=0.94\columnwidth]{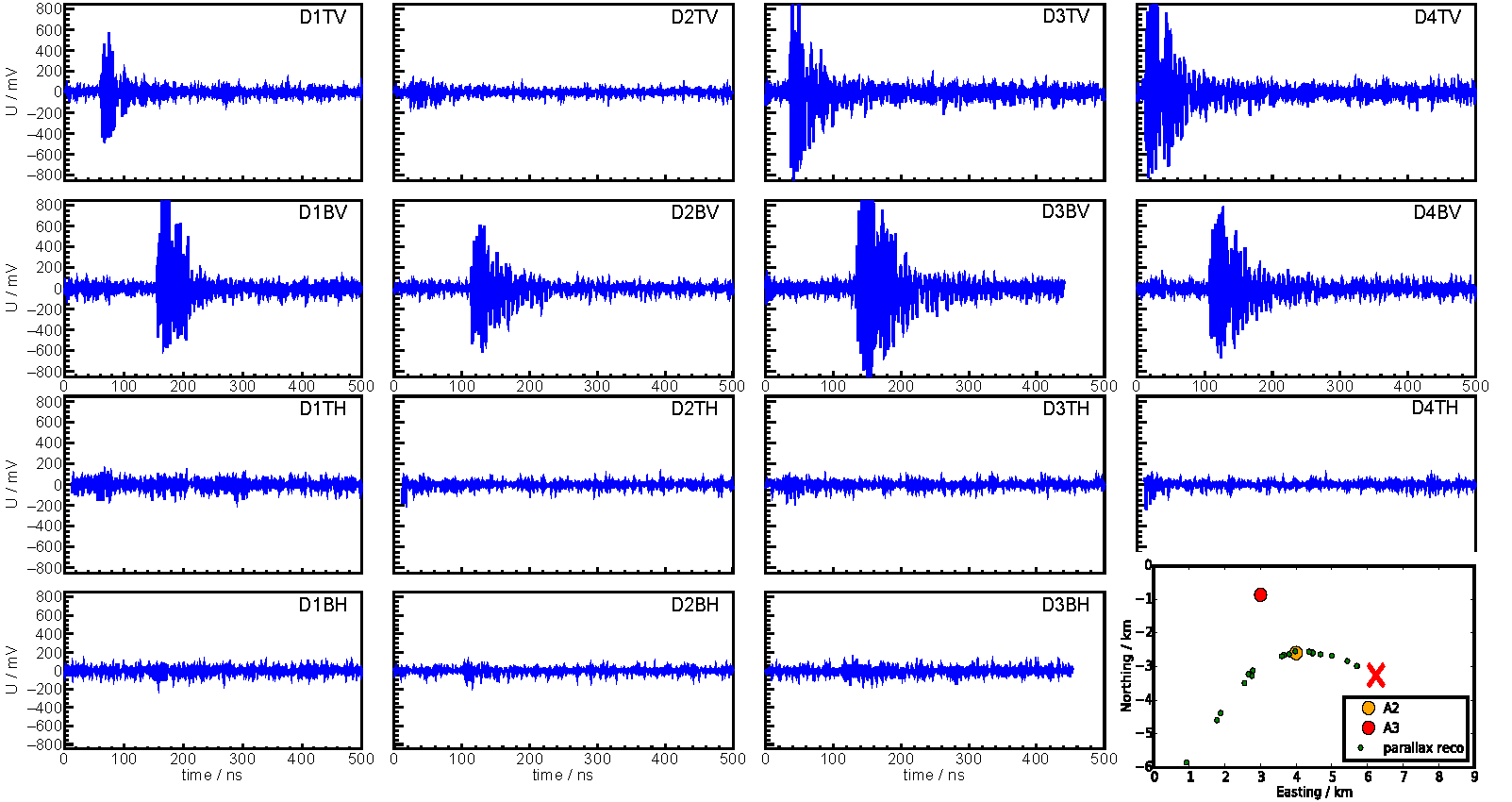}
\caption{Recorded waveforms for an A2 event, which was part of the event sequence presented in Figure \ref{fig_parallaxReco}. The event shown corresponds specifically to the source at the point indicated by the red X (bottom right plot), as it moves above the array. The reconstructed zenith angle in A2 is $\unit{51.6}{\degree}$. Note that channel D4BH on A2 is not operational.}
\label{fig_StartTrackEvent}
\end{figure*}

\begin{figure*}[h!]
\centering
\includegraphics[width=0.99\columnwidth]{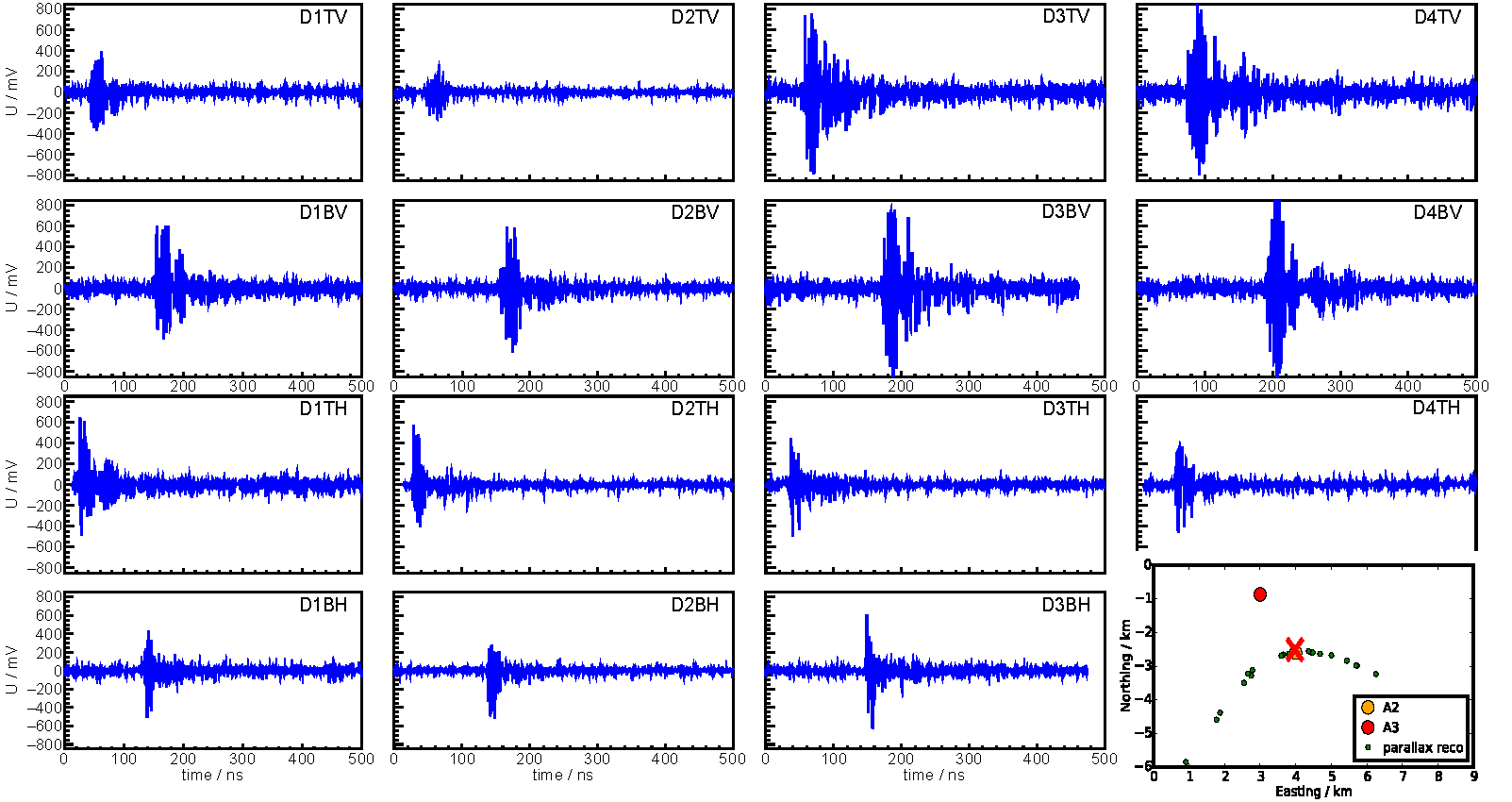}
\caption{Recorded waveforms for an A2 event, which was part of the event sequence presented in Figure \ref{fig_parallaxReco}. The event shown corresponds specifically to the source at the point indicated by the red X (bottom right plot), as it moves above the array. The reconstructed zenith angle in A2 is $\unit{87.0}{\degree}$.}
\label{fig_MidTrackEvent}
\end{figure*}

\begin{figure*}[h!]
\centering
\includegraphics[width=0.99\columnwidth]{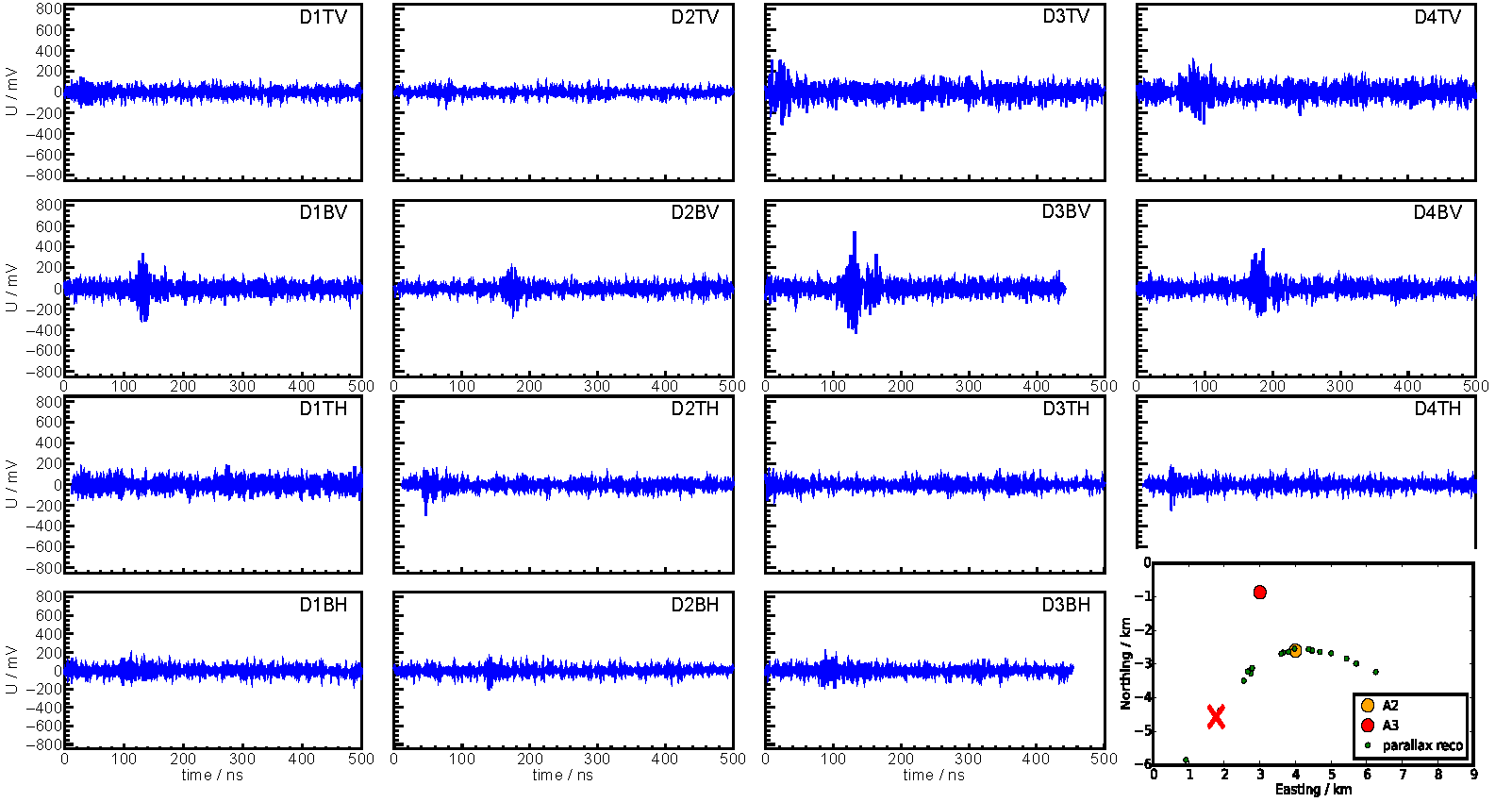}
\caption{Recorded waveforms for an A2 event, which was part of the event sequence presented in Figure \ref{fig_parallaxReco}. The event shown corresponds specifically to the source at the point indicated by the red X (bottom right plot), as it moves above the array. The reconstructed zenith angle in A2 is $\unit{57.5}{\degree}$.}
\label{fig_EndTrackEvent}
\end{figure*}

\end{widetext}


\bibliography{ARA2results}

\begin{thebibliography}{52}%
\makeatletter
\providecommand \@ifxundefined [1]{%
 \@ifx{#1\undefined}
}%
\providecommand \@ifnum [1]{%
 \ifnum #1\expandafter \@firstoftwo
 \else \expandafter \@secondoftwo
 \fi
}%
\providecommand \@ifx [1]{%
 \ifx #1\expandafter \@firstoftwo
 \else \expandafter \@secondoftwo
 \fi
}%
\providecommand \natexlab [1]{#1}%
\providecommand \enquote  [1]{``#1''}%
\providecommand \bibnamefont  [1]{#1}%
\providecommand \bibfnamefont [1]{#1}%
\providecommand \citenamefont [1]{#1}%
\providecommand \href@noop [0]{\@secondoftwo}%
\providecommand \href [0]{\begingroup \@sanitize@url \@href}%
\providecommand \@href[1]{\@@startlink{#1}\@@href}%
\providecommand \@@href[1]{\endgroup#1\@@endlink}%
\providecommand \@sanitize@url [0]{\catcode `\\12\catcode `\$12\catcode
  `\&12\catcode `\#12\catcode `\^12\catcode `\_12\catcode `\%12\relax}%
\providecommand \@@startlink[1]{}%
\providecommand \@@endlink[0]{}%
\providecommand \url  [0]{\begingroup\@sanitize@url \@url }%
\providecommand \@url [1]{\endgroup\@href {#1}{\urlprefix }}%
\providecommand \urlprefix  [0]{URL }%
\providecommand \Eprint [0]{\href }%
\providecommand \doibase [0]{http://dx.doi.org/}%
\providecommand \selectlanguage [0]{\@gobble}%
\providecommand \bibinfo  [0]{\@secondoftwo}%
\providecommand \bibfield  [0]{\@secondoftwo}%
\providecommand \translation [1]{[#1]}%
\providecommand \BibitemOpen [0]{}%
\providecommand \bibitemStop [0]{}%
\providecommand \bibitemNoStop [0]{.\EOS\space}%
\providecommand \EOS [0]{\spacefactor3000\relax}%
\providecommand \BibitemShut  [1]{\csname bibitem#1\endcsname}%
\let\auto@bib@innerbib\@empty
\bibitem [{\citenamefont {Greisen}(1966)}]{Greisen1966}%
  \BibitemOpen
  \bibfield  {author} {\bibinfo {author} {\bibfnamefont {K.}~\bibnamefont
  {Greisen}},\ }\href {\doibase 10.1103/PhysRevLett.16.748} {\bibfield
  {journal} {\bibinfo  {journal} {Phys. Rev. Lett.}\ }\textbf {\bibinfo
  {volume} {16}},\ \bibinfo {pages} {748} (\bibinfo {year} {1966})}\BibitemShut
  {NoStop}%
\bibitem [{\citenamefont {Zatsepin}\ and\ \citenamefont
  {Kuzmin}(1966)}]{Zatsepin1966}%
  \BibitemOpen
  \bibfield  {author} {\bibinfo {author} {\bibfnamefont {G.~T.}\ \bibnamefont
  {Zatsepin}}\ and\ \bibinfo {author} {\bibfnamefont {V.~A.}\ \bibnamefont
  {Kuzmin}},\ }\href@noop {} {\bibfield  {journal} {\bibinfo  {journal} {JETP
  Lett.}\ }\textbf {\bibinfo {volume} {4}},\ \bibinfo {pages} {78} (\bibinfo
  {year} {1966})},\ \bibinfo {note} {[Pisma Zh. Eksp. Teor.
  Fiz.4,114(1966)]}\BibitemShut {NoStop}%
\bibitem [{\citenamefont {Berezinsky}\ and\ \citenamefont
  {Zatsepin}(1969)}]{Beresinsky1968}%
  \BibitemOpen
  \bibfield  {author} {\bibinfo {author} {\bibfnamefont {V.~S.}\ \bibnamefont
  {Berezinsky}}\ and\ \bibinfo {author} {\bibfnamefont {G.~T.}\ \bibnamefont
  {Zatsepin}},\ }\href {\doibase 10.1016/0370-2693(69)90341-4} {\bibfield
  {journal} {\bibinfo  {journal} {Phys. Lett.}\ }\textbf {\bibinfo {volume}
  {B28}},\ \bibinfo {pages} {423} (\bibinfo {year} {1969})}\BibitemShut
  {NoStop}%
\bibitem [{\citenamefont {Tinyakov}(2014)}]{Tinyakov2014}%
  \BibitemOpen
  \bibfield  {author} {\bibinfo {author} {\bibfnamefont {P.}~\bibnamefont
  {Tinyakov}} (\bibinfo {collaboration} {Telescope Array}),\ }\bibfield
  {booktitle} {\emph {\bibinfo {booktitle} {{Proceedings, 4th Roma
  International Conference on Astro-Particle Physics (RICAP 13)}}},\ }\href
  {\doibase 10.1016/j.nima.2013.10.067} {\bibfield  {journal} {\bibinfo
  {journal} {Nucl. Instrum. Meth.}\ }\textbf {\bibinfo {volume} {A742}},\
  \bibinfo {pages} {29} (\bibinfo {year} {2014})}\BibitemShut {NoStop}%
\bibitem [{\citenamefont {Abraham}\ \emph {et~al.}(2010)\citenamefont {Abraham}
  \emph {et~al.}}]{Abraham2010}%
  \BibitemOpen
  \bibfield  {author} {\bibinfo {author} {\bibfnamefont {J.}~\bibnamefont
  {Abraham}} \emph {et~al.} (\bibinfo {collaboration} {Pierre Auger}),\ }\href
  {\doibase 10.1016/j.physletb.2010.02.013} {\bibfield  {journal} {\bibinfo
  {journal} {Phys. Lett.}\ }\textbf {\bibinfo {volume} {B685}},\ \bibinfo
  {pages} {239} (\bibinfo {year} {2010})},\ \Eprint
  {http://arxiv.org/abs/1002.1975} {arXiv:1002.1975 [astro-ph.HE]} \BibitemShut
  {NoStop}%
\bibitem [{\citenamefont {Allard}\ \emph {et~al.}(2006)\citenamefont {Allard}
  \emph {et~al.}}]{Allard2006}%
  \BibitemOpen
  \bibfield  {author} {\bibinfo {author} {\bibfnamefont {D.}~\bibnamefont
  {Allard}} \emph {et~al.},\ }\href {\doibase 10.1088/1475-7516/2006/09/005}
  {\bibfield  {journal} {\bibinfo  {journal} {JCAP}\ }\textbf {\bibinfo
  {volume} {0609}},\ \bibinfo {pages} {005} (\bibinfo {year} {2006})},\ \Eprint
  {http://arxiv.org/abs/astro-ph/0605327} {arXiv:astro-ph/0605327 [astro-ph]}
  \BibitemShut {NoStop}%
\bibitem [{\citenamefont {Aartsen}\ \emph {et~al.}(2013)\citenamefont {Aartsen}
  \emph {et~al.}}]{Aartsen:2013dsm}%
  \BibitemOpen
  \bibfield  {author} {\bibinfo {author} {\bibfnamefont {M.}~\bibnamefont
  {Aartsen}} \emph {et~al.} (\bibinfo {collaboration} {IceCube}),\ }\href@noop
  {} {\bibfield  {journal} {\bibinfo  {journal} {Phys.Rev.}\ }\textbf {\bibinfo
  {volume} {D88}},\ \bibinfo {pages} {112008} (\bibinfo {year} {2013})},\
  \Eprint {http://arxiv.org/abs/1310.5477} {arXiv:1310.5477 [astro-ph.HE]}
  \BibitemShut {NoStop}%
\bibitem [{\citenamefont {Fang}\ \emph {et~al.}(2014)\citenamefont {Fang},
  \citenamefont {Kotera}, \citenamefont {Murase},\ and\ \citenamefont
  {Olinto}}]{Fang:2013vla}%
  \BibitemOpen
  \bibfield  {author} {\bibinfo {author} {\bibfnamefont {K.}~\bibnamefont
  {Fang}}, \bibinfo {author} {\bibfnamefont {K.}~\bibnamefont {Kotera}},
  \bibinfo {author} {\bibfnamefont {K.}~\bibnamefont {Murase}}, \ and\ \bibinfo
  {author} {\bibfnamefont {A.~V.}\ \bibnamefont {Olinto}},\ }\href {\doibase
  10.1103/PhysRevD.90.103005} {\bibfield  {journal} {\bibinfo  {journal} {Phys.
  Rev.}\ }\textbf {\bibinfo {volume} {D90}},\ \bibinfo {pages} {103005}
  (\bibinfo {year} {2014})},\ \Eprint {http://arxiv.org/abs/1311.2044}
  {arXiv:1311.2044 [astro-ph.HE]} \BibitemShut {NoStop}%
\bibitem [{\citenamefont {Murase}\ \emph {et~al.}(2014)\citenamefont {Murase},
  \citenamefont {Inoue},\ and\ \citenamefont {Dermer}}]{Murase:2014foa}%
  \BibitemOpen
  \bibfield  {author} {\bibinfo {author} {\bibfnamefont {K.}~\bibnamefont
  {Murase}}, \bibinfo {author} {\bibfnamefont {Y.}~\bibnamefont {Inoue}}, \
  and\ \bibinfo {author} {\bibfnamefont {C.~D.}\ \bibnamefont {Dermer}},\
  }\href {\doibase 10.1103/PhysRevD.90.023007} {\bibfield  {journal} {\bibinfo
  {journal} {Phys. Rev.}\ }\textbf {\bibinfo {volume} {D90}},\ \bibinfo {pages}
  {023007} (\bibinfo {year} {2014})},\ \Eprint {http://arxiv.org/abs/1403.4089}
  {arXiv:1403.4089 [astro-ph.HE]} \BibitemShut {NoStop}%
\bibitem [{\citenamefont {Waxman}\ and\ \citenamefont
  {Bahcall}(2000)}]{Waxman:1999ai}%
  \BibitemOpen
  \bibfield  {author} {\bibinfo {author} {\bibfnamefont {E.}~\bibnamefont
  {Waxman}}\ and\ \bibinfo {author} {\bibfnamefont {J.~N.}\ \bibnamefont
  {Bahcall}},\ }\href {\doibase 10.1086/309462} {\bibfield  {journal} {\bibinfo
   {journal} {Astrophys. J.}\ }\textbf {\bibinfo {volume} {541}},\ \bibinfo
  {pages} {707} (\bibinfo {year} {2000})},\ \Eprint
  {http://arxiv.org/abs/hep-ph/9909286} {arXiv:hep-ph/9909286 [hep-ph]}
  \BibitemShut {NoStop}%
\bibitem [{\citenamefont {Murase}(2007)}]{PhysRevD.76.123001}%
  \BibitemOpen
  \bibfield  {author} {\bibinfo {author} {\bibfnamefont {K.}~\bibnamefont
  {Murase}},\ }\href {\doibase 10.1103/PhysRevD.76.123001} {\bibfield
  {journal} {\bibinfo  {journal} {Phys. Rev. D}\ }\textbf {\bibinfo {volume}
  {76}},\ \bibinfo {pages} {123001} (\bibinfo {year} {2007})}\BibitemShut
  {NoStop}%
\bibitem [{\citenamefont {Ahlers}\ and\ \citenamefont
  {Halzen}(2012)}]{AhlersMin2012}%
  \BibitemOpen
  \bibfield  {author} {\bibinfo {author} {\bibfnamefont {M.}~\bibnamefont
  {Ahlers}}\ and\ \bibinfo {author} {\bibfnamefont {F.}~\bibnamefont
  {Halzen}},\ }\href {\doibase 10.1103/PhysRevD.86.083010} {\bibfield
  {journal} {\bibinfo  {journal} {Phys. Rev. D}\ }\textbf {\bibinfo {volume}
  {86}},\ \bibinfo {pages} {083010} (\bibinfo {year} {2012})}\BibitemShut
  {NoStop}%
\bibitem [{\citenamefont {Gandhi}\ \emph {et~al.}(1996)\citenamefont {Gandhi},
  \citenamefont {Quigg}, \citenamefont {Reno},\ and\ \citenamefont
  {Sarcevic}}]{Gandhi1996}%
  \BibitemOpen
  \bibfield  {author} {\bibinfo {author} {\bibfnamefont {R.}~\bibnamefont
  {Gandhi}}, \bibinfo {author} {\bibfnamefont {C.}~\bibnamefont {Quigg}},
  \bibinfo {author} {\bibfnamefont {M.~H.}\ \bibnamefont {Reno}}, \ and\
  \bibinfo {author} {\bibfnamefont {I.}~\bibnamefont {Sarcevic}},\ }\href
  {\doibase 10.1016/0927-6505(96)00008-4} {\bibfield  {journal} {\bibinfo
  {journal} {Astroparticle Physics}\ }\textbf {\bibinfo {volume} {5}},\
  \bibinfo {pages} {81 } (\bibinfo {year} {1996})}\BibitemShut {NoStop}%
\bibitem [{\citenamefont {Askaryan}(1962)}]{Askaryan1962}%
  \BibitemOpen
  \bibfield  {author} {\bibinfo {author} {\bibfnamefont {G.~A.}\ \bibnamefont
  {Askaryan}},\ }\href@noop {} {\bibfield  {journal} {\bibinfo  {journal}
  {JETP}\ }\textbf {\bibinfo {volume} {41}},\ \bibinfo {pages} {616 } (\bibinfo
  {year} {1962})}\BibitemShut {NoStop}%
\bibitem [{\citenamefont {Askaryan}(1965)}]{Askaryan1965}%
  \BibitemOpen
  \bibfield  {author} {\bibinfo {author} {\bibfnamefont {G.~A.}\ \bibnamefont
  {Askaryan}},\ }\href@noop {} {\bibfield  {journal} {\bibinfo  {journal}
  {JETP}\ }\textbf {\bibinfo {volume} {48}},\ \bibinfo {pages} {988 } (\bibinfo
  {year} {1965})}\BibitemShut {NoStop}%
\bibitem [{\citenamefont {Saltzberg}\ \emph {et~al.}(2001)\citenamefont
  {Saltzberg} \emph {et~al.}}]{AskaryanSand2001}%
  \BibitemOpen
  \bibfield  {author} {\bibinfo {author} {\bibfnamefont {D.}~\bibnamefont
  {Saltzberg}} \emph {et~al.},\ }\href {\doibase 10.1103/PhysRevLett.86.2802}
  {\bibfield  {journal} {\bibinfo  {journal} {Phys. Rev. Lett.}\ }\textbf
  {\bibinfo {volume} {86}},\ \bibinfo {pages} {2802} (\bibinfo {year}
  {2001})}\BibitemShut {NoStop}%
\bibitem [{\citenamefont {Gorham}\ \emph {et~al.}(2005)\citenamefont {Gorham}
  \emph {et~al.}}]{AskaryanSalt2005}%
  \BibitemOpen
  \bibfield  {author} {\bibinfo {author} {\bibfnamefont {P.~W.}\ \bibnamefont
  {Gorham}} \emph {et~al.},\ }\href {\doibase 10.1103/PhysRevD.72.023002}
  {\bibfield  {journal} {\bibinfo  {journal} {Phys. Rev. D}\ }\textbf {\bibinfo
  {volume} {72}},\ \bibinfo {pages} {023002} (\bibinfo {year}
  {2005})}\BibitemShut {NoStop}%
\bibitem [{\citenamefont {Gorham}\ \emph {et~al.}(2007)\citenamefont {Gorham}
  \emph {et~al.}}]{AskaryanIce2007}%
  \BibitemOpen
  \bibfield  {author} {\bibinfo {author} {\bibfnamefont {P.~W.}\ \bibnamefont
  {Gorham}} \emph {et~al.} (\bibinfo {collaboration} {ANITA}),\ }\href
  {\doibase 10.1103/PhysRevLett.99.171101} {\bibfield  {journal} {\bibinfo
  {journal} {Phys. Rev. Lett.}\ }\textbf {\bibinfo {volume} {99}},\ \bibinfo
  {pages} {171101} (\bibinfo {year} {2007})}\BibitemShut {NoStop}%
\bibitem [{\citenamefont {Zas}\ \emph {et~al.}(1992)\citenamefont {Zas},
  \citenamefont {Halzen},\ and\ \citenamefont {Stanev}}]{ZHS1992}%
  \BibitemOpen
  \bibfield  {author} {\bibinfo {author} {\bibfnamefont {E.}~\bibnamefont
  {Zas}}, \bibinfo {author} {\bibfnamefont {F.}~\bibnamefont {Halzen}}, \ and\
  \bibinfo {author} {\bibfnamefont {T.}~\bibnamefont {Stanev}},\ }\href
  {\doibase 10.1103/PhysRevD.45.362} {\bibfield  {journal} {\bibinfo  {journal}
  {Phys. Rev. D}\ }\textbf {\bibinfo {volume} {45}},\ \bibinfo {pages} {362}
  (\bibinfo {year} {1992})}\BibitemShut {NoStop}%
\bibitem [{\citenamefont {Alvarez-Muñiz}\ and\ \citenamefont
  {Zas}(1997)}]{AlvarezMuniz1997}%
  \BibitemOpen
  \bibfield  {author} {\bibinfo {author} {\bibfnamefont {J.}~\bibnamefont
  {Alvarez-Muñiz}}\ and\ \bibinfo {author} {\bibfnamefont {E.}~\bibnamefont
  {Zas}},\ }\href {\doibase 10.1016/S0370-2693(97)01009-5} {\bibfield
  {journal} {\bibinfo  {journal} {Physics Letters B}\ }\textbf {\bibinfo
  {volume} {411}},\ \bibinfo {pages} {218 } (\bibinfo {year}
  {1997})}\BibitemShut {NoStop}%
\bibitem [{\citenamefont {Landau}\ and\ \citenamefont
  {Pomeranchuk}(1953)}]{Landau1953}%
  \BibitemOpen
  \bibfield  {author} {\bibinfo {author} {\bibfnamefont {L.~D.}\ \bibnamefont
  {Landau}}\ and\ \bibinfo {author} {\bibfnamefont {I.~I.}\ \bibnamefont
  {Pomeranchuk}},\ }\href@noop {} {\bibfield  {journal} {\bibinfo  {journal}
  {Dokl. Akad. Nauk SSSR}\ }\textbf {\bibinfo {volume} {92}},\ \bibinfo {pages}
  {535} (\bibinfo {year} {1953})}\BibitemShut {NoStop}%
\bibitem [{\citenamefont {Migdal}(1956)}]{Migdal1956}%
  \BibitemOpen
  \bibfield  {author} {\bibinfo {author} {\bibfnamefont {A.~B.}\ \bibnamefont
  {Migdal}},\ }\href {\doibase 10.1103/PhysRev.103.1811} {\bibfield  {journal}
  {\bibinfo  {journal} {Phys. Rev.}\ }\textbf {\bibinfo {volume} {103}},\
  \bibinfo {pages} {1811} (\bibinfo {year} {1956})}\BibitemShut {NoStop}%
\bibitem [{\citenamefont {Price}\ \emph {et~al.}(2002)\citenamefont {Price}
  \emph {et~al.}}]{Price11062002}%
  \BibitemOpen
  \bibfield  {author} {\bibinfo {author} {\bibfnamefont {P.~B.}\ \bibnamefont
  {Price}} \emph {et~al.},\ }\href {\doibase 10.1073/pnas.082238999} {\bibfield
   {journal} {\bibinfo  {journal} {Proceedings of the National Academy of
  Sciences}\ }\textbf {\bibinfo {volume} {99}},\ \bibinfo {pages} {7844}
  (\bibinfo {year} {2002})}\BibitemShut {NoStop}%
\bibitem [{\citenamefont {Allison}\ \emph {et~al.}(2012)\citenamefont {Allison}
  \emph {et~al.}}]{Allison2012457}%
  \BibitemOpen
  \bibfield  {author} {\bibinfo {author} {\bibfnamefont {P.}~\bibnamefont
  {Allison}} \emph {et~al.} (\bibinfo {collaboration} {ARA}),\ }\href {\doibase
  10.1016/j.astropartphys.2011.11.010} {\bibfield  {journal} {\bibinfo
  {journal} {Astroparticle Physics}\ }\textbf {\bibinfo {volume} {35}},\
  \bibinfo {pages} {457 } (\bibinfo {year} {2012})}\BibitemShut {NoStop}%
\bibitem [{\citenamefont {Kravchenko}\ \emph {et~al.}(2004)\citenamefont
  {Kravchenko}, \citenamefont {Besson},\ and\ \citenamefont
  {Meyers}}]{Kravchenko2004}%
  \BibitemOpen
  \bibfield  {author} {\bibinfo {author} {\bibfnamefont {I.}~\bibnamefont
  {Kravchenko}}, \bibinfo {author} {\bibfnamefont {D.}~\bibnamefont {Besson}},
  \ and\ \bibinfo {author} {\bibfnamefont {J.}~\bibnamefont {Meyers}},\ }\href
  {\doibase 10.3189/172756504781829800} {\bibfield  {journal} {\bibinfo
  {journal} {Journal of Glaciology}\ }\textbf {\bibinfo {volume} {50}},\
  \bibinfo {pages} {522} (\bibinfo {year} {2004})}\BibitemShut {NoStop}%
\bibitem [{\citenamefont {Allison}\ \emph {et~al.}(2015)\citenamefont {Allison}
  \emph {et~al.}}]{TestBed2014}%
  \BibitemOpen
  \bibfield  {author} {\bibinfo {author} {\bibfnamefont {P.}~\bibnamefont
  {Allison}} \emph {et~al.} (\bibinfo {collaboration} {ARA}),\ }\href {\doibase
  10.1016/j.astropartphys.2015.04.006} {\bibfield  {journal} {\bibinfo
  {journal} {Astropart. Phys.}\ }\textbf {\bibinfo {volume} {70}},\ \bibinfo
  {pages} {62} (\bibinfo {year} {2015})},\ \Eprint
  {http://arxiv.org/abs/1404.5285} {arXiv:1404.5285 [astro-ph.HE]} \BibitemShut
  {NoStop}%
\bibitem [{\citenamefont {Javaid}(2012)}]{Javaid2012}%
  \BibitemOpen
  \bibfield  {author} {\bibinfo {author} {\bibfnamefont {A.}~\bibnamefont
  {Javaid}},\ }\emph {\bibinfo {title} {Monte Carlo simulation for radio
  detection of ultra high energy air shower cores by ANITA-II}},\ \href@noop {}
  {Ph.D. thesis},\ \bibinfo  {school} {University of Delaware} (\bibinfo {year}
  {2012})\BibitemShut {NoStop}%
\bibitem [{\citenamefont {ITU}(2013)}]{itu372.2013}%
  \BibitemOpen
  \bibfield  {author} {\bibinfo {author} {\bibnamefont {ITU}},\ }\href@noop {}
  {\emph {\bibinfo {title} {Radio noise}}},\ \bibinfo {type} {Recommendation}\
  \bibinfo {number} {P.372-11}\ (\bibinfo  {institution} {International
  Telecommunication Union},\ \bibinfo {address} {Geneva},\ \bibinfo {year}
  {2013})\BibitemShut {NoStop}%
\bibitem [{\citenamefont {Gary}()}]{Gary2014}%
  \BibitemOpen
  \bibfield  {author} {\bibinfo {author} {\bibfnamefont {D.~E.}\ \bibnamefont
  {Gary}},\ }\href@noop {} {\enquote {\bibinfo {title} {Radio astronomy:
  Lecture 3},}\ }\bibinfo {howpublished}
  {\url{https://web.njit.edu/~gary/728/Lecture3.html}},\ \bibinfo {note}
  {accessed: 2015-04-09}\BibitemShut {NoStop}%
\bibitem [{\citenamefont {Shaw}(2013)}]{Shaw2013}%
  \BibitemOpen
  \bibfield  {author} {\bibinfo {author} {\bibfnamefont {J.~A.}\ \bibnamefont
  {Shaw}},\ }\href@noop {} {\bibfield  {journal} {\bibinfo  {journal} {American
  Journal of Physics}\ }\textbf {\bibinfo {volume} {81}} (\bibinfo {year}
  {2013})}\BibitemShut {NoStop}%
\bibitem [{XFD(2015)}]{XFDTD2015}%
  \BibitemOpen
  \href@noop {} {\enquote {\bibinfo {title} {Xfdtd documentation},}\ }\bibinfo
  {howpublished} {\url{http://www.remcom.com/xf7}} (\bibinfo {year}
  {2015})\BibitemShut {NoStop}%
\bibitem [{\citenamefont {Hong}\ \emph {et~al.}(2013)\citenamefont {Hong},
  \citenamefont {Connolly},\ and\ \citenamefont {Pfendner for~the
  ARA~Collaboration}}]{Hong2013}%
  \BibitemOpen
  \bibfield  {author} {\bibinfo {author} {\bibfnamefont {E.~S.}\ \bibnamefont
  {Hong}}, \bibinfo {author} {\bibfnamefont {A.}~\bibnamefont {Connolly}}, \
  and\ \bibinfo {author} {\bibfnamefont {C.~G.}\ \bibnamefont {Pfendner for~the
  ARA~Collaboration}},\ }in\ \href@noop {} {\emph {\bibinfo {booktitle} {33rd
  International Cosmic Ray Conference (ICRC 2013), Rio de Janeiro, Brazil}}}\
  (\bibinfo {year} {2013})\BibitemShut {NoStop}%
\bibitem [{\citenamefont {Connolly}\ \emph {et~al.}(2011)\citenamefont
  {Connolly}, \citenamefont {Thorne},\ and\ \citenamefont
  {Waters}}]{Connolly2011}%
  \BibitemOpen
  \bibfield  {author} {\bibinfo {author} {\bibfnamefont {A.}~\bibnamefont
  {Connolly}}, \bibinfo {author} {\bibfnamefont {R.~S.}\ \bibnamefont
  {Thorne}}, \ and\ \bibinfo {author} {\bibfnamefont {D.}~\bibnamefont
  {Waters}},\ }\href {\doibase 10.1103/PhysRevD.83.113009} {\bibfield
  {journal} {\bibinfo  {journal} {Phys. Rev. D}\ }\textbf {\bibinfo {volume}
  {83}},\ \bibinfo {pages} {113009} (\bibinfo {year} {2011})}\BibitemShut
  {NoStop}%
\bibitem [{\citenamefont {Alvarez-Mu\~niz}\ \emph {et~al.}(2010)\citenamefont
  {Alvarez-Mu\~niz}, \citenamefont {Romero-Wolf},\ and\ \citenamefont
  {Zas}}]{AlvarezTD2010}%
  \BibitemOpen
  \bibfield  {author} {\bibinfo {author} {\bibfnamefont {J.}~\bibnamefont
  {Alvarez-Mu\~niz}}, \bibinfo {author} {\bibfnamefont {A.}~\bibnamefont
  {Romero-Wolf}}, \ and\ \bibinfo {author} {\bibfnamefont {E.}~\bibnamefont
  {Zas}},\ }\href {\doibase 10.1103/PhysRevD.81.123009} {\bibfield  {journal}
  {\bibinfo  {journal} {Phys. Rev. D}\ }\textbf {\bibinfo {volume} {81}},\
  \bibinfo {pages} {123009} (\bibinfo {year} {2010})}\BibitemShut {NoStop}%
\bibitem [{\citenamefont {Meures for~the
  ARA~Collaboration}(2013)}]{Meures2013}%
  \BibitemOpen
  \bibfield  {author} {\bibinfo {author} {\bibfnamefont {T.}~\bibnamefont
  {Meures for~the ARA~Collaboration}},\ }in\ \href@noop {} {\emph {\bibinfo
  {booktitle} {33rd International Cosmic Ray Conference (ICRC 2013), Rio de
  Janeiro, Brazil}}}\ (\bibinfo {year} {2013})\BibitemShut {NoStop}%
\bibitem [{\citenamefont {de~Vries}\ \emph {et~al.}(2015)\citenamefont
  {de~Vries}, \citenamefont {Buitink}, \citenamefont {van Eijndhoven},
  \citenamefont {Meures}, \citenamefont {O'Murchadha},\ and\ \citenamefont
  {Scholten}}]{deVries2015}%
  \BibitemOpen
  \bibfield  {author} {\bibinfo {author} {\bibfnamefont {K.~D.}\ \bibnamefont
  {de~Vries}}, \bibinfo {author} {\bibfnamefont {S.}~\bibnamefont {Buitink}},
  \bibinfo {author} {\bibfnamefont {N.}~\bibnamefont {van Eijndhoven}},
  \bibinfo {author} {\bibfnamefont {T.}~\bibnamefont {Meures}}, \bibinfo
  {author} {\bibfnamefont {A.}~\bibnamefont {O'Murchadha}}, \ and\ \bibinfo
  {author} {\bibfnamefont {O.}~\bibnamefont {Scholten}},\ }\href@noop {} {\
  (\bibinfo {year} {2015})},\ \Eprint {http://arxiv.org/abs/1503.02808}
  {arXiv:1503.02808 [astro-ph.HE]} \BibitemShut {NoStop}%
\bibitem [{\citenamefont {Ahlers}\ \emph {et~al.}(2010)\citenamefont {Ahlers}
  \emph {et~al.}}]{Ahlers2010}%
  \BibitemOpen
  \bibfield  {author} {\bibinfo {author} {\bibfnamefont {M.}~\bibnamefont
  {Ahlers}} \emph {et~al.},\ }\href {\doibase
  10.1016/j.astropartphys.2010.06.003} {\bibfield  {journal} {\bibinfo
  {journal} {Astroparticle Physics}\ }\textbf {\bibinfo {volume} {34}},\
  \bibinfo {pages} {106 } (\bibinfo {year} {2010})}\BibitemShut {NoStop}%
\bibitem [{\citenamefont {Bai}\ \emph {et~al.}(2009)\citenamefont {Bai},
  \citenamefont {Chirkin}, \citenamefont {Gaisser}, \citenamefont {Stanev},\
  and\ \citenamefont {Seckel}}]{Bai2009}%
  \BibitemOpen
  \bibfield  {author} {\bibinfo {author} {\bibfnamefont {X.}~\bibnamefont
  {Bai}}, \bibinfo {author} {\bibfnamefont {D.}~\bibnamefont {Chirkin}},
  \bibinfo {author} {\bibfnamefont {T.}~\bibnamefont {Gaisser}}, \bibinfo
  {author} {\bibfnamefont {T.}~\bibnamefont {Stanev}}, \ and\ \bibinfo {author}
  {\bibfnamefont {D.}~\bibnamefont {Seckel}},\ }in\ \href@noop {} {\emph
  {\bibinfo {booktitle} {{31st International Cosmic Ray Conference (ICRC 2009)
  Lodz, Poland, July 7-15, 2009}}}}\ (\bibinfo {year} {2009})\BibitemShut
  {NoStop}%
\bibitem [{\citenamefont {Bancroft}(1985)}]{Bancroft1985}%
  \BibitemOpen
  \bibfield  {author} {\bibinfo {author} {\bibfnamefont {S.}~\bibnamefont
  {Bancroft}},\ }\href {\doibase 10.1109/TAES.1985.310538} {\bibfield
  {journal} {\bibinfo  {journal} {Aerospace and Electronic Systems, IEEE
  Transactions on}\ }\textbf {\bibinfo {volume} {AES-21}},\ \bibinfo {pages}
  {56} (\bibinfo {year} {1985})}\BibitemShut {NoStop}%
\bibitem [{\citenamefont {Press}(2007)}]{NR2007}%
  \BibitemOpen
  \bibfield  {author} {\bibinfo {author} {\bibfnamefont {W.~H.}\ \bibnamefont
  {Press}},\ }\href@noop {} {\emph {\bibinfo {title} {{Numerical recipes 3rd
  edition: The art of scientific computing}}}}\ (\bibinfo  {publisher}
  {Cambridge University Press, New York},\ \bibinfo {year} {2007})\BibitemShut
  {NoStop}%
\bibitem [{Eig(2014)}]{Eigen2014}%
  \BibitemOpen
  \href@noop {} {\enquote {\bibinfo {title} {Eigen3 documentation},}\ }\bibinfo
  {howpublished} {\url{http://eigen.tuxfamily.org}} (\bibinfo {year}
  {2014})\BibitemShut {NoStop}%
\bibitem [{\citenamefont {Gorham}\ \emph {et~al.}(2010)\citenamefont {Gorham}
  \emph {et~al.}}]{Gorham2010}%
  \BibitemOpen
  \bibfield  {author} {\bibinfo {author} {\bibfnamefont {P.~W.}\ \bibnamefont
  {Gorham}} \emph {et~al.},\ }\href {\doibase 10.1103/PhysRevD.82.022004}
  {\bibfield  {journal} {\bibinfo  {journal} {Phys. Rev. D}\ }\textbf {\bibinfo
  {volume} {82}},\ \bibinfo {pages} {022004} (\bibinfo {year}
  {2010})}\BibitemShut {NoStop}%
\bibitem [{\citenamefont {Kravchenko}\ \emph {et~al.}(2012)\citenamefont
  {Kravchenko} \emph {et~al.}}]{RICE2011}%
  \BibitemOpen
  \bibfield  {author} {\bibinfo {author} {\bibfnamefont {I.}~\bibnamefont
  {Kravchenko}} \emph {et~al.},\ }\href {\doibase 10.1103/PhysRevD.85.062004}
  {\bibfield  {journal} {\bibinfo  {journal} {Phys. Rev. D}\ }\textbf {\bibinfo
  {volume} {85}},\ \bibinfo {pages} {062004} (\bibinfo {year}
  {2012})}\BibitemShut {NoStop}%
\bibitem [{\citenamefont {Aab}\ \emph {et~al.}(2015)\citenamefont {Aab} \emph
  {et~al.}}]{Aab:2015kma}%
  \BibitemOpen
  \bibfield  {author} {\bibinfo {author} {\bibfnamefont {A.}~\bibnamefont
  {Aab}} \emph {et~al.} (\bibinfo {collaboration} {Pierre Auger}),\ }\href@noop
  {} {\bibfield  {journal} {\bibinfo  {journal} {Phys.Rev.D}\ } (\bibinfo
  {year} {2015})},\ \Eprint {http://arxiv.org/abs/1504.05397} {arXiv:1504.05397
  [astro-ph.HE]} \BibitemShut {NoStop}%
\bibitem [{\citenamefont {Aartsen}\ \emph {et~al.}(2014)\citenamefont {Aartsen}
  \emph {et~al.}}]{Aartsen:2014gkd}%
  \BibitemOpen
  \bibfield  {author} {\bibinfo {author} {\bibfnamefont {M.~G.}\ \bibnamefont
  {Aartsen}} \emph {et~al.} (\bibinfo {collaboration} {IceCube}),\ }\href
  {\doibase 10.1103/PhysRevLett.113.101101} {\bibfield  {journal} {\bibinfo
  {journal} {Phys. Rev. Lett.}\ }\textbf {\bibinfo {volume} {113}},\ \bibinfo
  {pages} {101101} (\bibinfo {year} {2014})},\ \Eprint
  {http://arxiv.org/abs/1405.5303} {arXiv:1405.5303 [astro-ph.HE]} \BibitemShut
  {NoStop}%
\bibitem [{\citenamefont {Kotera}\ \emph {et~al.}(2011)\citenamefont {Kotera}
  \emph {et~al.}}]{Kotera2011}%
  \BibitemOpen
  \bibfield  {author} {\bibinfo {author} {\bibfnamefont {K.}~\bibnamefont
  {Kotera}} \emph {et~al.},\ }\href {\doibase
  10.1146/annurev-astro-081710-102620} {\bibfield  {journal} {\bibinfo
  {journal} {Annual Review of Astronomy and Astrophysics}\ }\textbf {\bibinfo
  {volume} {49}},\ \bibinfo {pages} {119} (\bibinfo {year} {2011})},\ \Eprint
  {http://arxiv.org/abs/http://dx.doi.org/10.1146/annurev-astro-081710-102620}
  {http://dx.doi.org/10.1146/annurev-astro-081710-102620} \BibitemShut
  {NoStop}%
\bibitem [{\citenamefont {Feldman}\ and\ \citenamefont
  {Cousins}(1998)}]{Feldman1998}%
  \BibitemOpen
  \bibfield  {author} {\bibinfo {author} {\bibfnamefont {G.~J.}\ \bibnamefont
  {Feldman}}\ and\ \bibinfo {author} {\bibfnamefont {R.~D.}\ \bibnamefont
  {Cousins}},\ }\href {\doibase 10.1103/PhysRevD.57.3873} {\bibfield  {journal}
  {\bibinfo  {journal} {Phys. Rev. D}\ }\textbf {\bibinfo {volume} {57}},\
  \bibinfo {pages} {3873} (\bibinfo {year} {1998})}\BibitemShut {NoStop}%
\bibitem [{\citenamefont {Conrad}\ \emph {et~al.}(2003)\citenamefont {Conrad},
  \citenamefont {Botner}, \citenamefont {Hallgren},\ and\ \citenamefont {Perez
  de~los Heros}}]{Conrad2002}%
  \BibitemOpen
  \bibfield  {author} {\bibinfo {author} {\bibfnamefont {J.}~\bibnamefont
  {Conrad}}, \bibinfo {author} {\bibfnamefont {O.}~\bibnamefont {Botner}},
  \bibinfo {author} {\bibfnamefont {A.}~\bibnamefont {Hallgren}}, \ and\
  \bibinfo {author} {\bibfnamefont {C.}~\bibnamefont {Perez de~los Heros}},\
  }\href {\doibase 10.1103/PhysRevD.67.012002} {\bibfield  {journal} {\bibinfo
  {journal} {Phys. Rev.}\ }\textbf {\bibinfo {volume} {D67}},\ \bibinfo {pages}
  {012002} (\bibinfo {year} {2003})},\ \Eprint
  {http://arxiv.org/abs/hep-ex/0202013} {arXiv:hep-ex/0202013 [hep-ex]}
  \BibitemShut {NoStop}%
\bibitem [{\citenamefont {Hill}(2003)}]{Hill2003}%
  \BibitemOpen
  \bibfield  {author} {\bibinfo {author} {\bibfnamefont {G.~C.}\ \bibnamefont
  {Hill}},\ }\href {\doibase 10.1103/PhysRevD.67.118101} {\bibfield  {journal}
  {\bibinfo  {journal} {Phys. Rev.}\ }\textbf {\bibinfo {volume} {D67}},\
  \bibinfo {pages} {118101} (\bibinfo {year} {2003})},\ \Eprint
  {http://arxiv.org/abs/physics/0302057} {arXiv:physics/0302057 [physics]}
  \BibitemShut {NoStop}%
\bibitem [{\citenamefont {Barwick}\ \emph {et~al.}(2005)\citenamefont
  {Barwick}, \citenamefont {Besson}, \citenamefont {Gorham},\ and\
  \citenamefont {Saltzberg}}]{Barwick2005}%
  \BibitemOpen
  \bibfield  {author} {\bibinfo {author} {\bibfnamefont {S.}~\bibnamefont
  {Barwick}}, \bibinfo {author} {\bibfnamefont {D.}~\bibnamefont {Besson}},
  \bibinfo {author} {\bibfnamefont {P.}~\bibnamefont {Gorham}}, \ and\ \bibinfo
  {author} {\bibfnamefont {D.}~\bibnamefont {Saltzberg}},\ }\href {\doibase
  10.3189/172756505781829467} {\bibfield  {journal} {\bibinfo  {journal}
  {Journal of Glaciology}\ }\textbf {\bibinfo {volume} {51}},\ \bibinfo {pages}
  {231} (\bibinfo {year} {2005})}\BibitemShut {NoStop}%
\bibitem [{\citenamefont {Bahcall}\ and\ \citenamefont
  {Waxman}(2001)}]{WB2001}%
  \BibitemOpen
  \bibfield  {author} {\bibinfo {author} {\bibfnamefont {J.}~\bibnamefont
  {Bahcall}}\ and\ \bibinfo {author} {\bibfnamefont {E.}~\bibnamefont
  {Waxman}},\ }\href {\doibase 10.1103/PhysRevD.64.023002} {\bibfield
  {journal} {\bibinfo  {journal} {Phys. Rev. D}\ }\textbf {\bibinfo {volume}
  {64}},\ \bibinfo {pages} {023002} (\bibinfo {year} {2001})}\BibitemShut
  {NoStop}%
\bibitem [{\citenamefont {Varner}\ \emph {et~al.}(2009)\citenamefont {Varner}
  \emph {et~al.}}]{IRS_spec}%
  \BibitemOpen
  \bibfield  {author} {\bibinfo {author} {\bibfnamefont {G.~S.}\ \bibnamefont
  {Varner}} \emph {et~al.},\ }\href
  {http://www.phys.hawaii.edu/~varner/IRS_spec_v01.pdf} {\enquote {\bibinfo
  {title} {{Specifications for the IceRay Sampler (IRS) ASIC}},}\ }\bibinfo
  {howpublished} {\url{http://www.phys.hawaii.edu/~varner/IRS_spec_v01.pdf}}
  (\bibinfo {year} {2009})\BibitemShut {NoStop}%
\end{thebibliography}%

\end{document}